\RequirePackage{fix-cm}
\documentclass[smallextended]{svjour3}       
\smartqed  
\usepackage{graphicx}
\usepackage{amsfonts}

\newtheorem{algorithm}{Algorithm}

\usepackage{mathptmx}      

\begin{document}

\title{Evaluation of Gaussian approximations for data assimilation in reservoir models
}


\author{Marco A. Iglesias    \and
     Kody J.H. Law \and Andrew M. Stuart.
}


\institute{A.M. Stuart \at
              University of Warwick \\
              Tel.: +44 (0)24 7652 2685\\
             \email{A.M.Stuart@warwick.ac.uk}           
           \and
M. A. Iglesias \at
              University of Warwick \\
              Tel.: +44 (0)24 7657 4827\\
             \email{M.A.Iglesias-Hernandez@warwick.ac.uk}   
\and 
K.J.H. Law \at
              University of Warwick \\
              Tel.: +44 (0)24 7652 8332\\
             \email{K.J.H.Law@warwick.ac.uk}   
}

\date{}

\maketitle

\begin{abstract}
The Bayesian framework is the standard approach for data assimilation in reservoir modeling. This framework consists mainly in characterizing the posterior distribution of geologic parameters given a prior distribution and data from the reservoir dynamics. Since the posterior distribution quantifies the uncertainty in the geologic parameters of the reservoir, the characterization of the posterior is fundamental for the optimal management of reservoirs. Unfortunately, due to the large-scale highly-nonlinear properties of standard reservoir models, characterizing the posterior is computationally prohibitive. Instead, more affordable {\em ad hoc} techniques, based on Gaussian approximations, are often used for characterizing the posterior distribution. Evaluating the performance of those Gaussian approximations is typically conducted by assessing their ability at reproducing the truth within the confidence interval provided by the {\em ad hoc} technique under consideration.  This has the disadvantage of mixing-up 
the approximation properties of the history matching algorithm
employed with the information content of the particular observations
used, making it hard to evaluate the effect of the
{\em ad hoc} approximations alone. 

In this paper we propose to numerically assess the performance of standard Gaussian approximations to probe the Bayesian posterior distribution. 
In particular we assess the performance of (i) the linearization around the maximum a posterior estimate, (ii) the randomized maximum likelihood and (iii) standard ensemble Kalman filter-type methods. In order to fully resolve the posterior distribution we implement a state-of-the art MCMC method that scales well with respect to the dimension of the parameter space. Our implementation of the MCMC method provides the gold standard against which to assess the aforementioned Gaussian approximations. We present numerical synthetic experiments where we quantify the capability of each of the {\em ad hoc} Gaussian approximation in reproducing the mean and the variance of the posterior distribution (characterized via MCMC) associated to a data assimilation problem. Both single-phase and two-phase (oil-water) reservoir models are considered so that fundamental differences in the resulting forward operators are highlighted. The main objective of our controlled experiments is to exhibit the substantial discrepancies of the approximation properties of standard {\em ad hoc} Gaussian approximations. Numerical investigations of the type we present here will
lead to greater understanding of the cost-efficient, but
{\em ad hoc}, Bayesian techniques used for data assimilation in petroleum reservoirs, and
hence ultimately to improved techniques with more
accurate uncertainty quantification.

\end{abstract}

\section{Introduction}\label{Introduction}
Simulating the dynamics of a reservoir involves solving a large-scale numerical model that depends on parameters related to petrophysical properties of the reservoir. These properties need to be known at each discretization point of the physical domain of the reservoir. Unfortunately, direct measurements of petrophysical properties are only available at a small number of locations within the domain of interest. Therefore, a statistical description of the reservoir parameters is required to properly account for the uncertainty in the petrophsyical properties caused by the lack of information. A \textit{prior distribution} of geologically consistent reservoir parameters can be generated, for example, from a variogram analysis conducted on static data from core samples. Geostatistical techniques can then be used to generate realizations of reservoir parameters conditioned to static data \cite{Geos}. In some cases, information concerning geologic facies may also be incorporated \cite{Caers}. On the other hand, with the aid of downhole permanent sensors, measurements of reservoir flow can be continuously acquired. In a Bayesian framework, these flow measurements, the reservoir model and the prior distribution of the petrophysical properties are combined to characterize the \textit{posterior distribution} of the reservoir parameters given dynamic (flow) data. This posterior distribution quantifies the uncertainty in the reservoir predictions and it is essential for assessing the economical and environmental risk of oil recovery procedures.

Markov Chain Monte Carlo (MCMC) methods are the standard techniques for sampling the posterior distribution described above. In particular, the Metropolis- Hasting variant of MCMC has been typically used for data assimilation in reservoir models \cite{Barker,Emerick,Ma,Oliver1,Oliver2,Efendiev1,Emerick2}. In general, the posterior distribution that arises from Bayesian data assimilation does not admit a finite-dimensional parametrization, with the exception of very few particular cases such as the linear and Gaussian case. Therefore, strictly speaking, an infinite number of samples are required to define it. This implies, in practice, that hundreds of thousands or even millions of reservoir simulations may be required for standard MCMC methods to accurately characterize the posterior distribution. This computational disadvantage of standard MCMC approaches has given rise to the development of more computationally efficient MCMC techniques \cite{David,Efendiev1,Ma,Emerick2}. With increasing advance of computational power, the aforementioned MCMC methods may potentially become viable tools for reservoir management in the decades to come. At the present time, however, it is essential to develop techniques that provide a reasonable characterization of the posterior with a computational cost that involves a limited number of reservoir model runs. It then comes as no surprise that, in the previous years, research on data assimilation for uncertainty quantification (UQ) applications has mainly focused on improving the efficiency and accuracy of \textit{ad-hoc} ensemble-based techniques that provide approximations of the posterior distribution based on Gaussian assumptions.

As pointed out in the recent literature review of \cite{OliverReview}, there are three main approaches that have been consistently adopted for sampling approximations of the posterior distribution: (i) linearization around the maximum a posteriori (MAP) estimate (LMAP), (ii) randomized maximum likelihood (RML) method and (iii) ensemble Kalman filter (EnKF). Under a Gaussian prior and a linear model, it can be shown that all these techniques provide samples of the posterior distribution  \cite{Oliver,IterativeEnKF}. We therefore refer to those methods as \textit{Gaussian approximations} of the posterior. For standard (nonlinear) reservoir models, the mathematical structure of the approximation provided by the three approaches is still unknown. Nonetheless, the aforementioned methods are widely applied for generating model parameters conditioned to dynamics data, which are then used for statistical analysis of reservoir performance. Consequently, in the Bayesian framework, optimal decision-making and risk management depend on  the quality of the underlying Gaussian approximations of the posterior. It is therefore, fundamental to understand the accuracy and convergence properties of those Gaussian approximations in order to interpret predictions of
uncertainty made using them, and in order to develop improved methodologies
from them. Rigorous numerical studies of these Gaussian approximate
algorithms can shed light on these issues and can point us towards 
theoretical analyses. In this paper we therefore provide a numerical evaluation of the performance of the {\em ad hoc} Gaussian
approximate algorithms LMAP, RML, and variants of the EnKF 
methodology, by using an expensive, but full resolved, MCMC simulation
as our gold standard. This approach is analogous to the recent study of similar Gaussian approximate algorithms arising in the context of atmospheric data assimilation \cite{lawstuart}.

\subsection{ Literature review}
Although the theoretical aspects of the approximation properties of LMAP, RML and EnKF are unknown for the case of nonlinear models, some numerical investigations have been performed \cite{Oliver2,Barker}. To our best knowledge, only the work in \cite{Oliver2} provides an evaluation of approximate methods for sampling the posterior. To accomplish this goal, Liu et-al \cite{Oliver2} use a standard random walk MCMC method to generate accurate samples of the posterior. These, in turn, are used as gold-standard against to which compare the performance of the approximate methods: LMAP, RML and pilot point methods. In their evaluation,  Liu et-al use synthetic data from a single-phase one-dimensional reservoir discretized with 20 gridblocks. The main conclusion of \cite{Oliver2} is that RML provides the best uncertainty quantification when compared against the MCMC gold standard. In particular, RML outperforms pilot point methods whose application to reservoir data assimilation problems has been lately abandoned. 

Within the context of evaluating the uncertainty quantification properties of data assimilation techniques, it is relevant to mention the work of \cite{Barker} where several methods were compared for the synthetic PUNQ-S3 reservoir model. The main goal of \cite{Barker}, however, is to evaluate the ability of the corresponding techniques to provide confidence of interval that contain the truth estimate. Among those techniques, MCMC is the only one that can potentially provide accurate samples from the Bayesian posterior. Unfortunately, as stated in \cite{Barker}, the MCMC results are not conclusive due to the small chain used for their experiments. Thus, the evaluation in \cite{Barker} does not provide an evaluation in the strict sense of a Bayesian framework. It is worth mentioning that the work of \cite{Barker} and \cite{Oliver2} appeared almost a decade ago when EnKF had just been introduced to the history matching community \cite{EnKFReview}, and so EnKF was not assessed in \cite{Oliver2,Barker}. Almost a decade later and after hundreds of publications, EnKF is now perhaps the only computationally feasible technique for real-time data assimilation in petroleum reservoirs. For a comprehensive review of EnKF for reservoir applications, we refer the reader to \cite{EnKFReview}.

In the recent work of \cite{Emerick2}, the EnKF is combined with an MCMC algorithm to improve the efficiency of standard MCMC methods for sampling the posterior in a Bayesian framework. Although the analysis of the approximate properties of EnKF is not the main goal of \cite{Emerick2}, an implicit evaluation of EnKF is displayed.  Indeed, under the assumption that the MCMC samples of \cite{Emerick2} provide an accurate characterization of the posterior, then \cite{Emerick2} provides a partial assessment of EnKF  for approximating the posterior. While the aforementioned work exposes severe limitations of the EnKF for sampling the posterior, an evaluation with respect to other approximate methods remains nonexistent. 

In the context of evaluating the uncertainty quantification properties of Gaussian approximations of the posterior, we highlight the work of \cite{Reynolds,svdRML}. For the PUNQ-S3 model mentioned above, \cite{Reynolds} compares the performance of EnKF and RML. In \cite{svdRML}, a new SVD-based RML is introduced and compared against EnKF. It is important to mention, however, that \cite{Reynolds,svdRML} evaluate the capability of these Gaussian approximations for reproducing the truth within the confidence of interval of the technique under consideration. While this is a natural strategy for assessing uncertainty quantification properties, it is an insufficient evaluation from the perspective of the Bayesian framework. In other words, capturing the truth within the spread of model predictions obtained with a Gaussian approximation does not ensure that the spread correctly represents the uncertainty quantified by the posterior distribution of Bayesian data assimilation. It is therefore essential to develop a controlled experiment where standard Gaussian approximation can be tested against the solution to the Bayesian data assimilation problem: the posterior distribution.

\subsection{The proposed work}
In this paper we propose the numerical evaluation of LMAP, RML and some standard versions of ensemble Kalman filter-type methods for approximating the posterior distribution within the Bayesian framework of data assimilation. We characterize the posterior distribution by using a state-of-the art MCMC method that provides a gold-standard against which to compare the aforementioned Gaussian approximations. In this sense, our work has a similar goal to the one of \cite{Oliver2}. However, there are two recent algorithmic developments
which motivate our desire to revisit the perspective introduced
in \cite{Oliver2}. The first, discussed in detail below, is that
MCMC methodology has evolved significantly, enabling the study of
considerably more sophisticated forward models and more
high dimensional parameterizations of the unknown petro-physical
quantities, leading to greater realism. The second new aspect of our work is the assessment of ensemble Kalman filter-type methods. In particular we consider the most standard EnKF implementations, namely:
 (i) the perturbed observation EnKF; and (ii) the square root filter EnSRF of \cite{Sakov,Emerick}. In both cases we also evaluate the effect of performing distance-based localization \cite{OliverLocal,Emerick3}. 

We emphasize that, in contrast to other approaches \cite{Reynolds,svdRML,Barker,OliverLocal,Emerick3} where the aim is to recover the truth within the confidence interval of relevant quantities and to history-match data, here we are interested in assessing the performance for characterizing the posterior distribution. Evaluation of algorithms
by their ability to recover the truth within
a confidence interval has the disadvantage of entangling 
the approximation properties of the history matching algorithm
employed with the information content of the particular observations
used, making it hard to evaluate the effect of the
{\em ad hoc} approximations alone. Our assessment of the
ability to probe the Bayesian posterior distribution is conducted by quantifying the capability of each Gaussian approximation in reproducing the mean and the variance of the posterior distribution associated to a data assimilation problem. Two prototypical reservoir models are used, both in two spatial dimensions: (i) slightly compressible single-phase Darcy flow model and (ii) incompressible oil-water reservoir model. In both models, the unknown is the logarithm of the absolute permeability of the reservoir $u = \log{K}$. For the single-phase model, pressure data is collected from production wells. For the oil-water model, total flow rate is measured at the production wells while bottom-hole pressure is collected at the injection wells. For both models considered here, the corresponding parameter-to-output map $G(u)$ is nonlinear. Thus, even when the prior distribution is Gaussian, the Bayesian posterior is non-Gaussian. This constitutes the ideal scenario to evaluate approximation properties of the techniques of interest, provided that a gold standard is obtained from accurately sampling the posterior as we describe below.

As we indicated earlier, some MCMC methods have been used for sampling the Bayesian posterior distribution in reservoir models. However, some of these methods \cite{Efendiev1,Ma} rely on reducing the parameter space (e.g. via truncating Karhunen-Loeve expansion) and/or upscaling the model to reduce the computational cost of the algorithm. In the present experiment, however, we are interested in the more general case where no reduction of the parameter space is possible. In other words, we assume that the petrophysical property is unknown at each at the location of the physical domain of the reservoir. In this case, a standard MCMC technique like the one used in \cite{Oliver2} is computationally prohibitive for larger size problems like the ones considered here. To overcome this difficulty, we take advantage of
recent developments in MCMC methodology and sample the posterior by applying the  preconditioned Crank-Nicolson (pCN) MCMC method described in \cite{David}, and derived from the infinite-dimensional Bayesian framework developed in \cite{Andrew}.
In contrast to standard MCMC methods, the acceptance probability in the pCN-MCMC method is invariant with respect to the dimension of the parameter space, therefore making pCN-MCMC ideal for large-scale problems like the one studied here. The advantage of using pCN-MCMC over standard MCMC for data assimilation in some geophysical problems has been shown in \cite{David}. For petroleum reservoir applications, the computational efficiency of pCN-MCMC with respect to other existing methods deserves further investigation. Nevertheless, in the present work we apply pCN-MCMC and provide numerical evidence of convergence so that the corresponding realizations generated with pCN-MCMC are samples from the Bayesian posterior. These, in turn, provide a gold-standard against to which compare LMAP, RML and standard versions of ensemble Kalman filter-type of methods. The proposed numerical evaluation of the approximation techniques has two concrete goals: (i) assess the capability to recover the mean and variance of the posterior and (ii) evaluate the performance for reproducing the uncertainty (quantified by the posterior) in the reservoir model predictions. 

In Section \ref{ReservoirModels} we describe the prototypical reservoir models that define the forward operators that we use for data assimilation. The Bayesian framework for data assimilation as well as the MCMC methodology for sampling the posterior are introduced in Section \ref{BayesianFramework}. Methodologies based on Gaussian approximations of the posterior are described in Section \ref{Gaussian}. In Section \ref{NumericalResults} we report and discuss the numerical results and comparisons of our synthetic experiments. The summary and final remarks are presented in Section \ref{Conclusions}.

\section{ Forward Reservoir Models}\label{ReservoirModels}

In this section we briefly outline the forward (reservoir) models that we use for the evaluation of Gaussian approximations of the posterior. On the one hand, we consider simplified two-dimensional models for which a forward model run is computationally inexpensive and therefore feasible for the highly computationally challenging MCMC method. On the other hand, by sharing the mathematical structure of more sophisticated models, the models we describe below are ideal for prototyping and evaluating performance in a controlled fashion.
For each of the following models, we consider a two-dimensional reservoir whose physical domain, absolute permeability and porosity are denoted by $D$, $K$ and $\phi$ respectively. The interval $[0,T]$ ($T > 0$) is the time interval of interest for the flow simulation. For each reservoir model, we define the forward operator $G:X\to \mathbb{R}^{N}$ that maps the parameter space $X$ into the observation space $\mathbb{R}^{N}$. In other words, $G(u)$ is the model predictions corresponding to the parameter $u \in X$. For simplicity we assume that the only unknown parameter is $u=\log{K}$. Nevertheless, all the techniques and implementations that we describe in subsequent sections can be extended to include additional parameters, such as porosity which is routinely estimated alongside permeability in many practical scenarios.

\subsection{Single-phase Darcy flow}
We consider a single-phase reservoir where oil is produced at $N_w$ production wells operated under prescribed production rates $\{q^l(t)\}_{l=1}^{N_{w}}$ ($t\in [0,T]$). The flow in the reservoir is described in terms of the (state variable) fluid pressure $p(x,t)$ ($(x,t) \in D\times [0,T]$) which is governed by the following equation \cite{Chen}
\begin{eqnarray}\label{eq:2.1}
c\phi \frac{\partial p}{\partial t} -\nabla \cdot e^{u}\nabla p=\sum_{l=1}^{N_{w}}q^l\delta (x-x^l)\qquad \textrm{in }  D\times(0,T]
\end{eqnarray}
where $u\equiv  \log{K}$, $c$ is the total compressibility and $\delta (x-x^l)$ is (a possibly mollified) Dirac delta centered at the $l$-th well with location denoted by $x^l$. In addition, we consider the following boundary and initial conditions
\begin{eqnarray}\label{eq:2.2}
- e^{u}\nabla p\cdot \mathbf{n}&=&0 ~~~~~~~~~~\textrm{on }  \partial D\times (0,T],\\
p&=&p_0~~~~~~~~~~\textrm{in }   D\times\{0\}. \label{eq:2.3}
\end{eqnarray}
As we indicated earlier, the only uncertain parameter in (\ref{eq:2.1})-(\ref{eq:2.3}) is the log-permeability $u$. Therefore, the additional model parameters $c$, $\phi$, $\nu$, $p_0$ and the geometry $D$ in (\ref{eq:2.1})-(\ref{eq:2.3}) are prescribed. 

In order to construct the forward operator, we first define the model predictions of measurements. Let us then assume that $N_{M}$ measurements of pressure from wells are collected at times $t_1 ,\dots, t_M$. We define the measurement functional 
\begin{eqnarray}\label{eq:2.4}
M_n^l(p)=p(x^l,t_{n})
\end{eqnarray}
that corresponds to the fluid pressure at time $t_n$ and well location $x^l$. For the exposition of subsequent sections we also define the vector
\begin{eqnarray}\label{eq:2.5}
M_n(p)=(M_n^1(p), \dots ,M_n^{N_{w}}(p)).
\end{eqnarray}
We finally define $N=N_{w}N_{M}$, i.e. the total number of observations from wells, and construct the forward operator
\begin{eqnarray}\label{eq:2.6}
G(u)=(M_1(p), \dots, M_{N_{M}}(p))
\end{eqnarray}
Note that $p$ in (\ref{eq:2.6}) depends on $u$ via (\ref{eq:2.1})-(\ref{eq:2.3}). 

\subsection{Oil-water reservoir model}
We consider an oil-water reservoir model initially saturated with oil and irreducible
water. Let us index by $\gamma= w$ and $\gamma = o$ the water and oil phase respectively. We
assume that both fluids and the rock are incompressible. We are interested in a waterflood process where water is injected at $N_I$ injection wells located at $\{x_I^{l}\}_{l=1}^{N_{I}}$ . Water and oil are produced at $N_P$ production wells located at$\{x_P^{l}\}_{l=1}^{N_{P}}$. Additionally, we assume that injection wells are operated under prescribed rates $\{q^l(t)\}_{l=1}^{N_I}$ while production wells are constrained to bottom hole pressure denoted by $\{P_{bh}^{l}(t)\}_{l=1}^{N_P}$  . The reservoir dynamics in $[0,T]$ are described by the (state variables) water saturation and the pressure denoted by $s(x,t)$ and $p(x,t)$ respectively ($(x, t ) \in D \times [0, T ]$). From standard arguments it can be shown that $(s, p)$ is the solution to the following system \cite{Chen}
\begin{eqnarray}\label{eq:2.7}
-\nabla \cdot \lambda(s) e^{u}\nabla p= \sum_{l=1}^{N_{I}}q^l \delta(x-x_I^l)+\sum_{l=1}^{N_{w}}\omega^l\lambda(s)[P_{bh}^{l}-p]\delta(x-x_P^l),\\
\phi \frac{\partial s}{\partial t} -\nabla \cdot \lambda_w(s) e^{u}\nabla p= \sum_{l=1}^{N_{I}}q^l \delta (x-x_I^l)+\sum_{l=1}^{N_{w}}\omega^l\lambda_w(s)[P_{bh}^{l}-p]\delta(x-x_P^l),\label{eq:2.7B}
\end{eqnarray}
in $D\times(0,T]$, where $\delta(x-x_P^l)$ and $\delta(x-x_I^l)$ are the (possibly mollified) Dirac deltas as defined before, $\{\omega^l\}_{l=1}^{N_{P}}$ are constants related to the well model \cite{Chen}. 
 Additionally $\lambda_w(s)$ and $\lambda(s)$ denote the water and total mobility defined by
\begin{eqnarray}\label{eq:2.8}
\lambda_w(s)=\frac{k_{rw}(s)}{\mu_{w}},\qquad \lambda(s)=\frac{k_{ro}(s)}{\mu_{o}}+\lambda_w(s)
\end{eqnarray}
where $k_{r\gamma}(s)$ and $\mu_{\gamma}$ denote the relative permeability and the viscosity of the $\gamma$-phase fluid, respectively. Furthermore, we assume that
\begin{eqnarray}
k_{rw}(s)=a_{w}\Bigg[\frac{s-s_{iw}}{1-s_{iw}-s_{or}}\Bigg]^2,\qquad 
k_{ro}(s)=a_{o}\Bigg[\frac{1-s-s_{or}}{1-s_{iw}-s_{or}}\Bigg]^2
\end{eqnarray}
where $a_{w},a_{o}\in (0,1]$, $s_{iw}$ is the irreducible water saturation and $s_{or}$ is the residual oil saturation. We
additionally prescribe initial conditions for pressure and water saturation
\begin{eqnarray}\label{eq:2.9}
p=p_0, \qquad s=s_{0}\qquad \textrm{in }   D\times\{0\} 
\end{eqnarray}
For simplicity, no-flow boundary conditions are prescribed on the reservoir boundary
\begin{eqnarray}
- e^{u}\lambda(s)\nabla p\cdot \mathbf{n}&=&0 ~~~~~~~~~~\textrm{on }  \partial D\times (0,T],\\
- e^{u}\lambda_w(s)\nabla p\cdot \mathbf{n}&=&0 ~~~~~~~~~~\textrm{on }  \partial D\times (0,T].\label{eq:2.10}
\end{eqnarray}
Let us assume that there are $N_{M}$ measurement times denoted as before $\{t_{n}\}_{n=1}^{N_{M}}$. We assume measurements of bottom-hole pressure are collected at the injection wells at $\{t_{n}\}_{n=1}^{N_{M}}$. This, according to Peaceman's well-model \cite{Chen}, is defined by 
\begin{eqnarray}\label{eq:2.11}
M_{n}^{l,I}(p,s)=P_{bh}^{l,I}(t_{n})=\Bigg[ \frac{q^l(t_{n})}{\omega^{l}\lambda(s(x_{I}^{l},t_{n}))}+p(x_{I}^{l},t_{n})\Bigg]
\end{eqnarray}
for $l=1,\dots,N_{I}$ and $n=1,\dots, N_M$. Analogously, we consider measurements of total flow rate at the
production wells
\begin{eqnarray}\label{eq:2.12}
M_{n}^{l,P}(p,s)=q^{l,P}(t_{n})=\omega^{l}\lambda(s(x_{P}^l,t_{n}))[P_{bh}^{l}(t_{n})-p(x_{P}^l,t_{n})]
\end{eqnarray}
for $l=1,\dots, N_{P}$ and $n=1,\dots, N_M$. Let us denote by $N_{w}=N_{P}+N_{I}$ the total number of wells, and define the $N_{w}$-dimensional vector 
\begin{eqnarray}\label{eq:2.13}
M_{n}(p,s)=(M_{n}^{1,I}(p,s), \dots, M_{n}^{N_{I},I}(p,s),M_{n}^{1,P}(p,s), \dots, M_{n}^{N_{P},P}(p,s))
\end{eqnarray}
The total number of measurements $N$ is defined as before and the forward map $G:X\to \mathbb{R}^{N}$ is then given by 
expression
\begin{eqnarray}\label{eq:2.14}
G(u)=(M_1(p,s), \dots, M_{N_{M}}(p,s))
\end{eqnarray}
which in this case comprises the production data obtained from production and injection wells at the measurement times.

With both forward models written in terms of the forward operator $G$, in the next section we describe the Bayesian inverse problem of finding $u$ given noisy observations of $G(u)$.

\section{The Bayesian framework}\label{BayesianFramework}

We assume that the unknown parameter $u$ and the 
data $y\in Y$ are related by
\begin{eqnarray}\label{eq:3.1}
y=G(u)+\eta
\end{eqnarray}
where $G$ is the forward map introduced in the previous section, and $\eta\in \mathbb{R}^{N}$ is a vector of random noise. 
Informally the Bayesian approach to inversion proceeds by
placing a {\em prior} probability distribution ${\mathbb P}(u)$ 
on $u$ and assuming an independent probability distribution 
on $\eta$. The {\em likelihood}, namely the probability of
the observed data $y$ given a particular instance of the
unknown parameter $u$, is then denoted ${\mathbb P}(y|u).$ 
Bayes' rule then states that the {\em posterior} probability
of the unknown parameter $u$, given the observed data $y$,
denoted by ${\mathbb P}(u|y)$, is determined by
the formula
\begin{equation}
\label{eq:B}
\frac{{\mathbb P}(u|y)}{{\mathbb P}(u)} \propto {\mathbb P}(y|u).
\end{equation}
In this section we formulate this precisely in the case
where the unknown parameter is a function.

We define the norm $\vert\vert \cdot \vert\vert_{B} = \vert\vert B^{-1/2}(\cdot)\vert\vert$ for any covariance operator $B$ and we use this notation throughout the paper, in particular in the observation space, with $B = \Gamma$, and in the log-permeability space with $B = C$. For simplicity and following convention in the field, we will not distinguish notationally between the random variable and its realization, except in the case of the truth, which will be important to distinguish by $u^{\dagger}$ in subsequent sections in which it will be prescribed and known.

\subsection{An infinite-dimensional Bayesian framework}

We are interested in the inverse problem of characterizing the posterior distribution of the unknown log-permeability {\em function} $u$ 
given {\em finite-dimensional} observational data denoted by $y$. We approach this inverse problem by means of the infinite-dimensional Bayesian framework of \cite{Andrew} that we now briefly describe. Assume that $u\in X$ where $X$ is a Hilbert space, and denote by $\mu_0$ the prior probability measure on $u$. 
We assume that the unknown $u$ and the data $y\in Y$ are related by
(\ref{eq:3.1}). For simplicity, we assume that $\eta\sim N(0,\Gamma)$. Then, the rigorous interpretation of (\ref{eq:B})
is that the posterior distribution on $u\vert y$ is given by
measure $\mu$ satisfying 
\begin{eqnarray}\label{eq:3.3}
\frac{d\mu}{d\mu_{0}}(u) =\frac{\exp(-\Phi(u,y))}{\int_{X}\exp(-\Phi(u,y))\mu_0(du)} 
\end{eqnarray}
where the left hand side of (\ref{eq:3.3}) is the Radon-Nikodym derivative of the posterior distribution $\mu(u) =\mathbb{￼P}(u|y)$ with respect to the prior $\mu_0$ and 
$$\Phi (u, y) = \frac{1}{2} \vert\vert y - G(u)\vert\vert_{\Gamma}^2.$$ 
A sufficient condition for this to be well-defined is that $\Phi(\cdot,y)$ is 
continuous as a mapping from $X$ into ${\mathbb R}$ for each
fixed $y$, and that $\mu_0(X)=1$ so that functions drawn from
$\mu_0$ are in $X$ almost surely. The formula (\ref{eq:3.3})
then holds in infinite-dimensions, exhibiting the posterior 
density with respect to the prior; in practical
terms this means that posterior expectations can be found by
reweighting prior expectations by the right-hand side of
(\ref{eq:3.3}).

The posterior distribution $\mu$ quantifies the uncertainty of the 
logarithm of the absolute permeability given production data, 
normalized by the prior. 
Since $G$ is nonlinear, the posterior is non-Gaussian even when
the prior $\mu_0$ is Gaussian. Thus there is no useful closed-form 
expression for the posterior distribution and it must be 
characterized by means of sampling.

Before we describe the approach for sampling $\mu$, we note that, if we assume the prior $\mu_0$ is Gaussian with mean $\overline{u}$
and covariance $C$, it follows that the maximum a posteriori (MAP) estimate $u_{MAP}$ is the minimizer of the functional
\begin{eqnarray}\label{eq:3.4}
J(u) = \Phi(u,y)+ \frac{1}{2}\vert\vert u-\overline{u}\vert\vert_{C}^2
\end{eqnarray}
The MAP estimator is the typical estimate computed in standard history matching problems where the goal is to recover the truth by fitting historic production data. Note that the Bayesian approach thus
subsumes this classical approach to inversion, whilst also providing
rigorous quantification of uncertainty of predictions, given
clear assumptions on the prior and noise probabilities. 

\subsection{ Sampling the posterior with MCMC}
A  state-of-the-art class of MCMC methods that sample from $\mu$ defined in (\ref{eq:3.3}) has been proposed in \cite{David}. In particular, we consider the following preconditioned Crank-Nicolson (pCN) MCMC \cite[Section 5.2]{David}. 
\begin{algorithm}[pCN-MCMC ]\label{al:MCMC}
Take $u^{(0)}\sim N(\overline{u},C), n=1$, and $\beta\in (0,1)$. Then, 
\begin{itemize}
\item[(1)] pcN proposal. Generate $u$ from 
\begin{eqnarray}\label{eq:3.5}
u=\sqrt{1-\beta^2}u^{(n)}+(1-\sqrt{1-\beta^2})\overline{u}+\beta\xi, \qquad \textrm{with}~~ \xi\sim N(0,C)
\end{eqnarray} 
\item[(2)] Set $u^{n+1}=u$ with probability $a(u^n,u)$ and  $u^{n+1}=u^{n}$ with probability $1-a(u^n,u)$, where
\begin{eqnarray}\label{eq:3.6}
a(u,v)=\min\Big\{1, \exp{(\Phi(u,y)-\Phi(v,y))}\Big\}
\end{eqnarray} 
\item[(3)] $n \mapsto n+1$ and repeat.
\end{itemize}
\end{algorithm}

All the probabilities are generated independently of one another,
leading to a Markov chain which is invariant with respect to $\mu.$
Notice that the small change in proposal, when compared with the standard random walk MCMC \cite{Oliver2}, results in an acceptance probability defined via differences of $\Phi$ and not $J$. Because $\Phi$ is finite with respect to $\mu$, whilst $J$ is not, this leads to a considerably improved algorithm which has desirable $dim(X)$-independent properties when implemented on a sequence of approximating problems with $dim(X)\to \infty$. Therefore, for large $dim(X)$ like the one considered here, pCN-MCMC provides a more robust and efficient technique than the standard MCMC approaches. Numerical evidence of these scaling properties can be found in \cite{David}.

Based on the forward models of Section \ref{ReservoirModels}, the purpose of the present work is to design synthetic experiments for solving the Bayesian data assimilation problem, i.e. finding the posterior distribution. By implementing the pcN-MCMC algorithm, we characterize this posterior and generate a gold standard against to which compare the Gaussian approximations that we introduce in the following section.

\section{Gaussian approximations of the posterior}\label{Gaussian}

In this section we introduce some standard \textit{ad-hoc} methods that use Gaussian approximations to sample the posterior distribution (\ref{eq:3.3}). In particular, we consider LMAP, RML, EnKF and EnSRF which have been typically used for history matching and uncertainty quantification in the Bayesian framework of data assimilation of petroleum reservoirs. While many variants of the aforementioned techniques can be found in the literature \cite{OliverReview}, here we focus on the most standard and typical implementations used for history matching. For each of the aforementioned techniques, the objective of the subsequent description is twofold.  First, we introduce the algorithm and the associated computational cost. Second, we indicate the type of Gaussian approximation made for the definition of the technique under consideration.

￼

\subsection{Linearization around the MAP (LMAP)}

As described in Section \ref{BayesianFramework}, the minimizer of $J$ introduced in (\ref{eq:3.4}) defines the MAP estimator, i.e.
\begin{eqnarray}\label{eq:4.1}
u_{MAP} = {\rm argmin}_{u}\Big\{\Phi(u,y)+ \frac{1}{2}\vert\vert u-\overline{u}\vert\vert_{C}^{2}\Big\}
\end{eqnarray}
￼￼We can further define,
\begin{eqnarray}\label{eq:4.2}
￼￼C_{MAP} =C-CQ^{T}(QCQ^{T}+\Gamma)^{-1}QC
\end{eqnarray}
where $Q \equiv DG(u_{MAP})$ is the Frechet derivative of $G$ evaluated at $u=u_{MAP}$. The linearization around the MAP \cite[Section 10.5]{Oliver} consists of approximating the posterior $\mu$ in (\ref{eq:3.3}) by $\mu\approx N(u_{MAP},C_{MAP})$. 

The LMAP algorithm approximates the posterior with an ensemble of $N_{en}$ realizations from $N(u_{MAP},C_{MAP})$  \cite[Section 10.5]{Oliver}.  This ensemble can then be used to approximate integrals with respect to the posterior of {\it nonlinear} functions of $u$.
Note that when $G$ is linear, $\mu=N(u_{MAP},C_{MAP})$ and then $u_{MAP}$ and $C_{MAP}$ are the mean and covariance of the posterior.  The algorithm, however, is well-defined in general, and may thus be applied to cases in which $G$ is nonlinear.

\begin{algorithm}[LMAP]

\begin{itemize}
\item[(1)] Compute $u_{MAP}$  and $C_{MAP}$ from (\ref{eq:4.1}) and (\ref{eq:4.2})respectively.  
\item[(2)] Compute the Cholesky factor $L$ of $C_{MAP}$, i.e. $C_{MAP }=LL^T$. 
\item[(3)] For $j\in \{1,\dots N_{en}\}$, generate
\begin{eqnarray}\label{eq:4.3}
u^{(j)}=u_{MAP}+L^Tz^{(j) }
\end{eqnarray} 
where $z^{(j)} \sim N(0,I)$.
\end{itemize}
\end{algorithm}
Samples generated by (\ref{eq:4.3}) are draws from $N(u_{MAP},C_{MAP})$ and so the ensemble $\{u^{(j)}\}_{j=1}^{N_{en}}$ provides an approximation to $N(u_{MAP},C_{MAP})$ and, hence the posterior.

The computational cost of LMAP depends on the cost of computing the MAP estimator (\ref{eq:4.1}) and the factorization of $C_{MAP}$. For the present work, we develop implementations of the Levenberg-Marquardt algorithm of \cite[Section 8.4]{Oliver} with the stopping criteria given by (8.82) and (8.83) from \cite[Section 8.5]{Oliver}. It is worth mentioning that, within the context of reservoir characterizations, multiple techniques for computing the MAP estimator have been widely studied (e.g. BFGS, LBFGS, Gauss-Newton) \cite[Section 8]{Oliver}. It is of interest to evaluate the 
optimal minimization technique, but this is beyond the scope of our 
present work.

\subsection{Randomized Maximum Likelihood (RML)}
The RML technique was developed as an attempt to accelerate MCMC methods for sampling the posterior from Bayesian data assimilation in reservoir models \cite{Oliver1}. The main idea of RML is to construct an ensemble of MAP estimators from randomized objective functions (\ref{eq:4.1}). A standard implementation of RML is presented in the following algorithm
\begin{algorithm}[RML]
For $j\in \{1,\dots, N_{e}\}$
\begin{itemize}
\item[(1)] Generate $u^{(j)}\sim N(\overline{u},C)$
\item[(2)] Define $y^{(j)} =y+\eta^{(j)}$ with $\eta^{(j)} \sim N(0,\Gamma)$. 
\item [(3)]Compute
\begin{eqnarray}\label{eq:4.4}
 u_{RML}^{(j)} = {\rm argmin}_{u}\Big\{ \Phi(u,y^{(j)} )+\frac{1}{2}\vert\vert u-u^{(j)}\vert\vert_{C}^{2} \Big\}.
\end{eqnarray}
\end{itemize}
\end{algorithm}
In the case where $G$ is linear, the RML algorithm can be shown to sample the posterior distribution (i.e. from $\mu= N(u_{MAP},C_{MAP})$) \cite{IterativeEnKF}. For the nonlinear case of interest here, the RML algorithm provides an approximation the nature of which, 
to the best of our knowledge, has not been systematically understood. 


Note that for each ensemble member, RML requires the solution to the minimization problem (\ref{eq:4.4}). Nevertheless, since each minimization problem is independent from one another, RML is then embarrassingly parallelizable. Each of those minimization problems has the same structure as the one that we solve for the MAP estimator (\ref{eq:4.1}). For a relatively small problem the computational cost of computing $L$ in LMAP, given $Q$ which has already been constructed while solving (\ref{eq:4.1}), as well as the generation of (\ref{eq:4.3}), are negligible compared to the cost of one forward model evaluation. Thus the computational cost of RML is roughly $N_{e}$ times the computational cost 
of LMAP, although the effect of the multiplier $N_e$ can
be ameliorated in a parallel context. 

Similarly to our implementation of the MAP, for RML we consider the Levenberg-Marquardt method and the corresponding stopping criteria mentioned above. Improving the optimization technique required for (\ref{eq:4.4}) can reduce the overall computational cost of RML. Alternative methods to reduce the computational cost of RML by means of a truncated SVD approach can be found in \cite{svdRML}.

\subsection{Ensemble Kalman Filter (EnKF)}

Ensemble methods based on the Kalman filter have been extensively applied for Bayesian data assimilation in petroleum reservoir applications. For a complete review of most of the EnKF implementations we refer the reader to the monograph of \cite{EnKFReview}. In this section we briefly discuss some relevant aspects of EnKF in the context of history matching of petroleum reservoirs. These ensemble Kalman filter-type of algorithms, make Gaussian approximations in a sequential manner as we describe below. As a result, for the general case, those techniques do not provide correct sampling of the posterior (\ref{eq:3.3}). Nevertheless, due to its ease of implementation and low  computational cost, ensemble Kalman filter-type of methods are arguably the only feasible techniques for online data assimilation in subsurface applications. 

\subsubsection{Introduction and Main Algorithm
}
In order to introduce the algorithms, we first consider a sequential formulation of the reservoir model. In particular, let us define $v_{n}$ the state variable at time $t_{n}$ and $S$ the state space. For example, for the single-phase model of Section \ref{ReservoirModels}, $v_{n}=p(x,t_{n})$. We define the solution operator $\Psi_{n}:S\times X \to S$
\begin{eqnarray}\label{eq:4.5}
v_{n}=\Psi_{n}(v_{n-1},u)
\end{eqnarray}
which, for a given parameter $u$, maps the state variable from time $t=t_{n-1}$ to $t=t_{n}$. In practice, $\Psi_{n}$ is simply the numerical solver that arises from the time discretization of the reservoir model under consideration. In addition, we assume that data is given at each of these points in time and is correlated between times only through the state itself, i.e.
\begin{eqnarray}\label{eq:4.6}
y_{n} =M_{n}(v_{n})+\eta_{n}, 
\end{eqnarray}
where $\eta_{n}\sim N(0,\Gamma_{n})$ and $M_{n}:S\to \mathbb{R}^{N_{w}}$ is the measurement functional acting on the state variable at time $t = t_{n}$. For the models of Section \ref{ReservoirModels}, $M_{n}$ is defined by (\ref{eq:2.5}) and (\ref{eq:2.13}) respectively and $N_{w}$ is the number of total wells. Define,
\begin{eqnarray}\label{eq:4.6B}
z=\left(\begin{array}{c}
u\\
v\\
w\end{array}\right),\qquad \Xi_{n}(z)=\left(\begin{array}{c}
u\\
\Psi_{n}(v,u)\\
 M_{n}(\Psi_{n}(v,u))\end{array}\right).
\end{eqnarray}
Since the permeability in the forward reservoir model does not change in time, it follows that (\ref{eq:4.5})-(\ref{eq:4.6}) can be written as 
\begin{eqnarray}
\label{eq:mod}
z_{n}&=\Xi_{n}(z_{n-1}), \label{eq:4.7}\\
y_{n}&=Hz_{n}+\eta_{n}. \label{eq:4.8}
\end{eqnarray}
 where $H=(0,0,I)$. We now consider the following standard perturbed observation version of EnKF \cite{EnKFReview}.
\begin{algorithm}[EnKF]
Construct an initial ensemble
\begin{eqnarray}\label{eq:4.9}
z_{0}^{(j,a)}=\left(\begin{array}{c}
u_{0}^{(j)}\\
v_{0}\\
M_{0}(v_0)\end{array}\right)
\end{eqnarray}
 where 
$\{u_{0}^{(j)}\}_{j=1}^{N_{e}}\sim \mu_0$ and $v_{0}$ is the initial condition for the state variable. For $j=1,\dots,N_{m}$
\begin{itemize}
\item[(1)] Prediction Step: Propagate the ensemble of particles forward under (\ref{eq:4.7}) giving
\begin{eqnarray}\label{eq:4.10}
z_{n}^{(j,f)}&=\Xi_{n}(z_{n-1}^{(j,a)})\qquad j\in \{1,\dots,N_{e}\}
\end{eqnarray}
From this ensemble we define a sample mean and covariance as follows:
\begin{eqnarray}\label{eq:4.11}
\overline{z}_{n}^{f}&=\frac{1}{N_{e}}\sum_{j=1}^{N_{e}} z_{n}^{(j,f)}\\
C_{n}^{f}&=\frac{1}{(N_{e}-1)}\sum_{j=1}^{N_{e}} z_{n}^{(j,f)}(z_{n}^{(j,f)})^{T}-\overline{z}_{n}^{f}(\overline{z}_{n}^{f})^{T}\label{eq:4.11B}
\end{eqnarray}
\item[(2)] Analysis step: Compute the updated ensembles
\begin{eqnarray}\label{eq:4.12}
z_{n}^{(j,a) }=z_{n}^{(j,f)}+K_{n}(y_{n}^{(j)}-Hz_{n}^{(j,f)})
\end{eqnarray}
where
\begin{eqnarray}\label{eq:4.13}
K_{n} =C_{n}^{f} H^{T} \Big(  HC_{n}^fH^{T}+\Gamma_{n}\Big)^{-1}
\end{eqnarray}
and
\begin{eqnarray}\label{eq:4.14}
y_{n}^{(j)} =y_{n}+\eta_{n}^{(j)},\qquad \eta_{n}^{(j)}\sim N(0,\Gamma_{n})
\end{eqnarray}
\end{itemize}
\end{algorithm}
Here the $\eta_{n}^{(j)}$ are chosen i.i.d. In order to discuss the computational cost of the EnKF algorithm, let us first note that all the vectors and matrices involved have block structure inherited from the structure of the space $Z=X \times S\times \mathbb{R}^{N_{w}}$. For example, we have
\begin{eqnarray*}
z_{n}^{(j,f)}
=\left(
\begin{array}{c}
u_{n}^{(j,f)}\\
v_{n}^{(j,f)}\\
w_{n}^{(j,f)}
\end{array}
\right)
=\left(
\begin{array}{c}
u_{n-1}^{(j,a)}\\
\Psi_{n}(v_{n-1}^{(j,a)},u_{n-1}^{(j,a)})\\
M_{n}(\Psi_{n}(v_{n-1}^{(j,a)},u_{n-1}^{(j,a)}))\end{array}
\right),\qquad \overline{z}_{n}^{f}
=\left(
\begin{array}{c}
\overline{u}_{n}^{f}\\
\overline{v}_{n}^{f}\\
\overline{w}_{n}^{f}
\end{array}
\right)
\end{eqnarray*}
We also have
\begin{eqnarray*}
C_{n}^{zw,f}=\left(
\begin{array}{c}
C_{n}^{uw,f}\\
C_{n}^{vw,f}\\
C_{n}^{ww,f}\end{array}
\right), \quad
C_n^{f}=\left(
\begin{array}{ccc}
C^{uu,f}_n & C^{uv,f}_n &C^{uw,f}_n\\
(C^{uv,f}_n)^T & C^{vv,f}_n& C^{wv,f}_n\\
(C^{uw,f}_n)^{T}& (C^{wv,f}_n)^{T}&(C^{ww,f}_n)\end{array}\right).
\end{eqnarray*}
Then, expression (\ref{eq:4.12}) can be written as
\begin{eqnarray}\label{eq:4.15}
z_{n}^{(j,a) }=\Xi_{n}(z_{n-1}^{(j,a)})+C_{n}^{zw,f}(C_{n}^{ww,f} +\Gamma   )^{-1}(y_{n}^{(j)}-M_{n}(\Psi_{n}(v_{n-1}^{(j,a)},u_{n-1}^{(j,a)})))
\end{eqnarray}
The submatrices in $C_{n}^f$ needed for (\ref{eq:4.15}) are given by
\begin{eqnarray}\label{eq:4.16}
C^{uw,f}_n&=\frac{1}{N_{e}}\sum_{j=1}^{N_{e}} u_n^{(j,f)}(w_n^{(j,f)})^T-\overline{u}_n^{f}(\overline{w}_n^{f})^T,\\
C^{vw,f}_n&=\frac{1}{N_{e}}\sum_{j=1}^{N_{e}} v_n^{(j,f)}(w_n^{(j,f)})^T-\overline{v}_n^{f}(\overline{w}_n^{f})^T,\label{eq:4.16B}\\
C^{ww,f}_n&=\frac{1}{N_{e}}\sum_{j=1}^{N_{e}} w_n^{(j,f)}(w_n^{(j,f)})^T-\overline{w}_n^{f}(\overline{w}_n^{f})^T,\label{eq:16C}
\end{eqnarray}
We recall that $w\in \mathbb{R}^{N_{w}}$ where $N_{w}$ is the number of wells. Typically $N_{w}$ is much smaller than the dimensions of the (discretized) parameter space $X$. Consequently, the computational cost of constructing $C_{n}^{zw}$ and $C_{n}^{ww}$ and inverting the $(C_{n}^{ww} +\Gamma_{n})^{-1}$ in (\ref{eq:4.15}) is negligible compared to the cost of computing $\Xi_{n}(z_{n-1}^{(j,a)})$, which from (\ref{eq:4.6B}) we can see is mainly determined by the cost of $\Psi_{n}(v_{n-1}^{(a,j)},u_{n-1}^{(j,a)})$ (i.e. running the reservoir simulator in the time-interval
$[t_{n-1},t_{n}]$. Therefore, the computational cost of the EnKF is approximately $N_{e}$ times the cost of a forward model simulation.

\subsubsection{Derivation by Gaussian Approximation of the Filtering Distribution}
We now indicate how a Gaussian approximation gives rise to the EnKF algorithm presented above. We start by defining the conditional measures for $n_1,n_2 \leq N_{m}$
\begin{eqnarray}\label{eq:4.17}
\mu_{n_1\vert n_2}(z_{n_1})=\mathbb{￼P}(z_{n_1}\vert \{y_{k}\}_{k=1}^{n_{2}})
\end{eqnarray}
In the filtering approach, given the prior distribution $\mu_{n\vert n-1}$ of $z_{n}$ given data up to the previous time $t=t_{n-1}$ is combined with data provided at the current time time $t=t_{n}$ to define the posterior distribution ($\mu_{n\vert n}$) of $z_{n}$ given data up to the current time $t=t_{n}$. The latter can be obtained from Bayes rule:
\begin{eqnarray}\label{eq:4.18}
\frac{\mu_{n\vert n}(z)}{\mu_{n\vert n-1}(z)}\propto \exp{\{-\Phi_{n}(z)\} }
\end{eqnarray}
￼where
\begin{eqnarray}\label{eq:4.19}
\Phi_{n}(z) =\frac{1}{2}\vert\vert  y_{n}-Hz\vert\vert_{\Gamma}^{2}.
\end{eqnarray}
The EnKF approach then assumes that $\mu_{n\vert n-1}(z)$ is the Gaussian measure $N(\overline{z}_{n}^{f} ,C_{n}^{f})$ where $\overline{z}_{n}^{f}$ and $C_{n}^{f}$ are the ensemble mean and covariance defined in (\ref{eq:4.11})  and (\ref{eq:4.11B}) respectively. Given this Gaussian assumption, it is not difficult to see that (\ref{eq:4.18}) implies that $\mu_{n\vert n}(z) = N(\overline{z}_{n}^{a},C_{n}^{a} )$ with
\begin{eqnarray}\label{eq:4.20}
\overline{z}_{n}^{(a) }&=\overline{z}_{n}^{(f)}+K_{n}(y_{n}-H\overline{z}_{n}^{(f)})\\
C_{n}^{a}&=(I-K_{n})C_{n}^{f}\label{eq:4.20B}
\end{eqnarray}
and $K_{n}$ defined in (\ref{eq:4.13}).  In \cite[Appendix A]{IterativeEnKF} it has been shown that the ensemble updates defined in (\ref{eq:4.12}) are samples from $\mu_{n\vert n}(z) =  N(\overline{z}_{n}^{a},C_{n}^{a} )$. In fact, \cite[Appendix B]{IterativeEnKF} shows that the analysis step (\ref{eq:4.12}) can be derived from an application of RML under the Gaussian approximation $\mu_{n\vert n-1}(z)\approx N(\overline{z}_{n}^{f} ,C_{n}^{f})$. Indeed, it is straight forward to show that (\ref{eq:4.12}) can be obtained from 
\begin{eqnarray}\label{eq:4.20C}
z_{n}^{(j,a)}=\textrm{argmin}_{z}\Big(\vert\vert \Gamma_{n}^{-\frac{1}{2}}(y_{n}^{(j)}-Hz)\vert\vert^2+\vert\vert (C_{n}^{f})^{-\frac{1}{2}} (z-z_{n}^{(j,f)})\vert\vert^2\Big)
\end{eqnarray}
which is a sequential version of (\ref{eq:4.4}), for the augmented state $z$ with a prior $N(\overline{z}_{n}^{f} ,C_{n}^{f})$ and the linear measurement operator $H$. 

For our evaluation and comparison of techniques, we consider the outcome of the EnKF algorithm after all data has been assimilated in the time interval $[0,T]$. In other words, we are interested in $\mu_{n\vert N_{m}}(z_{n})=\mathbb{￼P}(z_{n}\vert \{y_{k}\}_{k=1}^{N_{m}})$ which corresponds to the probability of $z_{n}$ after all data has been assimilated (recall $N_{m}$ is the total number of assimilation times).  Then, the posterior $\mu_{n\vert N_{m}}(z_{n})$ computed via the EnKF algorithm provides an approximation to $\mu$ defined in (\ref{eq:3.3}). 

\subsubsection{Further Modifications}
While the standard version of EnKF has been successfully applied for some history matching problems, several shortcomings due to sampling error have been identified. In particular, when a small ensemble is used, spurious correlations often cause gross over-estimation of the physical variables that EnKF aims at recovering (e.g. permeability). In addition to the issues caused by small sample size, standard EnKF with a small ensemble is suboptimal when a large amount of data are assimilated. This can be easily  observed from the two following properties of EnKF. First, the ensemble updates (\ref{eq:4.12}), when projected into the parameter
space are a linear combination of the initial ensemble members \cite{IterativeEnKF,US}. Second,  the ensemble updates minimize (\ref{eq:4.20C}) which involves fitting data at each assimilation time. Therefore, when the prior ensemble is small, the EnKF updates cannot fit large amount of data within the subspace generated by the prior ensemble. These shortcomings of using standard EnKF have given rise to several EnKF variants designed to reduce the spurious correlations described above as well as increasing the number of degrees of freedom. In this work we focus on the application of distance-based covariance localization which has recently been investigated in \cite{OliverLocal,Emerick3}. In particular, the EnKF with localization that we implement for the forward models of Section \ref{ReservoirModels} is given by the same EnKF algorithm described before, except that (\ref{eq:4.13}) is replaced by
\begin{eqnarray}\label{eq:4.12B}
K_{n} =\rho\circ C_{n}^{f} H^{T} \Big(  H(\rho\circ  C_{n}^f) H^T+\Gamma_{n}\Big)^{-1}.
\end{eqnarray}
Here $\rho$, to be defined below, is a positive-definite matrix which induces
localization and the matrix $\rho \circ C_{n}$ is the Schur product between $\rho$ and $C_{n}$ with entries defined by $[\rho \circ C_{n}]_{ij}=[\rho]_{ij}[C_{n}]_{ij}$. Due to the spurious correlations described above, matrix $C_{n}^{f} $ may become positive semi-definite and the parameter update then lies in smaller subspace than the one generated by the prior ensemble. With properly chosen $\rho$ the matrix $\rho\circ C_{n}^{f}$ 
has full rank, and replacing $C_{n}^f$ with $\rho\circ C_{n}^{f}$ increases the dimension of the linear subspace where the parameter update is sought. This, in turn, results in a better estimation. In terms of the block structure previously described, covariance localization becomes
\begin{eqnarray}\label{eq:4.21}
z_{n}^{(j,a) }=z_{n}^{(j,f)}+\rho_{zw}\circ C_{n}^{zw}(\rho_{ww}\circ C_{n}^{ww} +\Gamma_{n}   )^{-1}(y_{n}^{(j)}-M_{n}(\Psi_{n}(v_{n-1}^{(j,a)},u_{n-1}^{(j,a)}) ))
\end{eqnarray}
As in the covariance localization approach of \cite{OliverLocal}, we consider only localization in the $u$-component (e.g. for the log-permeability updates). In other words, 
\begin{eqnarray}\label{eq:4.22}
u_{n}^{(j,a) }&=&u_{n}^{(j,f)}+\rho_{uw}\circ C_{n}^{uw}(\rho_{ww}\circ C_{n}^{ww} +\Gamma_{n}   )^{-1}(y_{n}^{(j)}-M_{n}(\Psi_{n}(v_{n-1}^{(j,a)},u_{n-1}^{(j,a)}) ))\\
v_{n}^{(j,a) }&=&v_{n}^{(j,f)}+ C_{n}^{vw}( C_{n}^{ww} +\Gamma_{n}   )^{-1}(y_{n}^{(j)}-M_{n}(\Psi_{n}(v_{n-1}^{(j,a)},u_{n-1}^{(j,a)}) ))\label{eq:4.23} \\
w_{n}^{(j,a) }&=&w_{n}^{(j,f)}+ C_{n}^{ww}(C_{n}^{ww} +\Gamma_{n}   )^{-1}(y_{n}^{(j)}-M_{n}(\Psi_{n}(v_{n-1}^{(j,a)},u_{n-1}^{(j,a)}) )) \label{eq:4.24}
\end{eqnarray}
Following the implementation of \cite{Emerick3},  each column of the localization matrix $\rho_{uw}$ is defined as the fifth order compact function of Gaspari-Cohn \cite{QJ:QJ49712555417} localized at the corresponding measurement location. Each row of the matrix $\rho_{ww}$ in (\ref{eq:4.22}) is obtained from $\rho_{uw}$ by projecting it on the corresponding measurement location. By construction, $\rho_{uw}$ and $\rho_{ww}$ are positive definite. 

Recent publications \cite{Emerick3,OliverLocal} have investigated optimal choices for the critical length of the correlation function used for distanced-based localization. The focus of those investigations is to improve the ability of the EnKF with localization to recover the truth within the confidence interval provided by the ensemble. In contrast to \cite{Emerick3,OliverLocal}, our goal is to assess the performance of EnKF with localization for reproducing the uncertainty quantified by the posterior. However, for the present work we consider a fixed critical length obtained from a simple trial-error procedure, that enables us to observe significant effect of covariance localization in characterizing the posterior distribution. While the optimal choice of covariance localization is beyond the scope of the present work, we recognize the importance for assessing optimal choices of covariance localization for providing better Gaussian approximation of the posterior distribution at a reasonable computational cost. Moreover, additional forms of covariance regularization (e.g. covariance inflation) should also be assessed.

\subsubsection{Ensemble square root filter (EnSRF)}

Sampling error that arises from perturbing the observations in standard EnKF has been often associated with a poor performance of history matching data. In order to avoid the aforementioned sampling error, an ensemble square root filter (EnSRF) is often
used. Here we consider the following EnSRF \cite{Emerick}:

\begin{algorithm}[EnSRF]
Construct an initial ensemble as in (\ref{eq:4.9}). For $j=1,\dots,N_{m}$
\begin{itemize}
\item[(1)] Prediction Step: Propagate the ensemble of particles forward under (\ref{eq:4.7}) yielding (\ref{eq:4.10}). Construct the sample mean and covariance from (\ref{eq:4.11})-(\ref{eq:4.11B}). Additionally define the deviations from the mean
\begin{eqnarray}\label{eq:4.25}
\Delta z_{n}^{(j,f) }=z_{n}^{(j,f)}-\overline{z}_{n}^{f}.
\end{eqnarray}
\item[(2)] Analysis step:  Compute the updated mean $\overline{z}_{n}^{(a)}$ via formula (\ref{eq:4.20}) with $K_{n}$ given by (\ref{eq:4.13}). Consider the matrices $
\Delta Z_n^f:=[\Delta z_{n}^{(1,f)} ~\Delta z_{n}^{(2,f)}~\cdots~ \Delta z_{n}^{(N_{e},f)}]$, with $j$th column $\Delta z_{n}^{(j,f) }$, and $\Delta Z_{n}^{a}$ defined analogously. 
Compute the matrix with updated deviations,
\begin{eqnarray}\label{eq:4.26}
\Delta Z_{n}^{a}=(I-\tilde{K}_{n}H) \Delta Z_{n}^{(f) }\Theta
\end{eqnarray}

with 
\begin{eqnarray}\label{eq:4.27}
\tilde{K}_{n}=C_{n}^{f}H^{T}\Big[ H_{n}C_{n}^{f}H_{n}^{T}+\Gamma_{n} \Big]^{-T/2}\Big[ (H_{n}C_{n}^{f}H_{n}^{T}+\Gamma_{n})^{1/2}+\Gamma_{n}^{1/2}\Big]^{-1}
\end{eqnarray}
The updated ensemble is then obtained from the expression
\begin{eqnarray}\label{eq:4.28}
z_{n}^{(j,a)}=\overline{z}_{n}^{(a)}+\Delta z_{n}^{(j,a)}.
\end{eqnarray}
\end{itemize}
\end{algorithm}

In expression (\ref{eq:4.26}), $\Theta$ is a $N_{e}\times N_{e}$ mean-preserving orthogonal random matrix constructed as suggested in \cite{Sakov}. The mean-preserving property of $\Theta$ ensures that $\sum_{j=1}^{N_{e}}\Delta z_{n}^{(j,a) } =0$ and so the analyzed ensemble (\ref{eq:4.28}) has mean $\overline{z}_{n}^{(a)}$ as required. In contrast to the EnKF, where (\ref{eq:4.20B}) is only exactly satisfied in the limit of arbitrarily
large ensemble size, the sample covariance computed from the 
EnSRF (finite) ensemble updates exactly satisfy (\ref{eq:4.20B}), therefore providing a better approximation of $\mu_{n\vert n}(z)$; it is then hoped that this will lead to a better approximation to the posterior distribution (\ref{eq:3.3}) itself. A block structure similar to the one introduced before applies to the EnSRF. From this structure it is easy to appreciate that only small matrices are involved in the square root computations.  Therefore, the computational cost of EnSRF is essentially the same as EnKF, i.e., $N_{e}$ times the number of forward model evaluations.

Even though the implementation of the EnSRF avoids the sampling error due to perturbing the observations, limitations related to the small ensemble size still apply. Distance-based localization can then be applied to the EnSRF as suggested in \cite{Emerick}. Concretely, we replace $K_{n}$ in (\ref{eq:4.13}) with (\ref{eq:4.12B}) and $\tilde{K}_{n}$ in (\ref{eq:4.25}) with
\begin{eqnarray}\label{eq:4.29}
\tilde{K}_{n}=(\rho\circ C_{n}^{f})H^{T}\Big[ H_{n}(\rho\circ C_{n}^{f}) H_{n}^{T}+\Gamma_{n} \Big]^{-T/2}\Big[ (H_{n}(\rho\circ C_{n}^{f})H_{n}^{T}+\Gamma_{n})^{1/2}+\Gamma_{n}^{1/2} \Big]^{-1}
\end{eqnarray}
Similar to the localization procedure for the EnKF, EnSRF is localized only in the $u$-component (log-permeability ) of (\ref{eq:4.20}) and (\ref{eq:4.24}) with the localization matrix $\rho$ described above.

\section{Numerical Results }\label{NumericalResults}

In this section we present the results of three numerical experiments for assessing the Gaussian approximations defined in Section \ref{Gaussian}. These experiments are described on the three subsections which follow, each of which is organized as follows: (i) details of the forward model and the generation of synthetic data are provided; (ii) numerical results from the gold-standard MCMC implementation described in Section \ref{BayesianFramework} are discussed; (iii) the numerical results of Gaussian approximations are presented and the results in (iii)
are compared against the results from (ii) in terms of their
ability to reproduce mean and variance; (iv) we assess the performance of the Gaussian approximation at quantifying the uncertainty in reservoir model forecast.

\subsection{Single-Phase flow}\label{singleResults}

\begin{figure}
\includegraphics[scale=0.25]{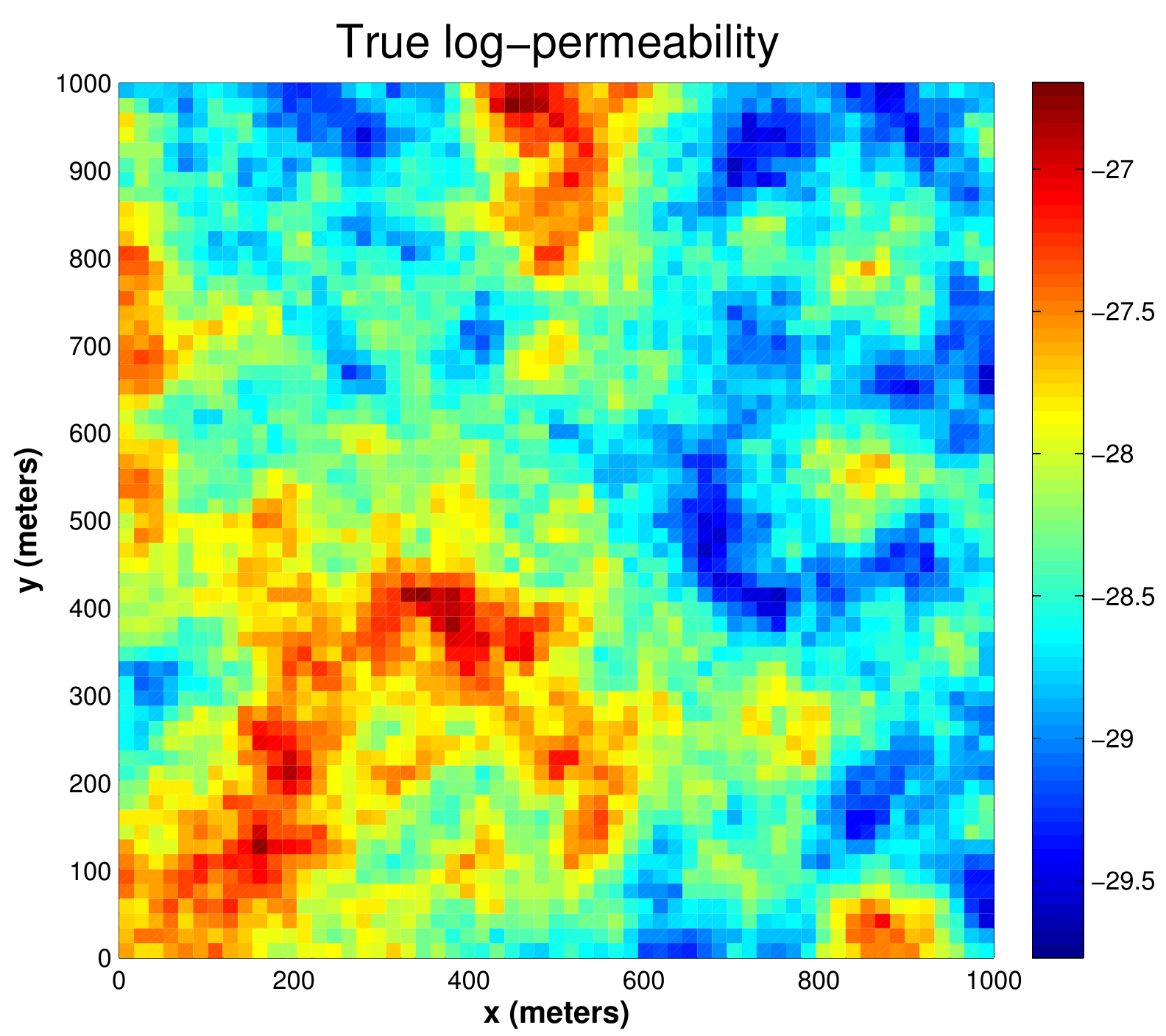}
\includegraphics[scale=0.25]{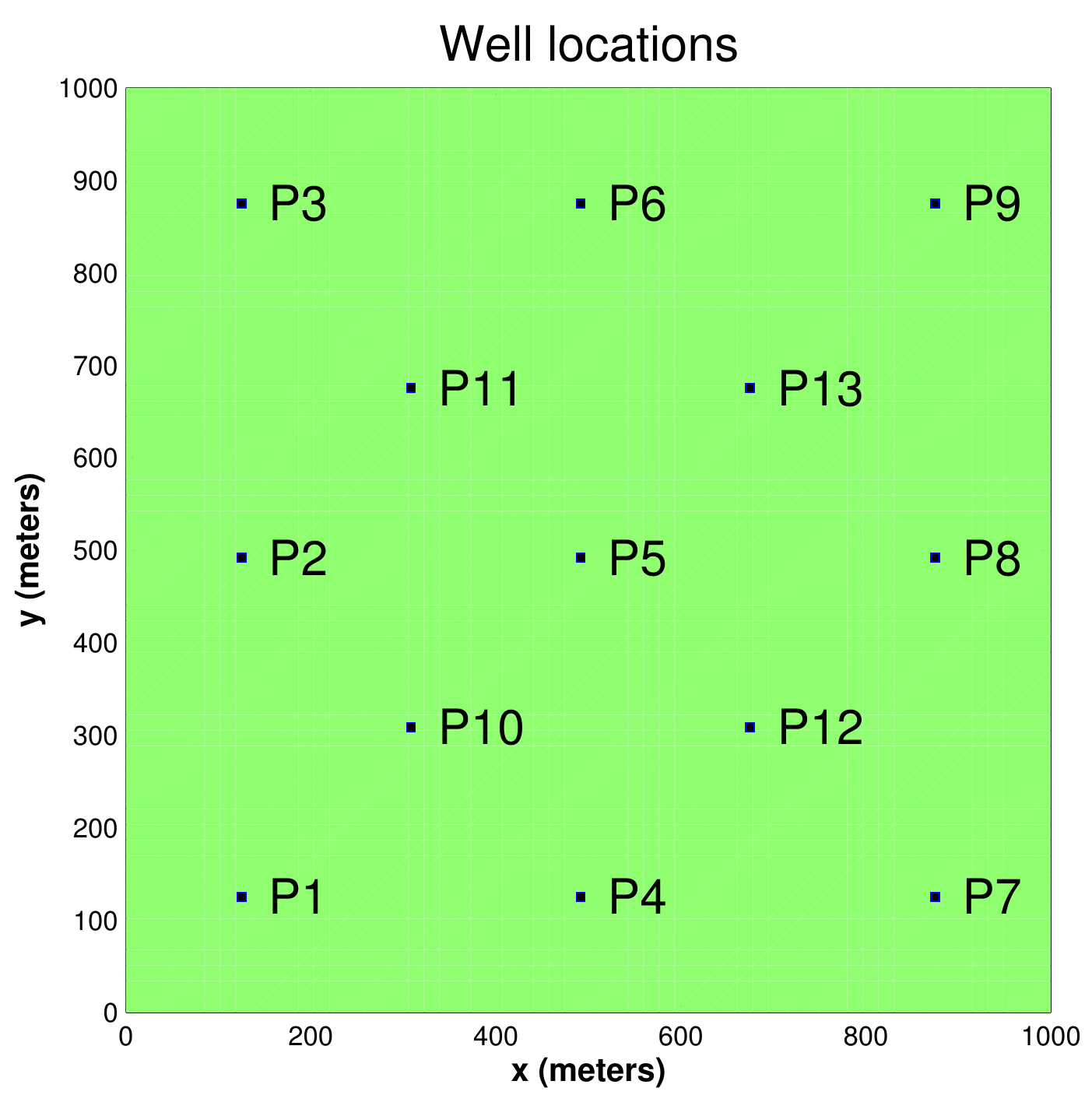}
\includegraphics[scale=0.25]{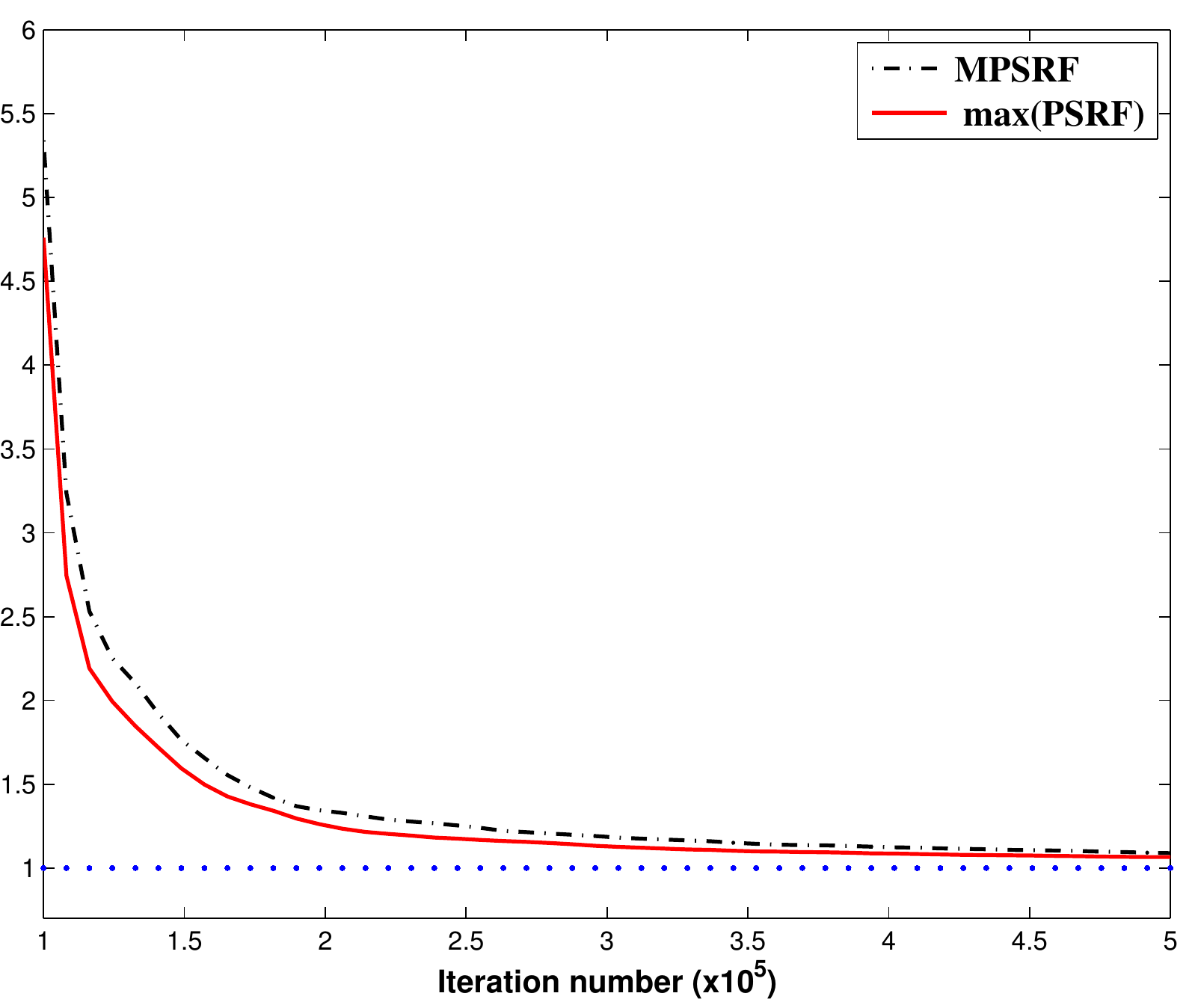}
\caption{Single-phase model. Left: True log-permeability [$\log{\textrm{m}^2}$]. Middle: Well configuration. Right: Gelman-Rubin diagnostic}  
\label{Figure1}
\end{figure}

For the first experiment we consider the single-phase reservoir model of Section \ref{ReservoirModels} on a square domain $D=[0,L]\times [0,L]$ with the production wells located at the points labeled by $P_{1},\dots, P_{9}$ in Figure \ref{Figure1} (middle). Relevant information of this model is displayed in the first column of Table \ref{Table1}. For each well term in the right hand side of (\ref{eq:2.1}) we prescribe a production rate of $85 \textrm{m}^{3}/\textrm{day}$ constant during the total simulation time of $50$ days. 

\begin{figure}
\includegraphics[scale=0.55]{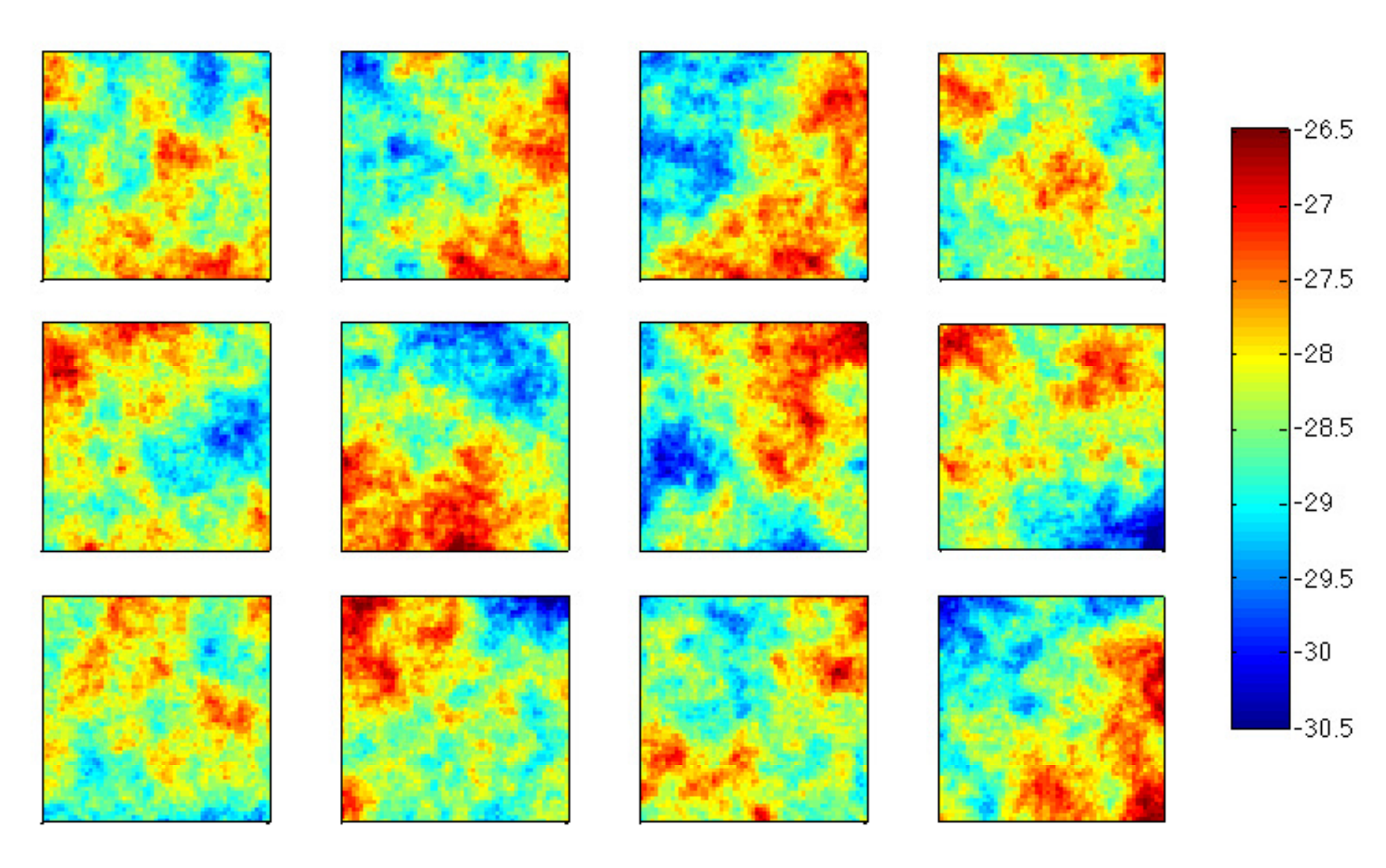}
\caption{Single-phase model. Samples from the prior distribution $[\log{\textrm{m}^2} ]$.}  
\label{Figure2}
\end{figure}

We consider a Gaussian prior distribution of log-permeability
\begin{eqnarray}\label{eq:5.1}
\mu_{0}(u) = N(\overline{u}, C)
\end{eqnarray}
where the covariance is defined by $C=\kappa A^{-\alpha}$, with the operator $A=-\Delta $ defined on
\begin{eqnarray}
D(A)=\{v\in H^{2}(D) \vert \nabla v \cdot \mathbf{n} =0 \textrm{   on  }\partial D, ~~\int_{D}v=0   \}
\end{eqnarray}
i.e. $A$ is the negative Laplacian with no-flow boundary
conditions and restricted to spatial average zero functions. The tunable parameters in (\ref{eq:5.1}) are defined as follows: $\overline{u}(x,y)=\log{(5\times 10^{13}\textrm{m}^2)}$ for all $(x,y)\in D$, $\kappa=2.0$ and $\alpha=1.3$. In Figure \ref{Figure2} we show some realizations of the prior distribution (\ref{eq:5.1}). It is important to mention that other choices of $C$ can also be used. In particular, $C$ can be defined in terms of a standard correlation function (e.g. spherical, exponential, etc). Our choice, however, has the advantage that $C$ becomes a diagonal operator in the spectral domain. Sampling from the prior on the spectral domain is straightforward and computationally inexpensive. This is a desirable property since at each iteration of MCMC (see equation (\ref{eq:3.5})), a draw from  the prior is generated for computing the proposal. The correlation function of draws from the prior is, of course, simply the Green's function of $C$.

\begin{table}
\caption{Reservoir model description}
\begin{center}

\label{Table1}       
\begin{tabular}{lccc}
\hline\noalign{\smallskip}
Variable  & single-phase reservoir &  water-oil reservoir &  water-oil reservoir\\
 & &(small number of wells) &(large number of wells)  \\
\noalign{\smallskip}\hline\noalign{\smallskip}
L  [$\textrm{m}^{3}$] & $10^{3}$ & $2\times 10^{3}$ & $5\times 10^{3}$\\
$c$  [$\textrm{Pa}^{-1}$] & $10^{-8}$ & 0.0 & 0.0\\
$\nu_{o}$  [$\textrm{Pa s}$] & $10^{-2}$ & $10^{-2}$  & $10^{-2}$ \\
$T$  [$\textrm{years}$] & $0.13$ & 5 & 3.5\\
$^{\rm a}$ $p_{0}$ [\textrm{Pa}] & $3.5\times 10^{7}$ & $2.5\times 10^{7}$ &  $2.5\times 10^{7}$ \\
$^{\rm a}$ $s_{0}$   &not applicable & 0.2 &  0.2\\
$\nu_{w}$  [$\textrm{Pa s}$] & not applicable & $5\times 10^{-4}$  & $5\times 10^{-4}$ \\
$s_{iw}$  &  not applicable& 0.2 & 0.2\\
$s_{ro}$   &not applicable  & 0.2 & 0.2\\
$^{\rm b}$ $P_{bh}^{l}$ [\textrm{Pa}] &not applicable & $2.7\times 10^{7}$ & $2.0\times 10^{7}$\\
$^{\rm b}$ $q_{w}^{l}$  [$\textrm{m}^{3}/\textrm{day}$] & not applicable & $2.6\times 10^3$  & $1.8\times 10^2$ \\
 $a_{w}$  & not applicable & $0.3$  & $0.3$ \\
 $a_{o}$  & not applicable & $0.9$  & $0.9$ \\
\noalign{\smallskip}\hline
\end{tabular}
\end{center}
$^{\rm a}$ Constant in $\Omega$. $^{\rm b}$ Constant in $[0,T]$.
\end{table}

For the generation of synthetic pressure data, we first define the ``true log-permeability'' denoted by $u^{\dagger}$ and displayed in Figure \ref{Figure1} (left). This ``true log-permeability'' is generated from the prior distribution defined above. Synthetic data is generated by first solving (\ref{eq:2.1})-(\ref{eq:2.3}) for $p$ with $u=u^{\dagger}$. Then, $G(u^{\dagger})$ is calculated from (\ref{eq:2.6}). Finally, we add random error, i.e. $y=G(u^{\dagger})+\eta$ with $\eta \sim N(0, \sigma^2 I)$ with $\sigma=4\times 10^5\textrm{Pa}$. The measurement times used in  (\ref{eq:2.4}) are $t_{1}=5$, $t_{n}=10n ~\textrm{days}$, $n=\{2,\dots,5\}$. 

\begin{figure}
\includegraphics[scale=0.55]{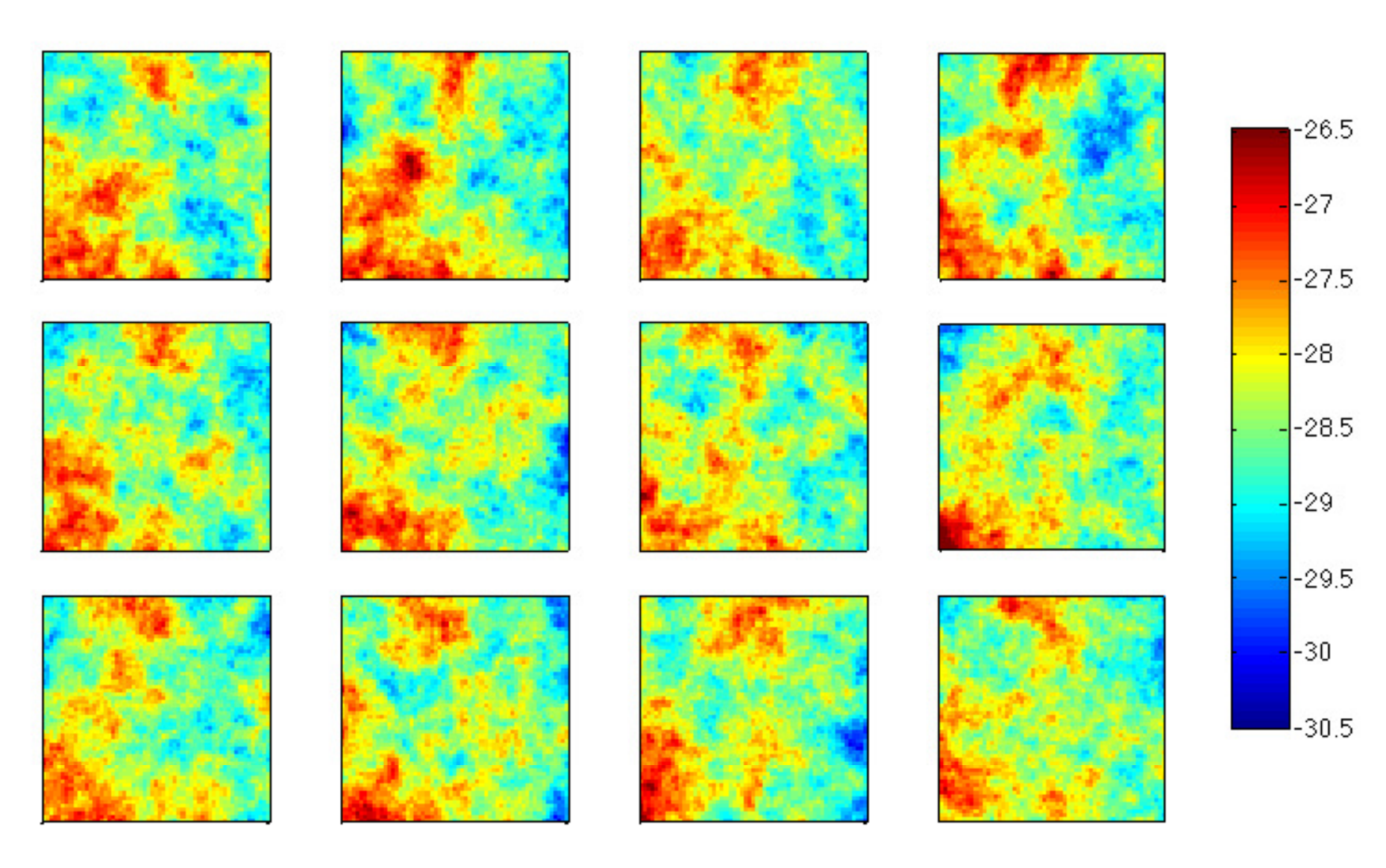}
\caption{Single-phase model. Samples from the posterior distribution (characterized with MCMC) [$\log{\textrm{m}^2}$]}  
\label{Figure3}
\end{figure}

Synthetic data are used in the pCN-MCMC Algorithm \ref{al:MCMC} with $\beta=0.015$. Our MCMC results consist of 110 chains starting  from independent draws from the prior distribution. After a burn-in period of $1\times 10^{4}$, each chain generates $5\times 10^5$ samples. For assessing the convergence of our chains, we consider the diagnostics suggested by Gelman and Rubin in \cite{Gelman}. In Figure \ref{Figure1} (right) we display the maximum of the potential scale reduction factor (PSRF) and the multivariate potential scale reduction factor (MPSRF) for the smallest $J=16$ frequencies that account for $76\%$ of the total prior energy defined by $e(J) = \sum_{j=1}^{J}\lambda_{k}/\sum_{j=1}^{\infty}\lambda_{k}$ where $\lambda_{k}$ are the eigenvalues of the prior covariance $C$ (ordered as $\lambda_{1}\ge \lambda_{2}\ge \dots$). Convergence of the chains is achieved when the maximum of the PSRF and the MPSRF are close to one. From Figure \ref{Figure1} (right), the max(PSRF) and the MPSRF have dropped below 1.1 after $5\times 10^{5}$ iterations where we stablish the convergence of our MCMC chains. From the numerical evidence of convergence of our chains, we conclude that the MCMC provides samples from the posterior. The associated mean and variance fields,
denoted by $u_{pos}(x)$ and $\sigma_{pos}(x)$, are used as gold standard for the assessment of the Gaussian approximations that we discuss below. In Figure \ref{Figure3} we display some samples from the independent chains (i.e. uncorrelated) obtained after convergence was achieved. Although there are substantial differences among those realizations, some common spatial features can be observed. For example, note the high permeability region around wells $P_{1}$ and $P_{6}$. Furthermore, the variability is considerably lower than under the prior, as exhibited in Figure \ref{Figure2};
this indicates that the data used is quite informative.

\begin{figure}
\includegraphics[scale=0.65]{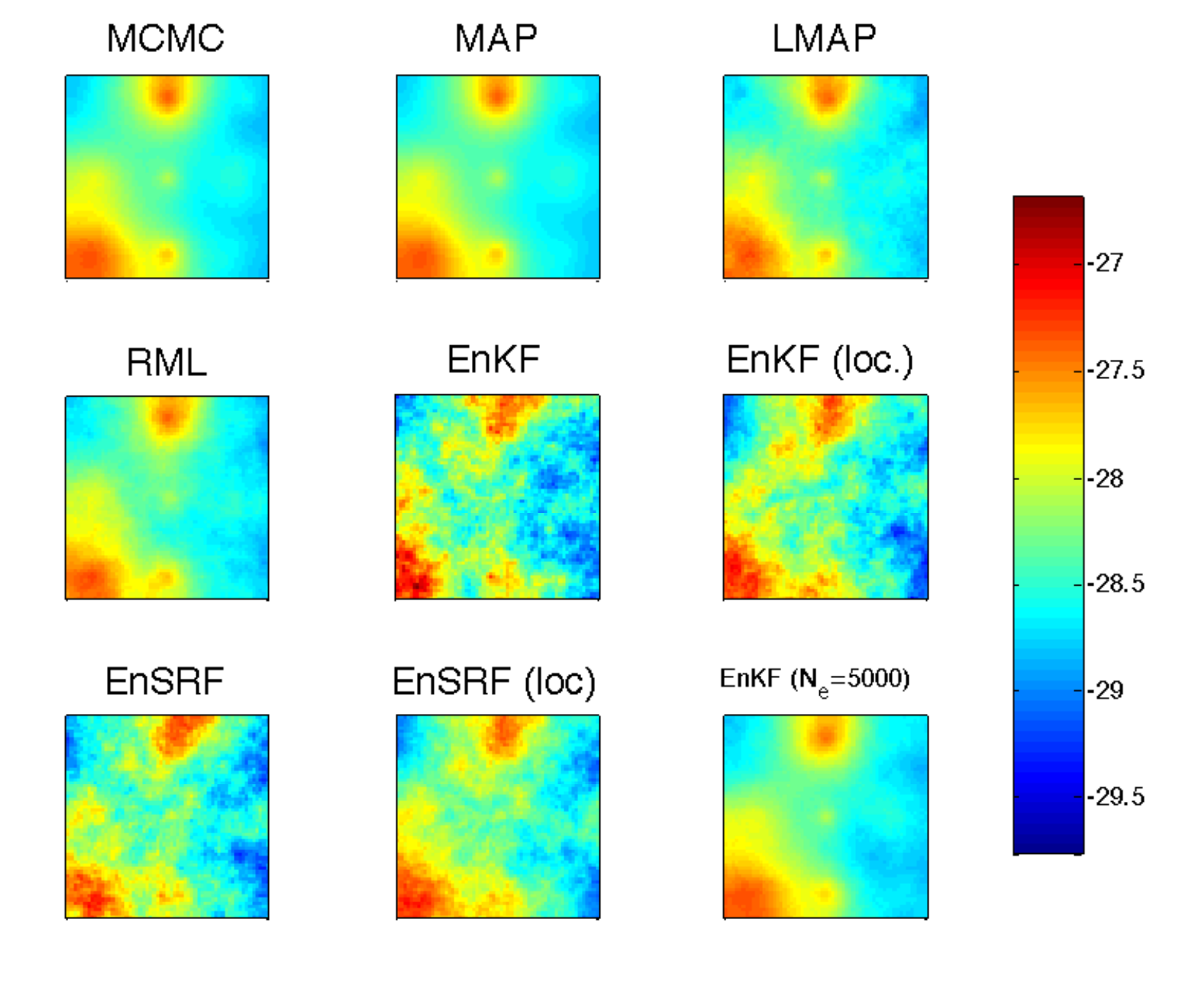}
\caption{Single-phase model. Mean of the posterior distribution (characterized with MCMC) and Gaussian approximations [$\log{\textrm{m}^2}$]}  
\label{Figure4}
\end{figure}

\begin{figure}
\includegraphics[scale=0.65]{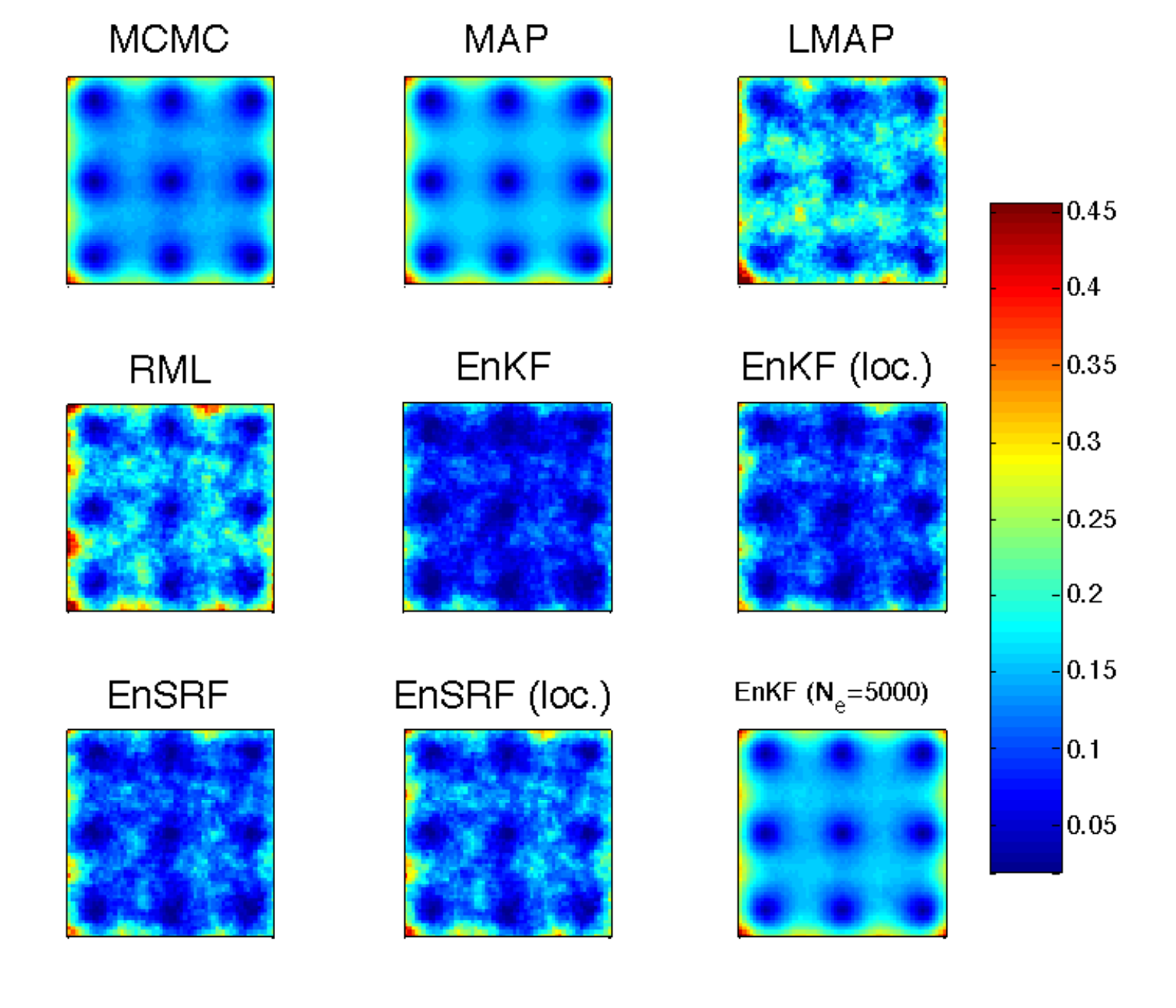}
\caption{Single-phase model. Variance of the posterior distribution (characterized with MCMC) and Gaussian approximations [$(\log{\textrm{m}^2})^2$]}  
\label{Figure5}
\end{figure}

The numerical implementation of the following Gaussian approximations are conducted for an ensemble of size $N_{e}=50$: LMAP, RML, EnKF, EnKF with localization, EnSRF and EnSRF with localization. It is important to emphasize that the mean and variance for (the sequential methods) EnKF, EnKF with localization, EnSRF and EnSRF with localization, are computed after all the measurements have been assimilated. In other words, exactly the same data used to sample the posterior via MCMC are also used for all the Gaussian approximations under consideration. From each of these approximations, we compute the mean and variance which we compare against the mean $u_{pos}$ and variance $\sigma_{pos}$ of the posterior distribution generated with the MCMC method. The mean an variance are shown in Figure \ref{Figure4} and Figure \ref{Figure5}, respectively. In Table \ref{Table2} we display the relative errors of the deviation of the mean with respect to $\overline{u}$ (the prior mean) and the relative error of the variance defined by
\begin{eqnarray}\label{eq:5.3}
\epsilon_{u}=\frac{\vert\vert (\hat{u}-\overline{u})-(u_{pos}-\overline{u})\vert\vert_{L^{2}(D)}}{\vert\vert (u_{pos}-\overline{u})\vert\vert_{L^{2}(D)}},\qquad \textrm{and}\qquad \epsilon_{\sigma}=\frac{\vert\vert \hat{\sigma}-\sigma_{pos}\vert\vert_{L^{2}(D)}}{\vert\vert \sigma_{pos}\vert\vert_{L^{2}(D)}},
\end{eqnarray}
respectively. In the previous expression, $\hat{u}$ and $\hat{\sigma}$ are the mean and variance of the Gaussian approximation under consideration. The right column of Table \ref{Table2} indicates the computational cost for computing each of the techniques in terms of forward model runs. As stated in Section \ref{Gaussian}, while the computational cost of the Kalman filter-type of methods is stable with respect to the details of the
implementation, the cost of RML and LMAP depends crucially
on the optimization technique used for solving (\ref{eq:4.1}) and (\ref{eq:4.4}). Our implementation of the Levenberg-Marquardt technique cost around $5$ forward model runs per iteration. Furthermore, in average, each of optimization problems (\ref{eq:4.4}) converged in $5$ iterations and so the average computational cost of  (\ref{eq:4.4}) is $25$ forward model runs. This computational cost can be potentially reduced by applying a more efficient optimization technique. 

Since we are assessing the Gaussian approximations only in terms of mean and variance, we can additionally measure the error (with respect to the posterior) of the exact mean and variance of $N(u_{MAP},C_{MAP})$ given directly  (\ref{eq:4.1}) and (\ref{eq:4.2}), respectively. The corresponding relative errors are provided in the ``MAP'' row in Table \ref{Table2}. From construction it is clear that the mean and covariance of LMAP are $u_{MAP}$ and $C_{MAP}$ for sufficiently large $N_{e}$. The results for this experiment indicate that $N(u_{MAP},C_{MAP})$ provides a good approximation to the posterior in terms of mean and variance. Therefore, more samples from LMAP can be generated at a negligible cost so that its mean and variance approaches $u_{MAP}$ and $C_{MAP}$, respectively. 

Table \ref{Table2} also indicates that, among the all the Gaussian approximation with $N_{e}=50$, RML provides the best approximation in terms of the mean. The worst performance in terms of mean and variance was obtained with EnKF. However, considerable improvement was obtained by applying the localization approach described in the preceding section  (see equation (\ref{eq:4.22})). We recall from exposition of Section \ref{Gaussian} that EnSRF reduces the sampling error that arises from standard implementations EnKF where data are perturbed with noise. From Table \ref{Table2} we observe that the effect of sampling error has a detrimental effect in the performance of EnKF for reproducing the posterior distribution. More precisely, EnSRF outperformed the EnKF both in terms of mean and variance with respect to the posterior distribution. Note that the application of localization in the EnSRF (expression (\ref{eq:4.25})) further reduces the relative errors in the mean and variance. In fact, among all techniques with $N_{e}=50$, the best approximation in terms of variance is given by the EnSRF with localization.

From the preceding comments it clear that sampling error causes severe limitations in the performance of EnKF and EnSRF. It is worth mentioning that issues of the EnKF and EnSRF due to sampling error have been often reported \cite{OliverLocal,OliverReview} and used as motivation for covariance regularization. However, this existing work is focused on (i) history matching production data (ii) recovering the true permeability and (iii) recovering data generated with the true permeability within the estimated confidence interval. In contrast, here we assess the performance in terms of the posterior distribution of the Bayesian framework.

Since sampling error due to the small ensemble size severely limits the ability of EnKF to produce reasonable approximations of the posterior, we consider three more additional implementations of EnKF for larger ensembles: $N_{e}=1250$, $N_{e}=2500$ and $N_{e}=5000$. These results appear at the end of Table \ref{Table2}. Note that $N_{e}=1250$ corresponds to the case where the computational cost of EnKF coincides with the cost of our implementation of RML. While the performance of EnKF improved significantly for $N_{e}=1250$, RML still provides a better approximation in terms of the mean. In addition, we observe that although increasing the size of the ensemble may reduce the sampling error, this is not associated to the convergence to the posterior. Actually, it is clear from this experiment that the variance of EnKF for large $N_{e}$ seems to diverge from the variance of the posterior. 

\begin{figure}
\includegraphics[scale=0.22]{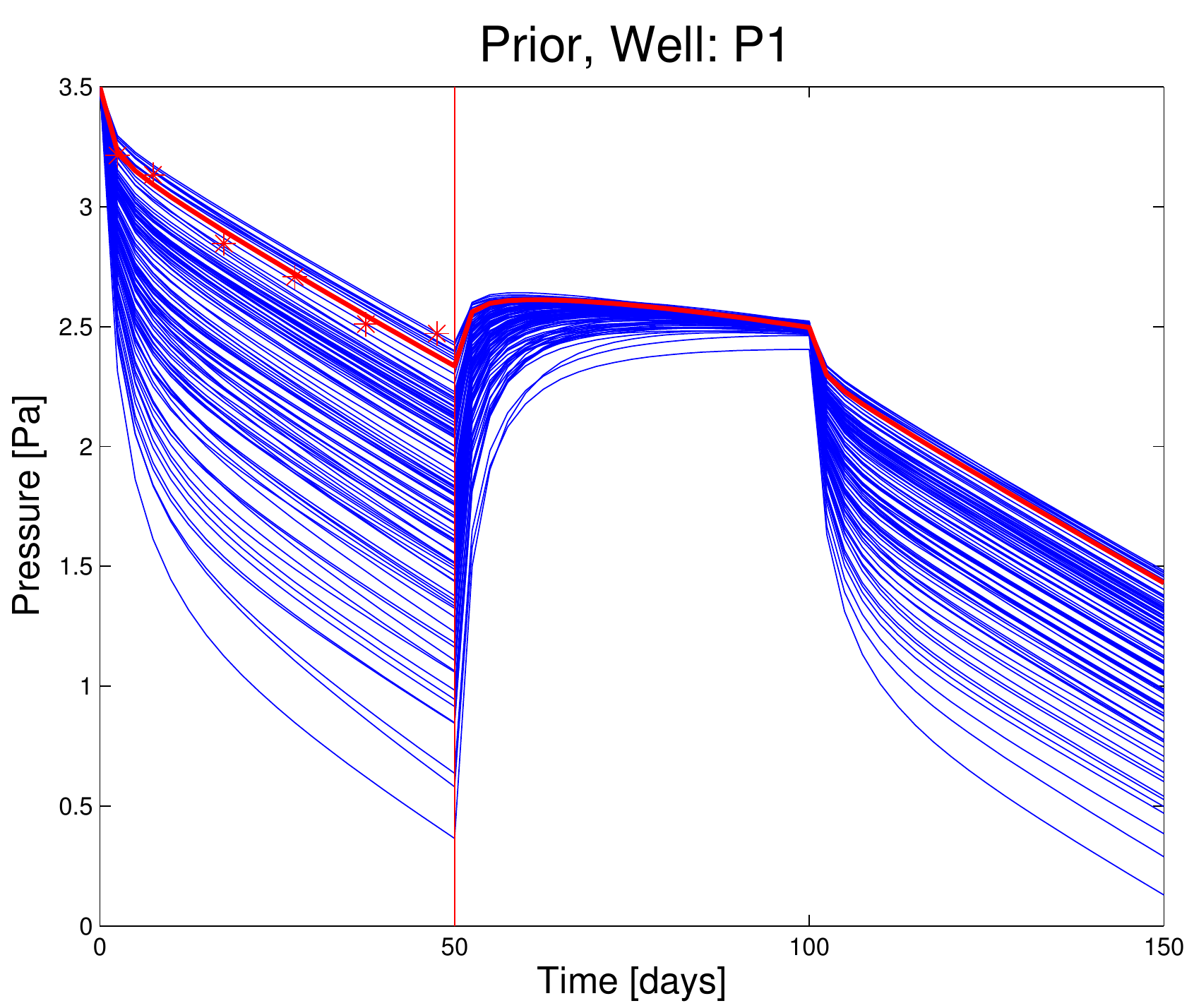}
\includegraphics[scale=0.22]{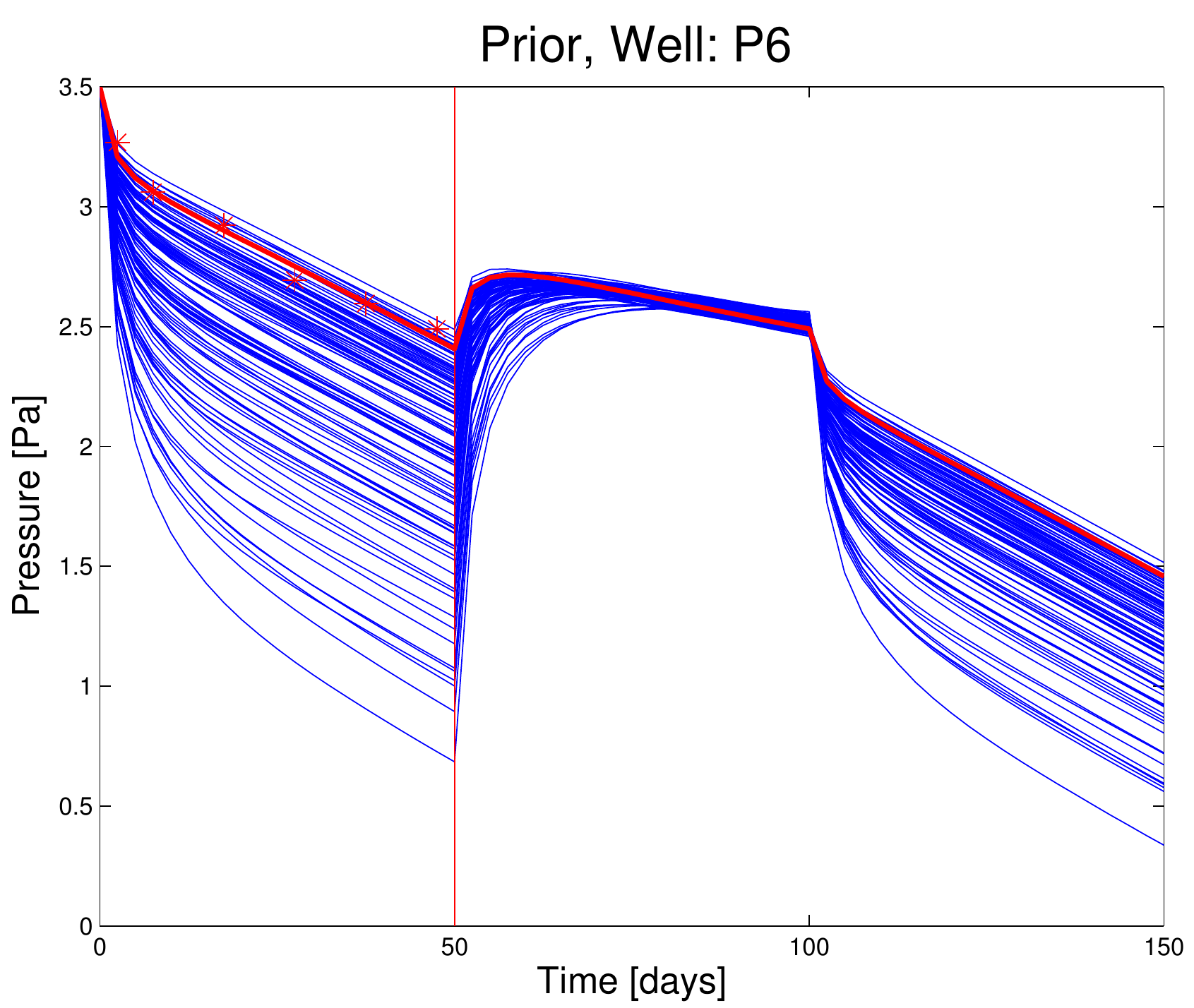}
\includegraphics[scale=0.22]{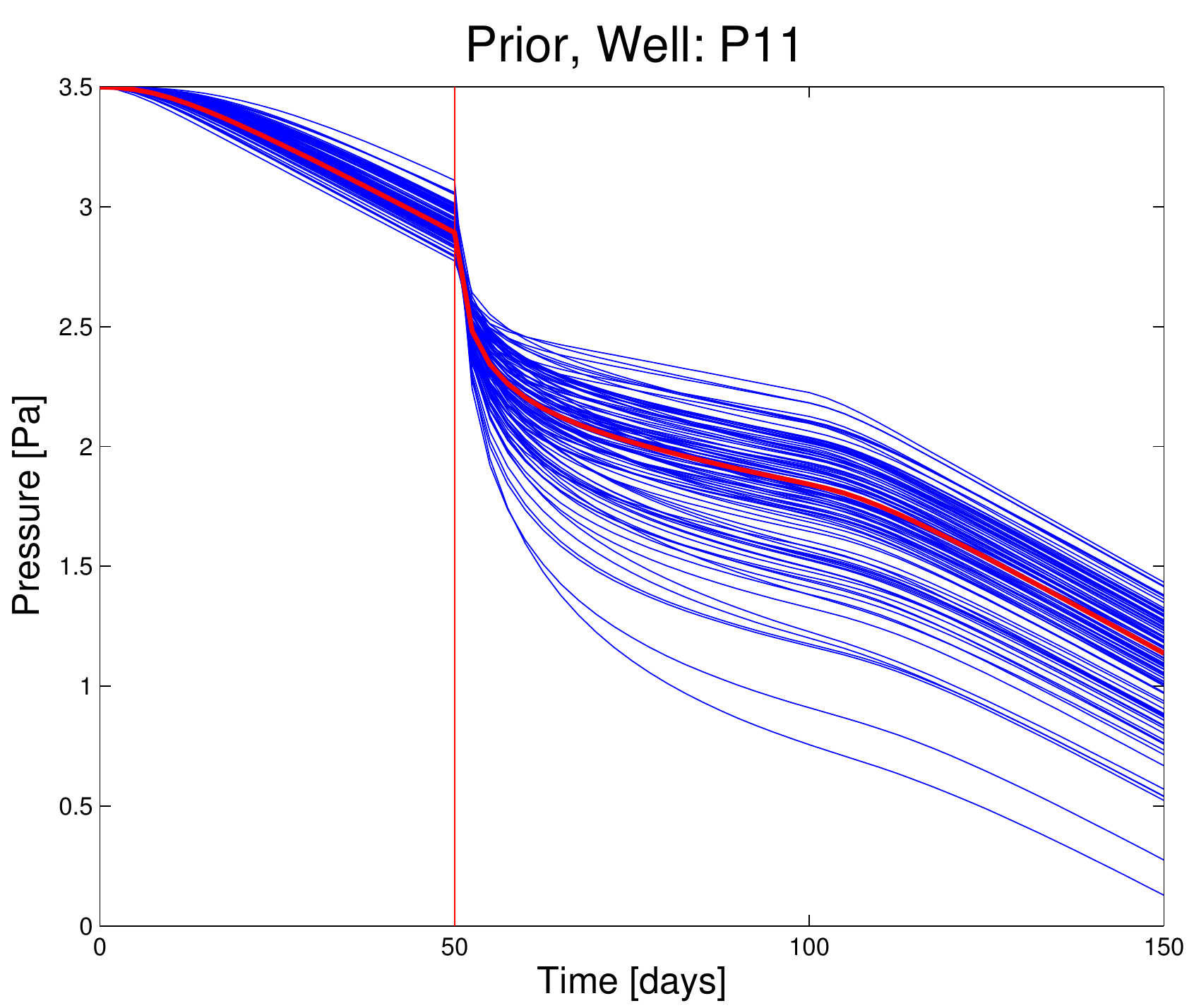}\\
\includegraphics[scale=0.22]{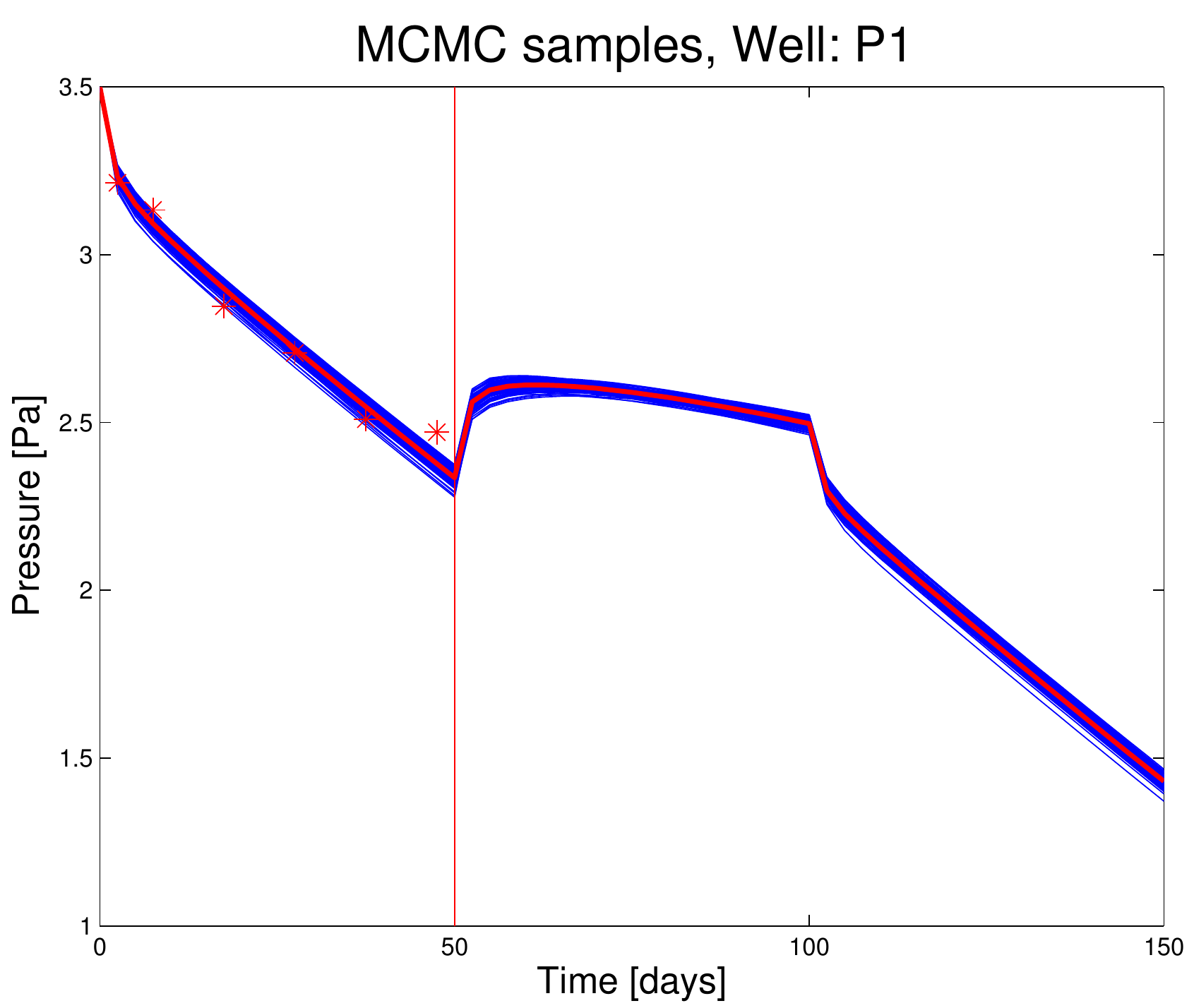}
\includegraphics[scale=0.22]{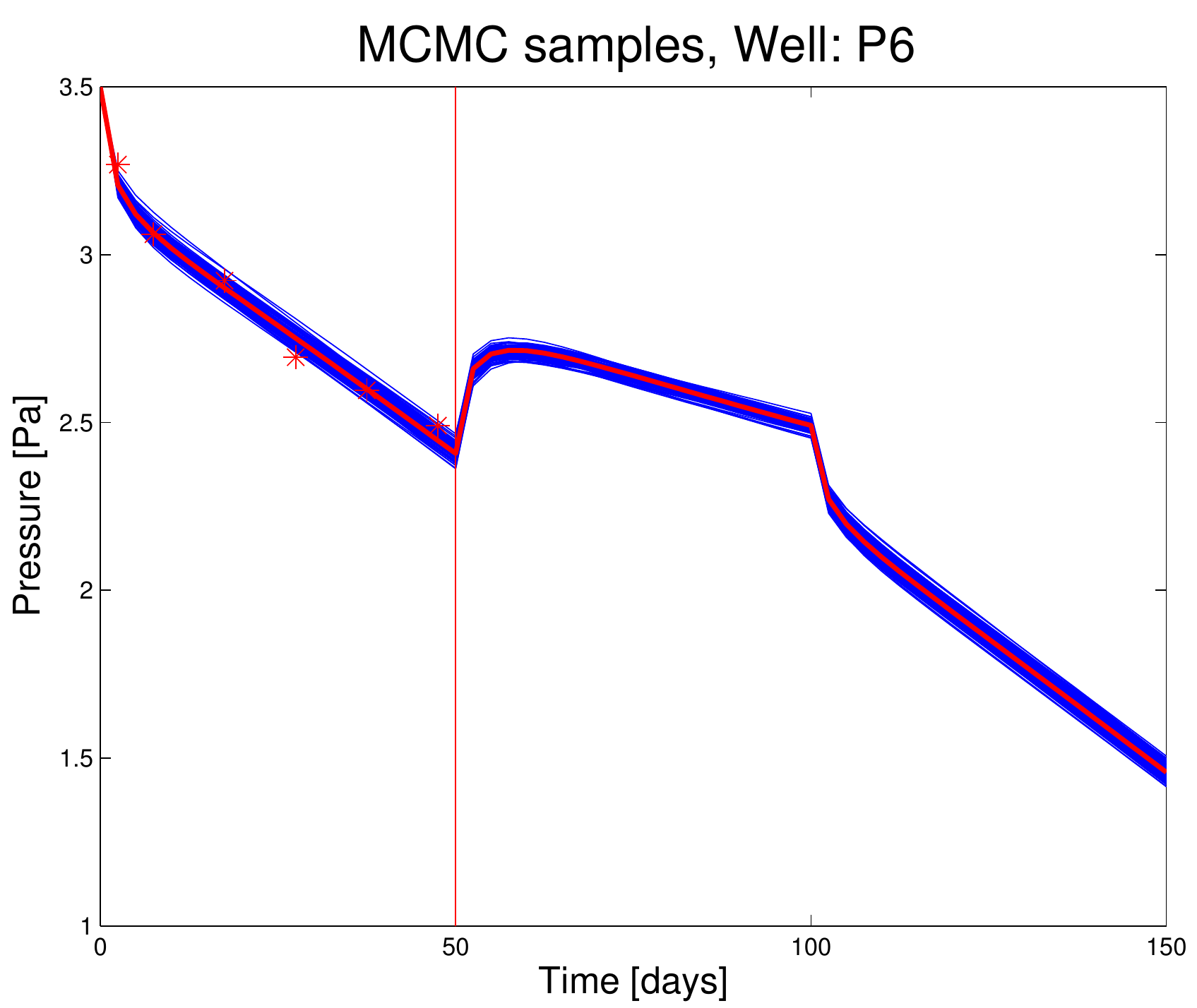}
\includegraphics[scale=0.22]{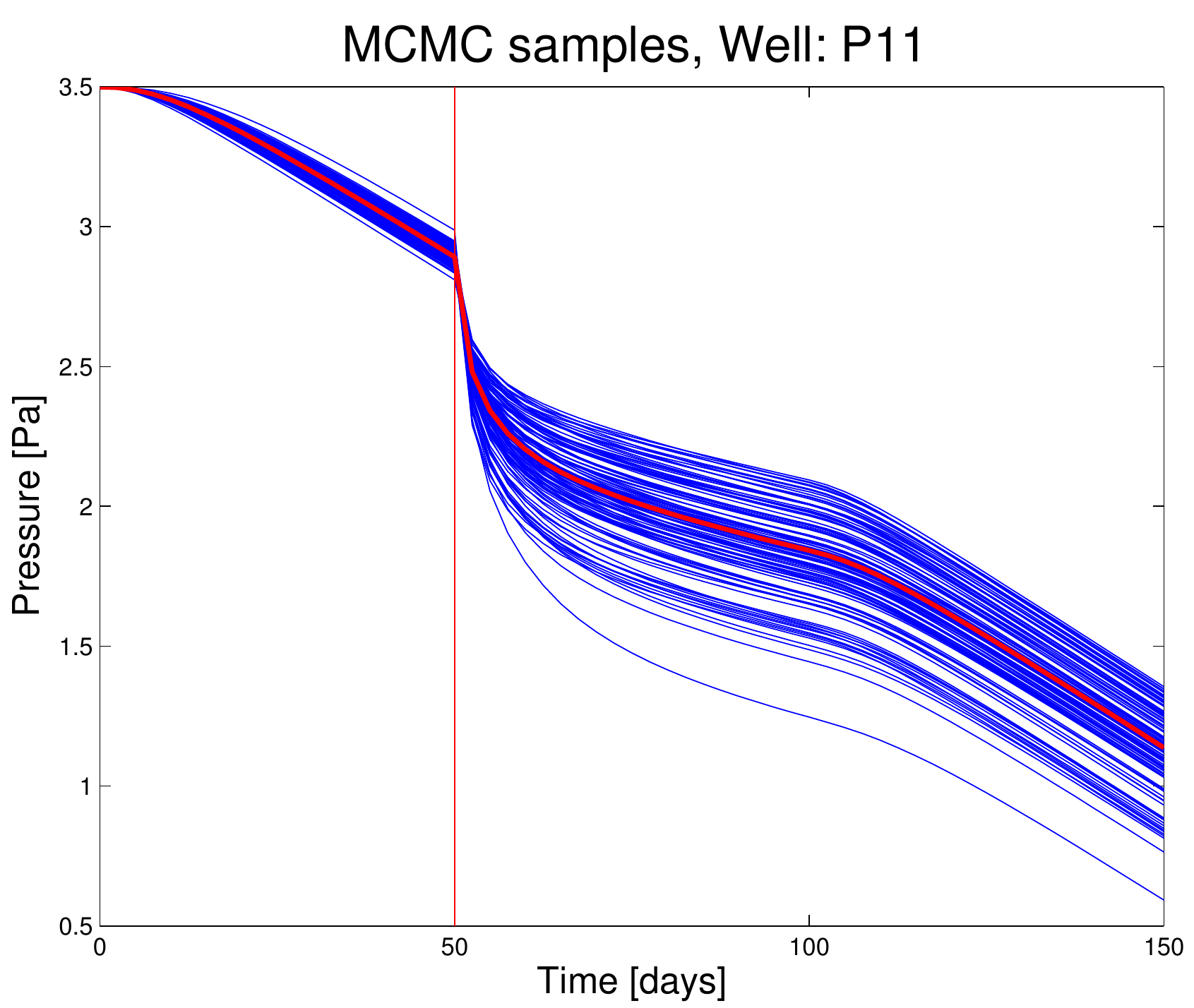}\\
\includegraphics[scale=0.22]{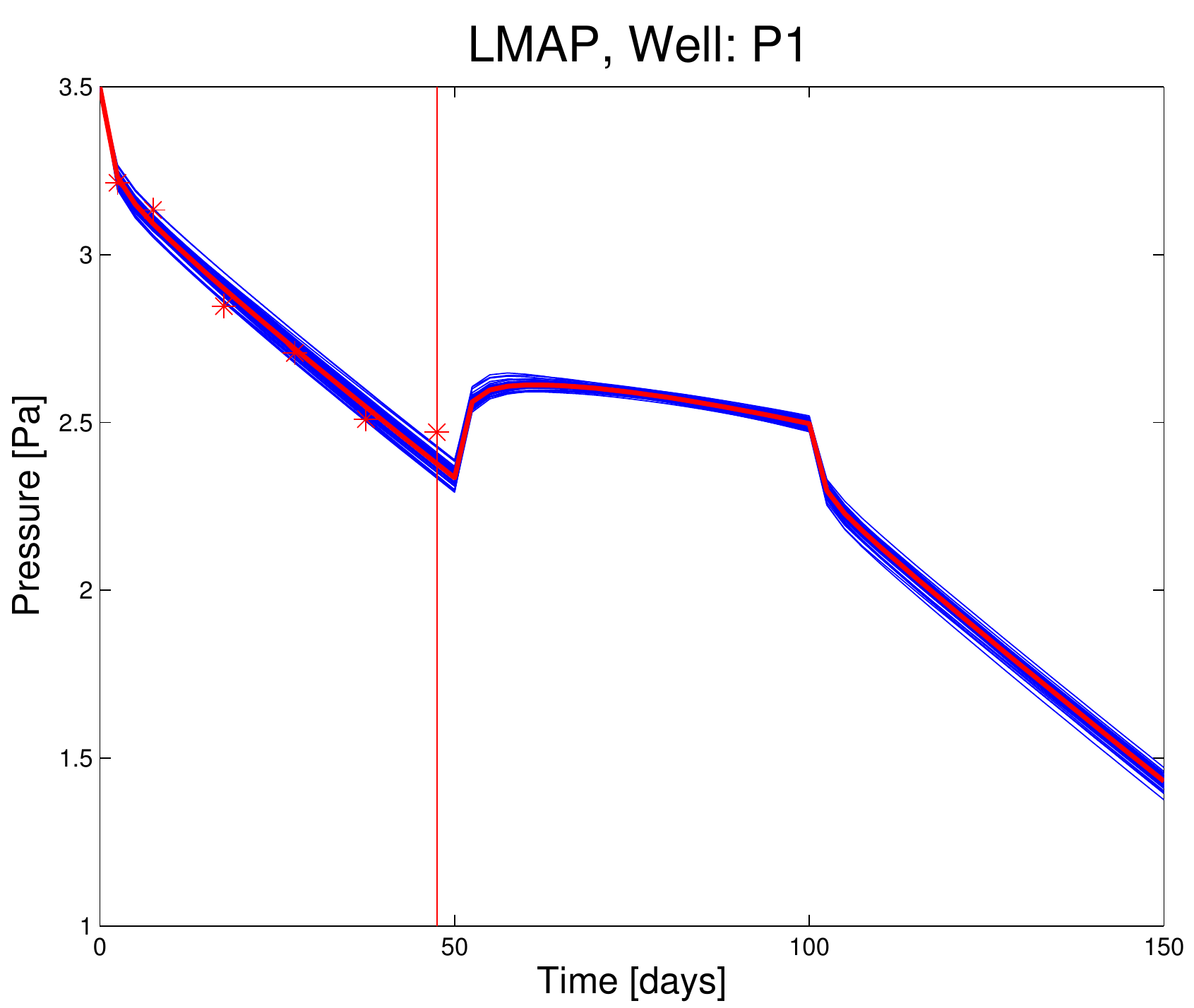}
\includegraphics[scale=0.22]{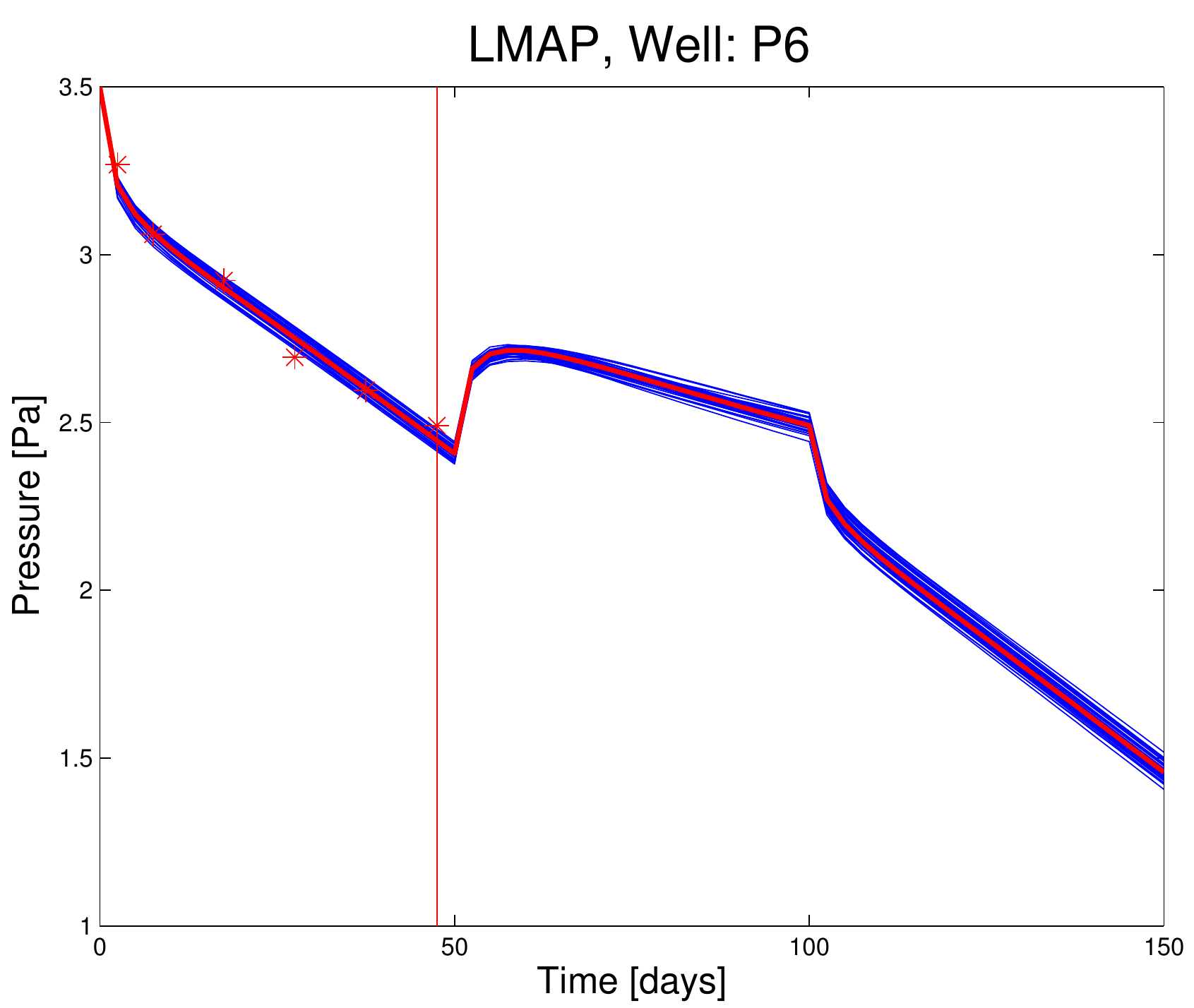}
\includegraphics[scale=0.22]{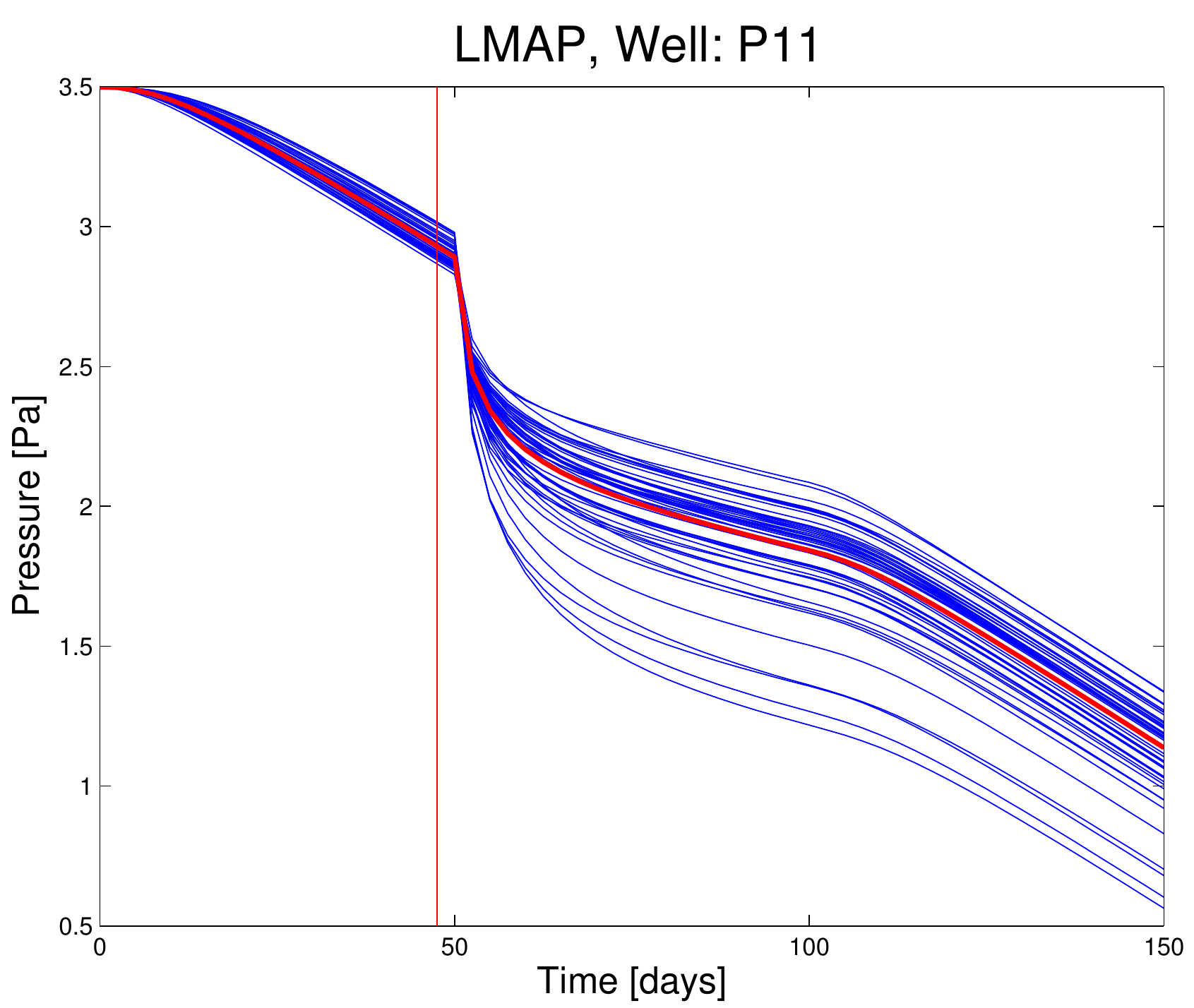}\\
\includegraphics[scale=0.22]{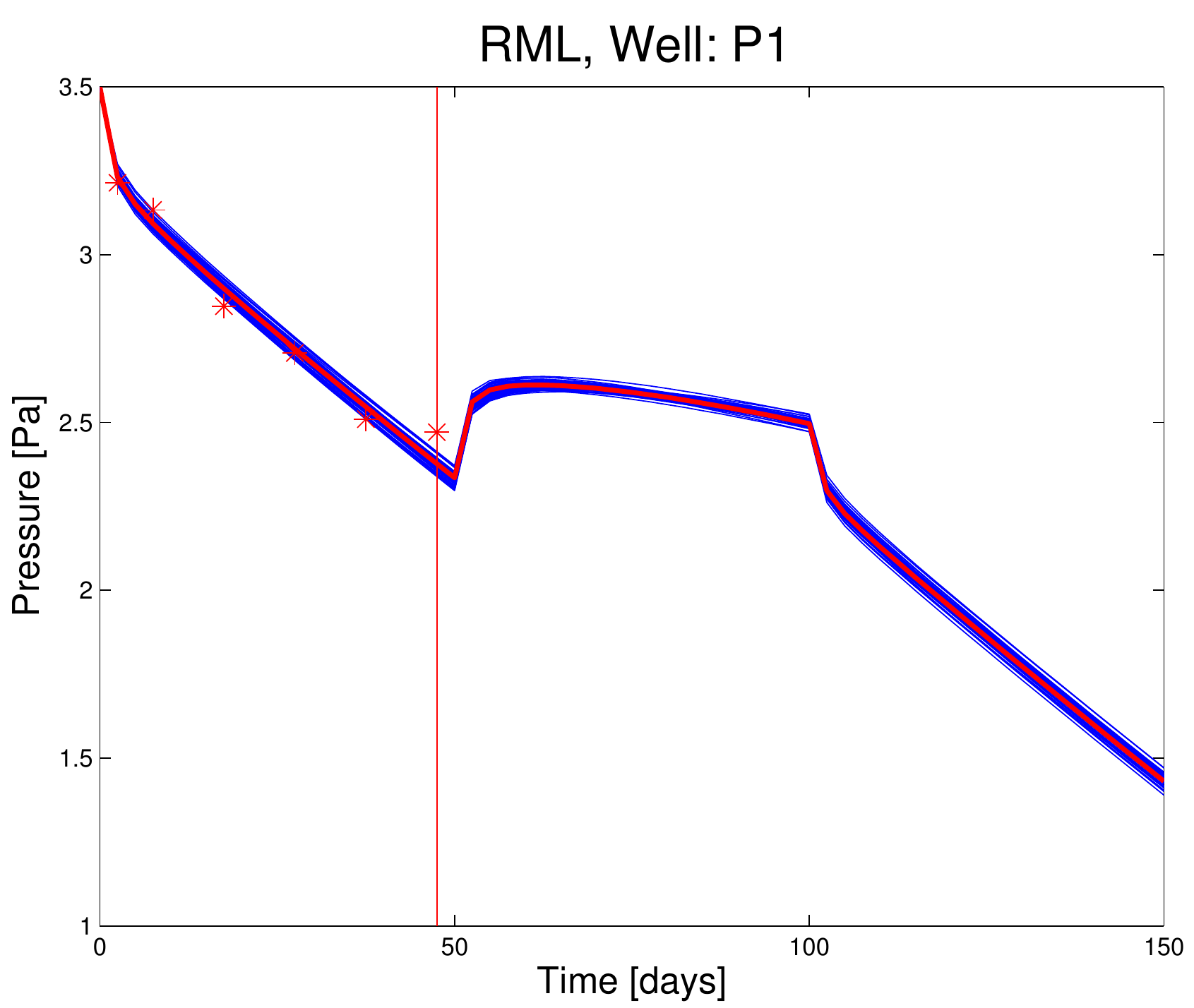}
\includegraphics[scale=0.22]{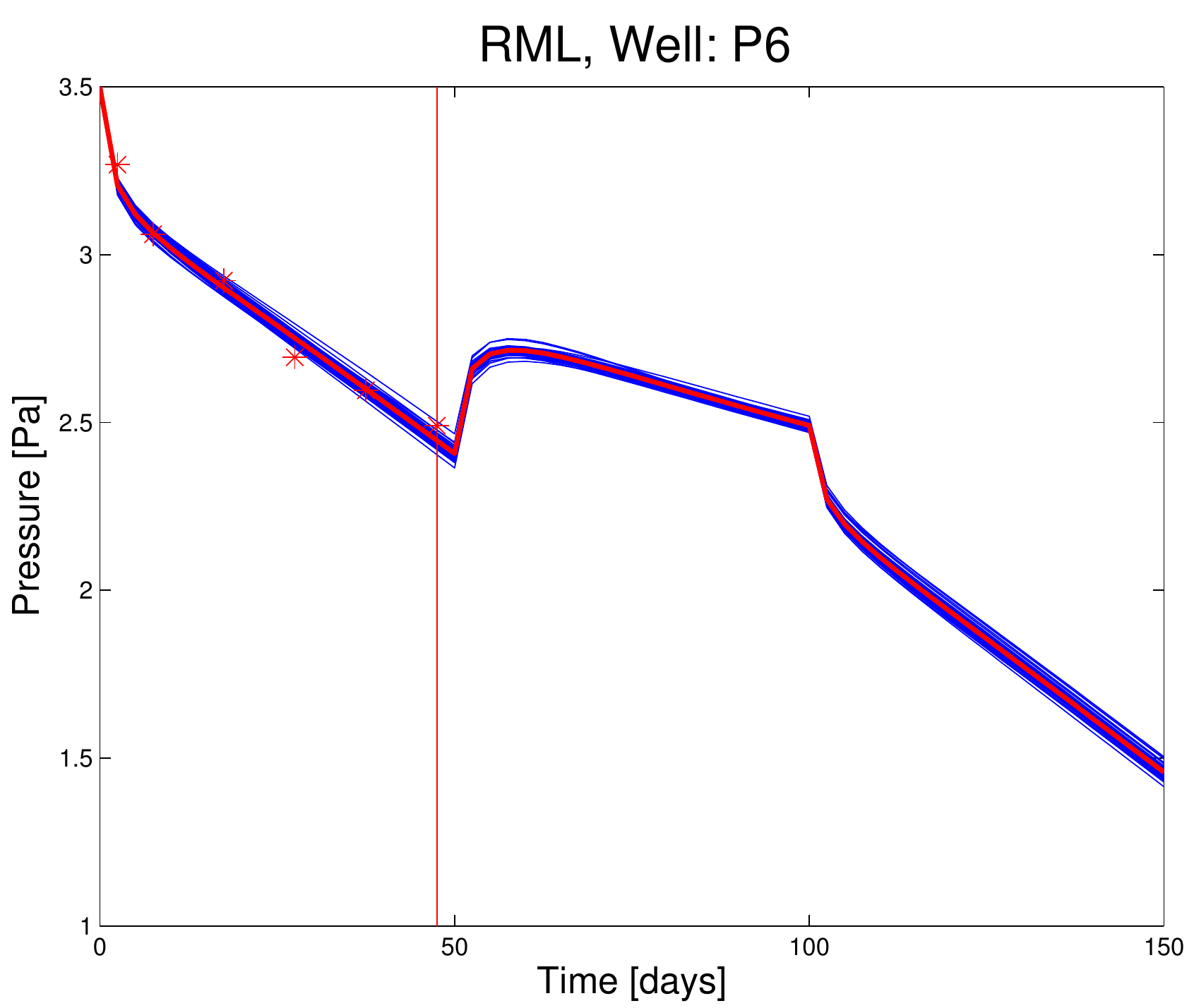}
\includegraphics[scale=0.22]{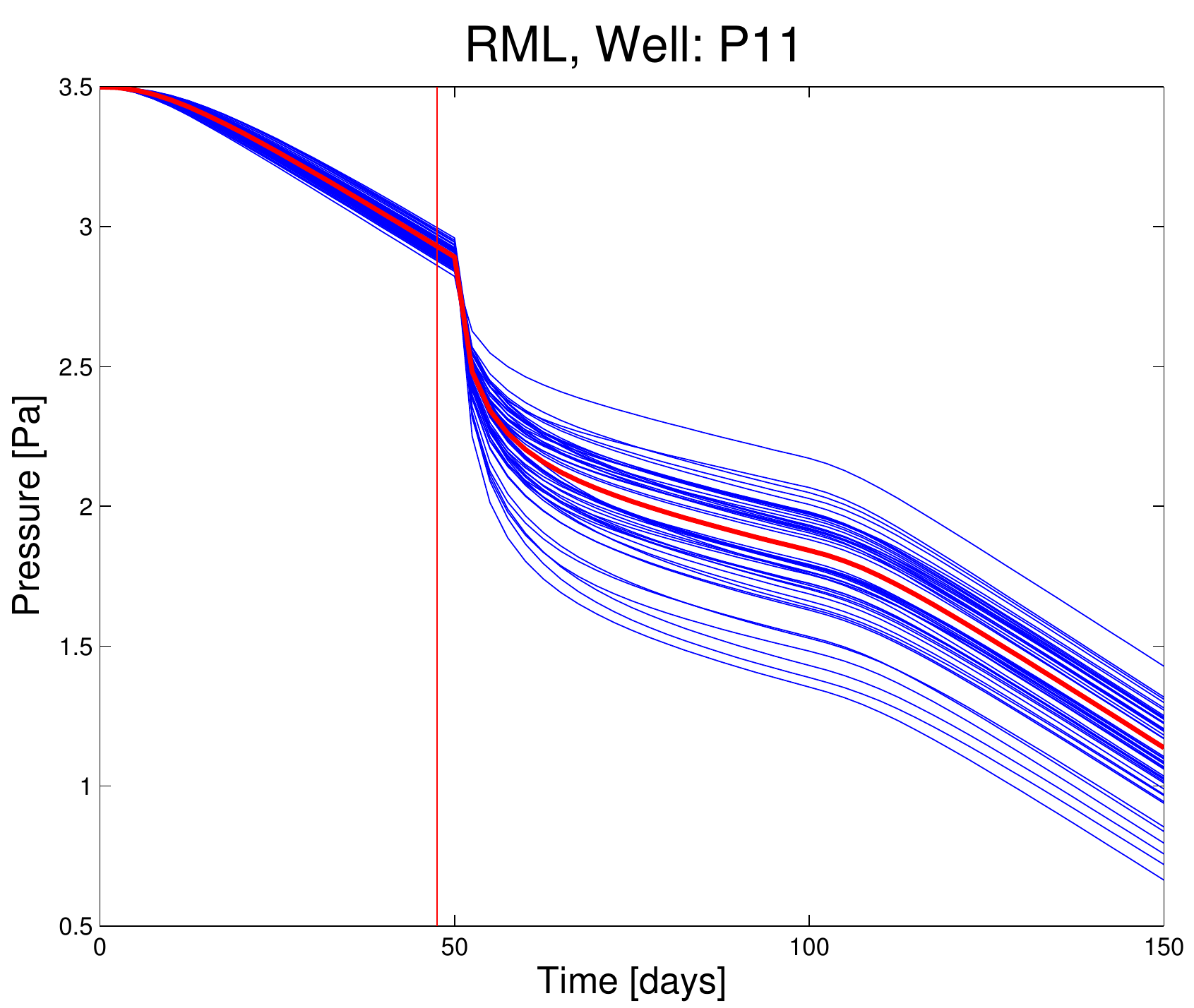}\\
\includegraphics[scale=0.22]{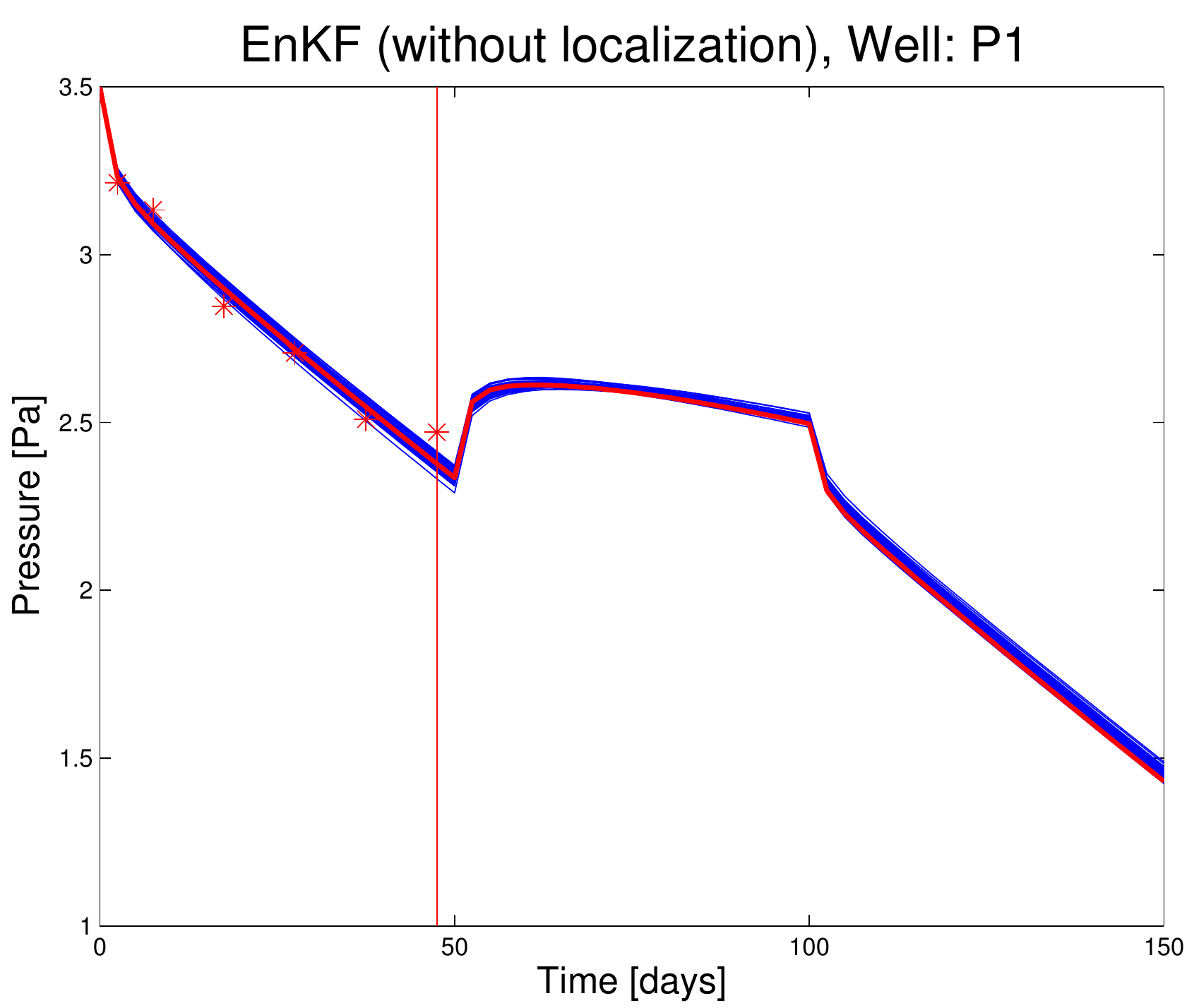}
\includegraphics[scale=0.22]{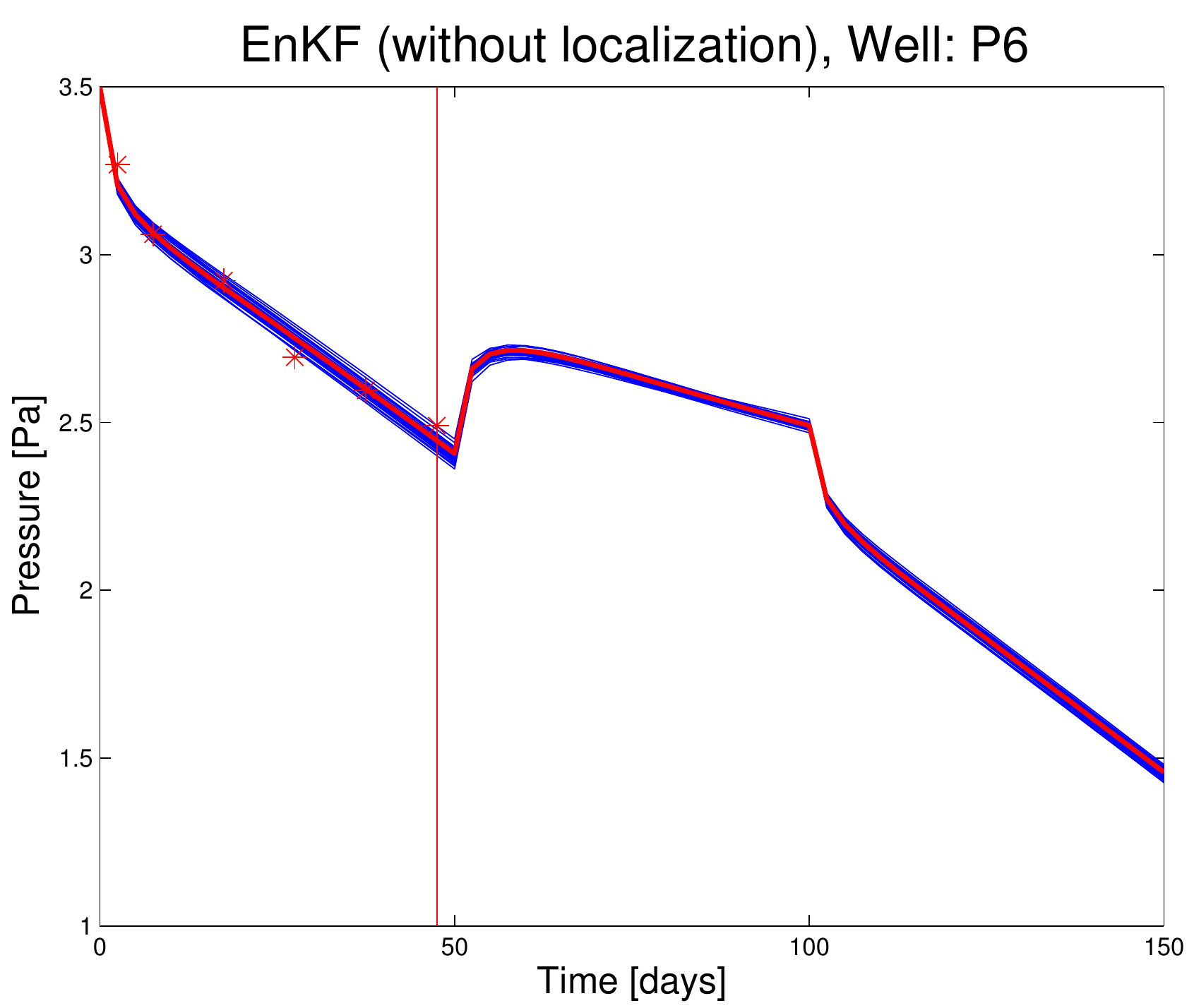}
\includegraphics[scale=0.22]{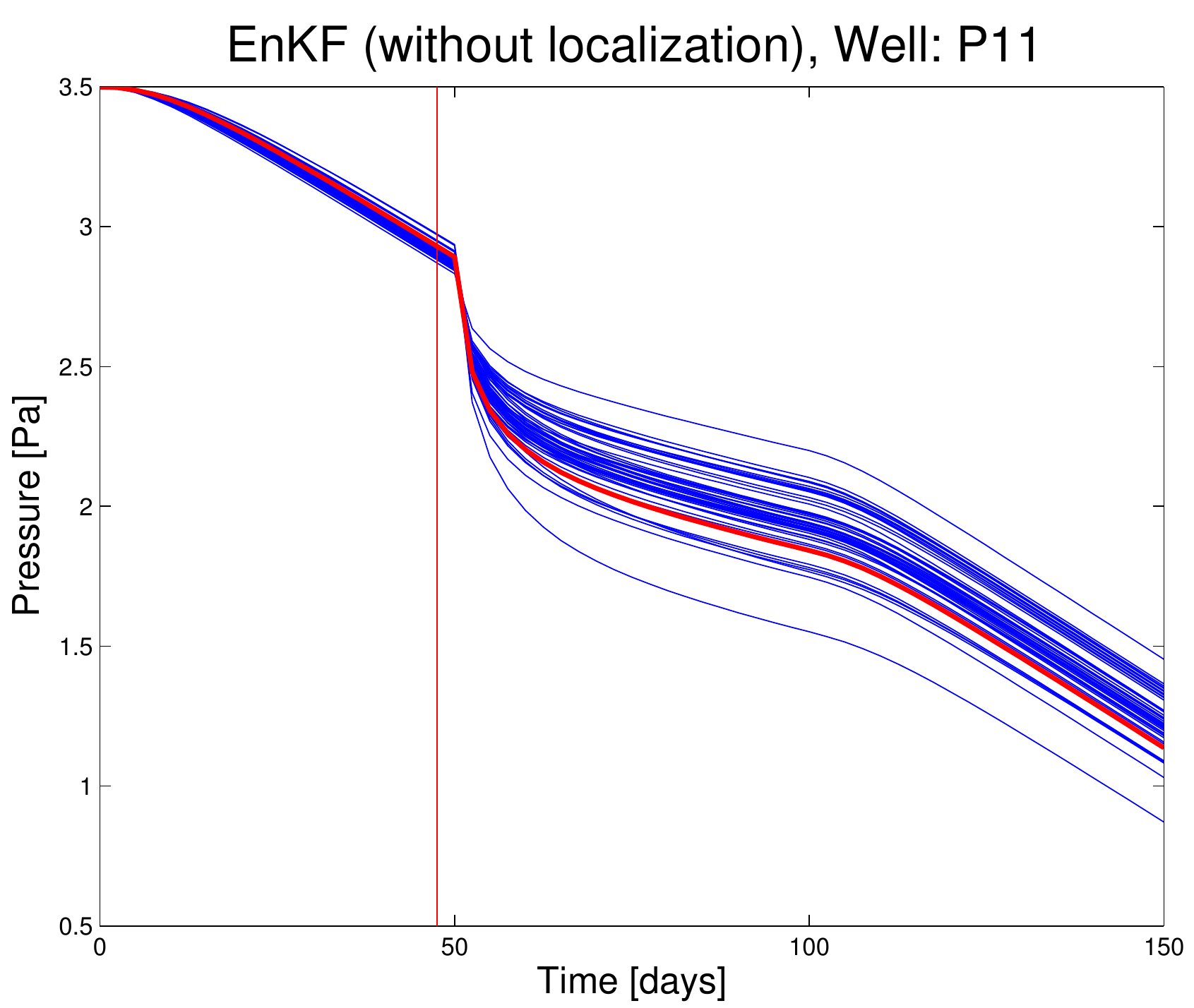}\\
\includegraphics[scale=0.22]{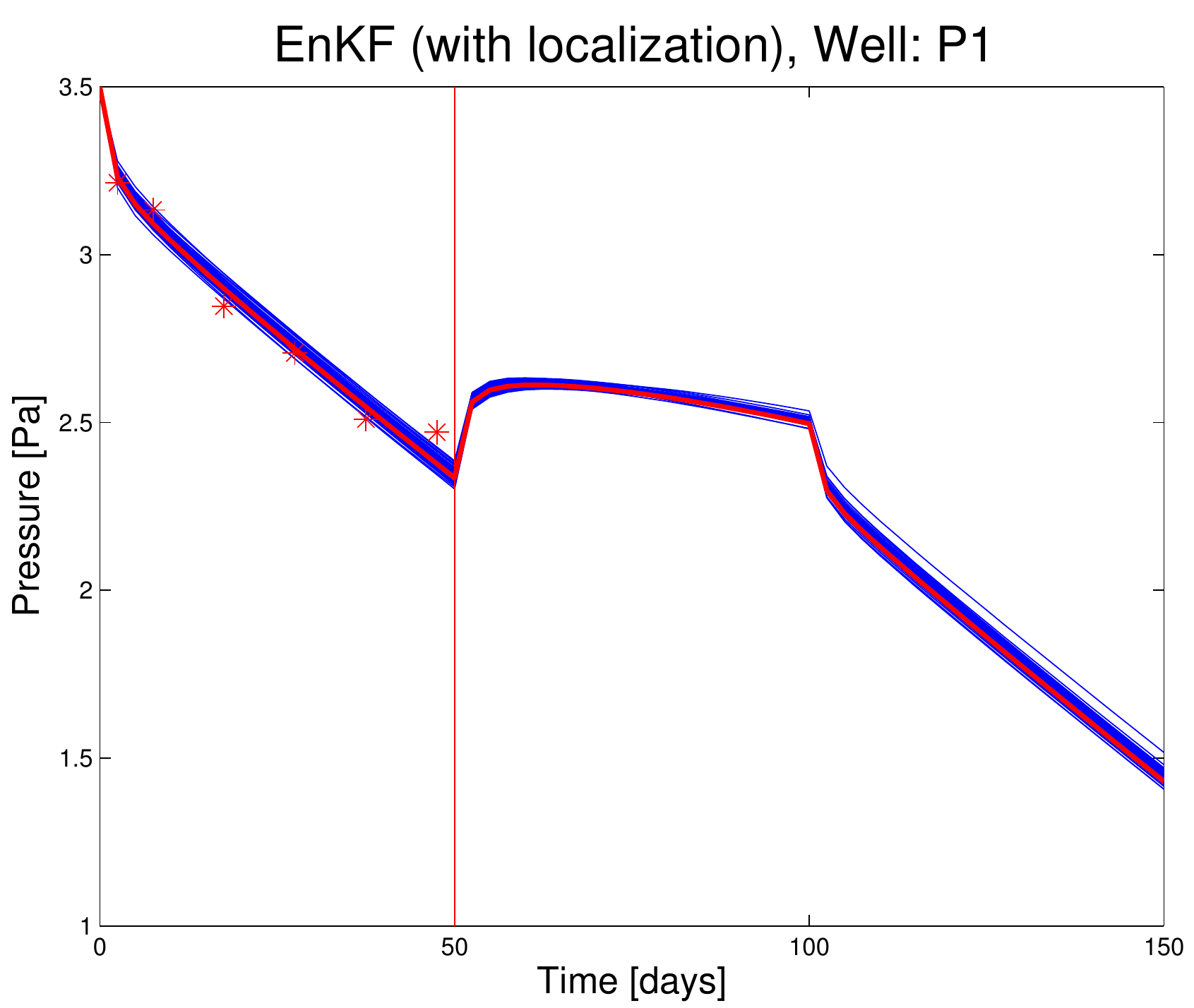}
\includegraphics[scale=0.22]{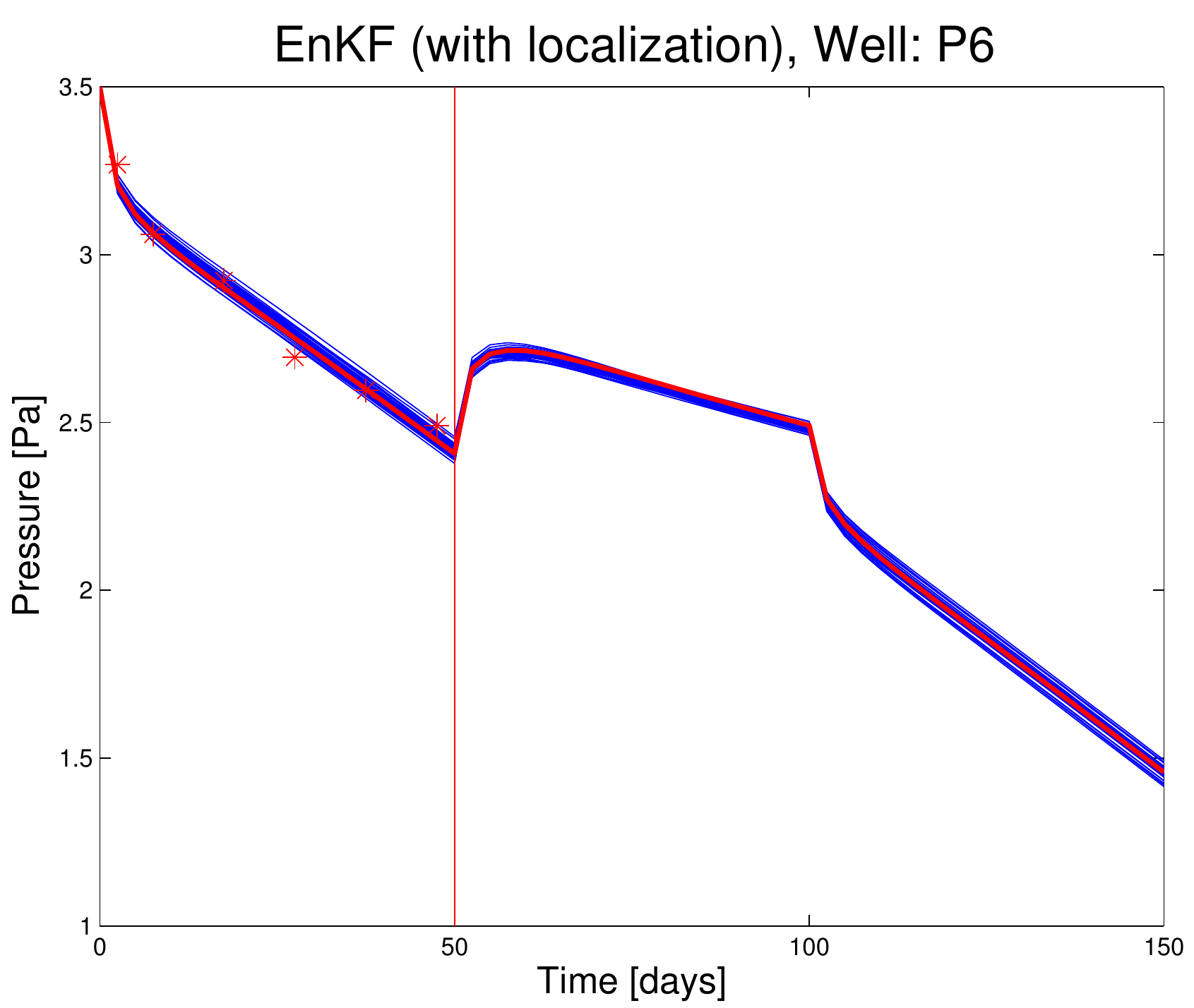}
\includegraphics[scale=0.22]{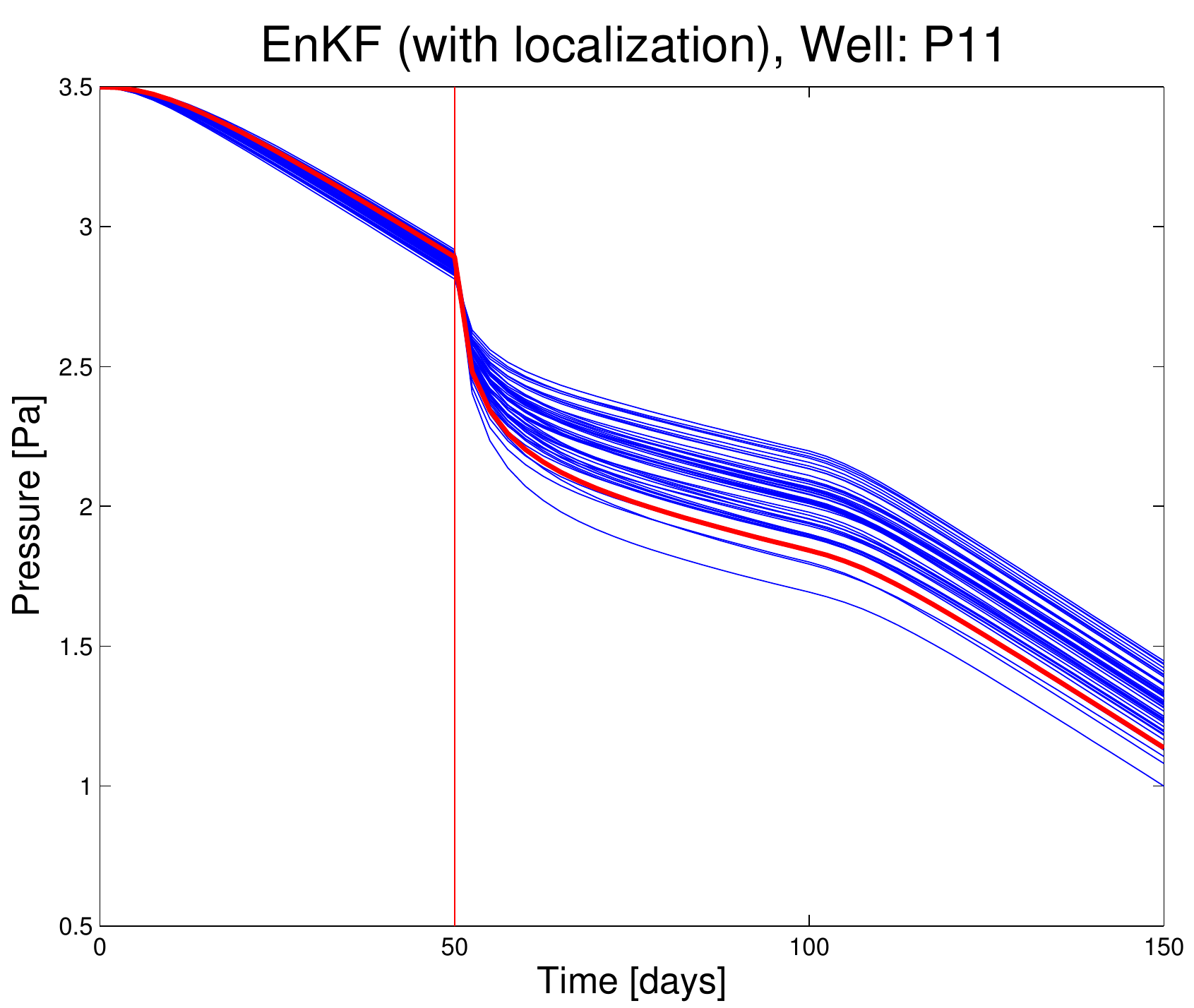}\\
\caption{Single-phase model. Pressure from wells $P_{11}$ (right column), $P_{6}$ (middle column) and $P_{1}$ (left column) simulated with permeabilities sampled from (top to bottom rows) the prior, the posterior, LMAP, RML, EnKF and EnKF with localization. }
\label{Figure6}
\end{figure}

With the previous results we are able to appreciate and evaluate the differences in the approximations provided by each of the Gaussian approximations of the posterior. This posterior is the conditional probability of the unknown (permeability) given production data collected during the $50$ days of simulation. For practical applications it is of particular interest to assess how different Gaussian approximations fare at reproducing the probability distribution of various predicted quantities, with respect to the posterior; in other words to assess how the approximate algorithms fare in the quantification of uncertainty in these predictions. The assessment of performance in terms of the distribution of predictions under the posterior is conducted by creating a new flow scenario as we now describe. Assume that after the initial $50$ days of simulation (that we used to generate synthetic data), we now drill new wells labeled by $P_{10}$, $P_{11}$, $P_{12}$ and $P_{13}$ in Figure \ref{Figure1}. These new wells are operated at constant production rate of $60 \textrm{m}^{3}/\textrm{day}$ during $100$ days. During this $100$ days of forecast, the old wells $P_{1},\dots,P_{9} $ are first shut-down for a pressure build-up time window of $50$ days, followed by a constant production of $60 \textrm{m}^{3}/\textrm{day}$ during the rest $50$ days. In Figure \ref{Figure6} we show, as a function of time, the pressure at the well locations $P_{9}$, $P_{10}$ and $P_{13}$. The first $50$ days corresponds to the data assimilation phase and the subsequent $100$ days are the prediction. In the first row of Figure \ref{Figure6} we display the pressure obtained from the reservoir simulation with 100 permeabilities obtained from the prior distribution (\ref{eq:5.1}). The second row corresponds to the pressure obtained from the simulation with the samples from the posterior obtained from independent MCMC chains. Subsequent rows of Figure \ref{Figure6} corresponds to pressure obtained from simulating the permeabilities obtained from some of the Gaussian approximations under consideration. The vertical line divides data assimilation phase from the forecast. Additionally, since we are interested in the performance with respect to the posterior, for each curve presented in Figure \ref{Figure6} we include a red curve of the pressure at the corresponding well location obtained by simulating the mean of the posterior distribution (i.e. top-left field of Figure \ref{Figure4}). 

Note that the uncertainty quantified by the posterior and the Gaussian approximations is considerably small at the wells $P_{1},\dots,P_{9}$ where measurements were collected. In contrast, large uncertainty in the forecast is observed for the new wells $P_{10},\dots,P_{13}$ for which data was not available during the data assimilation phase. Note for example that in $P_{11}$, the posterior is visually close to the prior, indicating the uninformative effect of the data at the location of $P_{11}$. For the new wells where uncertainty is larger, we can appreciate the performance on the Gaussian approximations. Note from Figure \ref{Figure6} that the EnKF (without localization) at well $P_{11}$ underestimates the uncertainty in the model predictions. 

\begin{figure}
\includegraphics[scale=0.335]{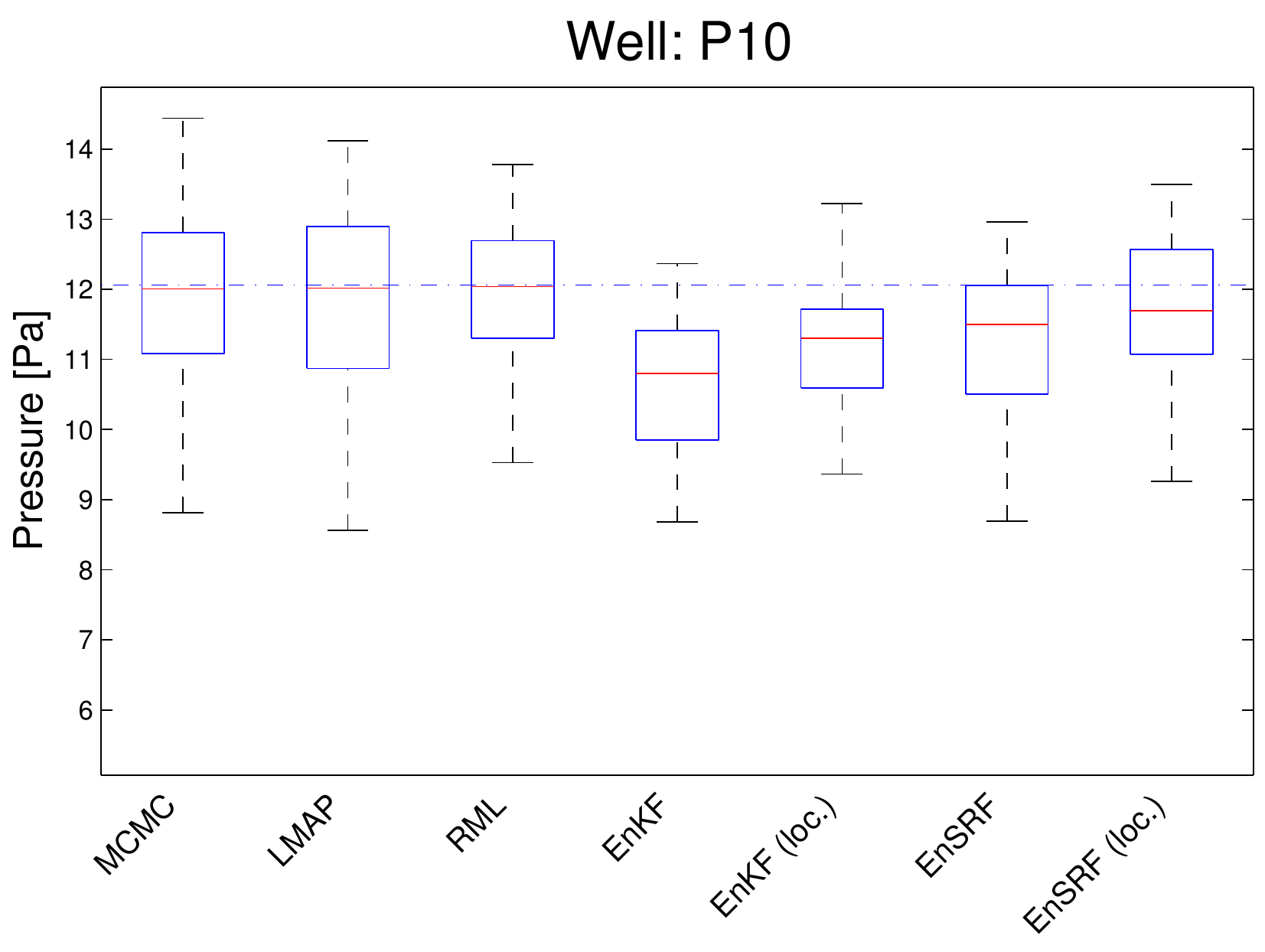}
\includegraphics[scale=0.335]{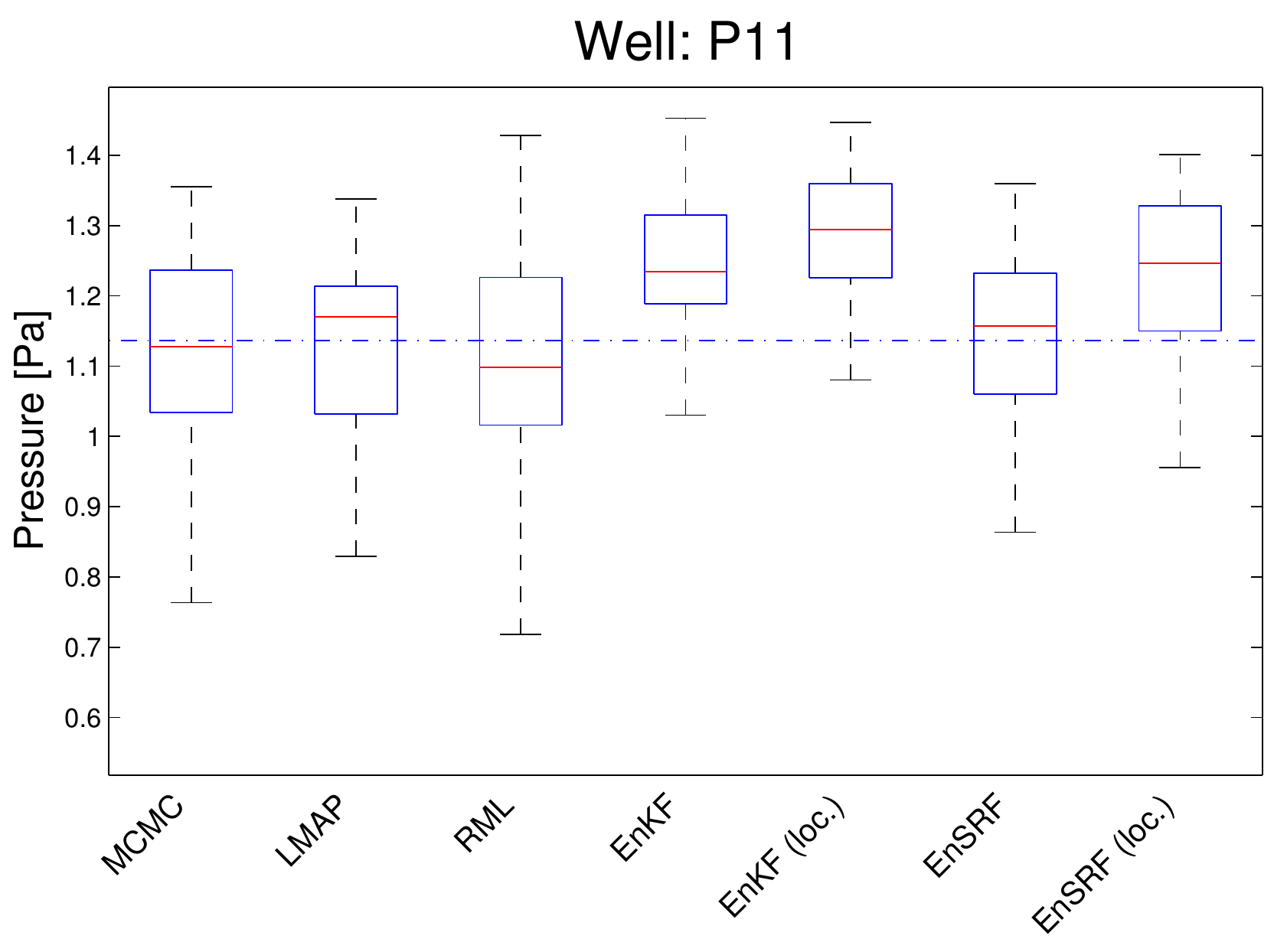}\\
\includegraphics[scale=0.335]{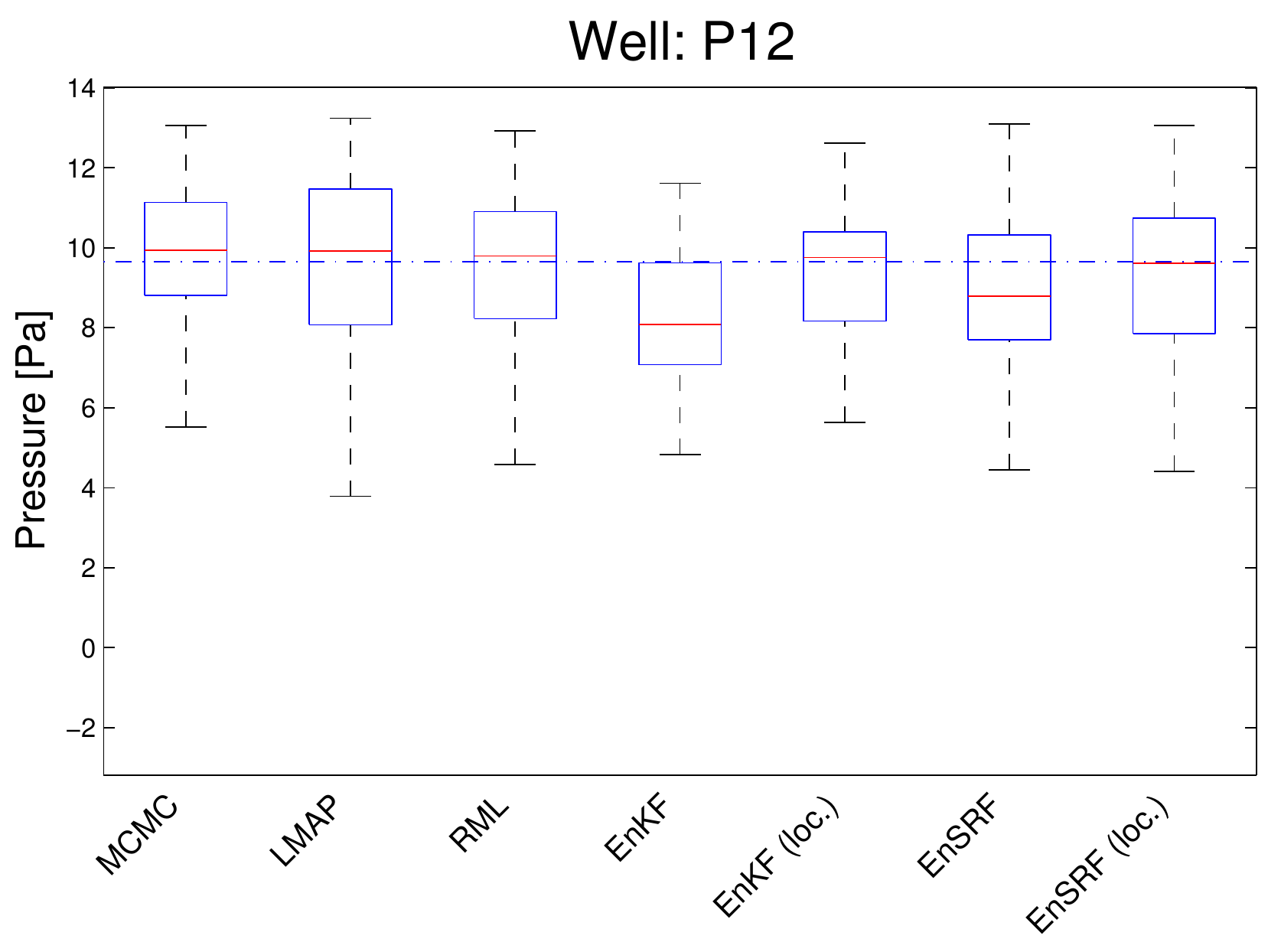}
\includegraphics[scale=0.335]{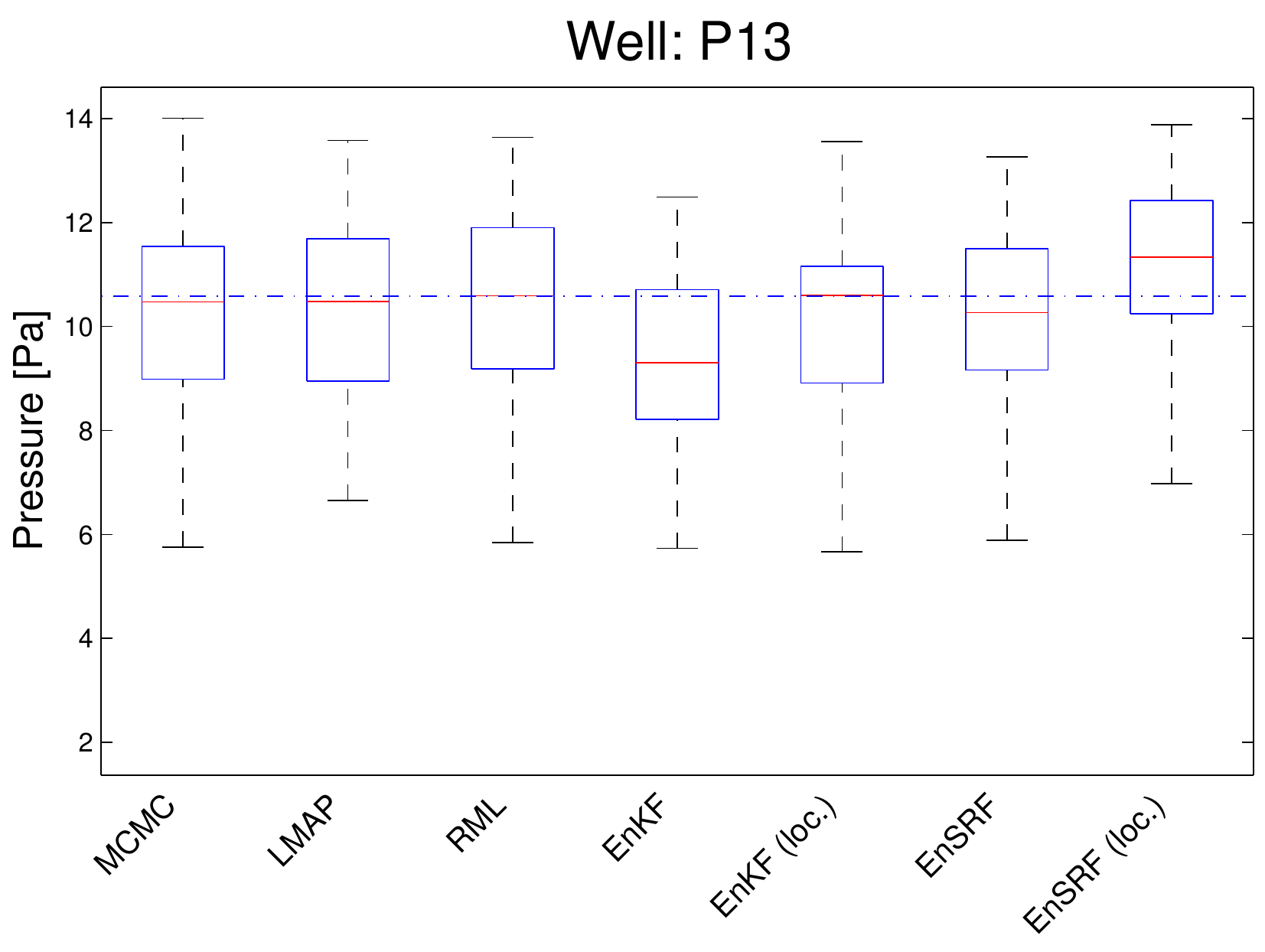}
\caption{Single-phase model. Distribution of pressure at wells $P_{10},\dots, P_{13}$ at the final time of simulation $t=150~ \textrm{days}$}  
\label{Figure7}
\end{figure}

In Figure \ref{Figure7} we display the distribution of the pressure at the new wells $P_{10},\dots,P_{13}$ at the final time $t=150~\textrm{days}$. The horizontal line correspond to the value of the pressure at the corresponding location obtained from simulating the model with the posterior mean. From \ref{Figure6}  and \ref{Figure7} we observe that LMAP and RML provide a better approximation to the predicting distribution than the EnKF-based methods. Since RML produces the best approximation of the posterior in terms of mean with a reasonable approximation of the variance, the associated approximation of the predicting distribution is the most accurate among the techniques considered here. Our results with respect to the optimality of RML for this experiment are similar to those reported in \cite{Oliver2} where a single-phase reservoir model was also utilized. As we will see in the next experiments, a less favorable performance of RML is 
observed for a two-phase reservoir model.

\begin{table}
\caption{Evaluation of Gaussian approximations on the single-phase model}
\label{Table2}       
\begin{tabular}{lccc}
\hline\noalign{\smallskip}
Method & Relative error &  Relative error & Computational cost\\
 &  in the mean $\epsilon_{u}$& in the variance $\epsilon_{\sigma}$ &  [Forward model runs]\\
\noalign{\smallskip}\hline\noalign{\smallskip}
MCMC &0.000 & 0.000 &  $5.5\times 10^{7}$\\
MAP &0.030 & 0.094 &  $2.5\times 10^{1}$\\
LMAP ($N_{e}=50$)&0.179 & 0.259 &  $2.5\times 10^{1}$\\
RML ($N_{e}=50$)&0.154 & 0.258 &  $1.25\times 10^{3}$\\
EnKF ($N_{e}=50$)&0.643& 0.417 &  $5.0\times 10^{1}$\\
EnKF (localization, $N_{e}=50$)  & 0.546 &0.288 &$5.0\times 10^{1}$\\
EnSRF ($N_{e}=50$) &0.519 & 0.267&  $5.0\times 10^{1}$\\
EnSRF (localization, $N_{e}=50$)  & 0.445 &0.208 &$5.0\times 10^{1}$\\ \hline
EnKF ($N_{e}=1250$)& 0.192 & 0.075&$1.25\times 10^{3}$\\
EnKF ($N_{e}=2500$)& 0.120 & 0.089&$2.5\times 10^{3}$\\
EnKF ($N_{e}=5000$)& 0.102 & 0.094&$5.0\times 10^{3}$\\
\noalign{\smallskip}\hline
\end{tabular}
\end{table}

\subsection{Oil-water reservoir: Small number of wells}

\begin{figure}
\includegraphics[scale=0.25]{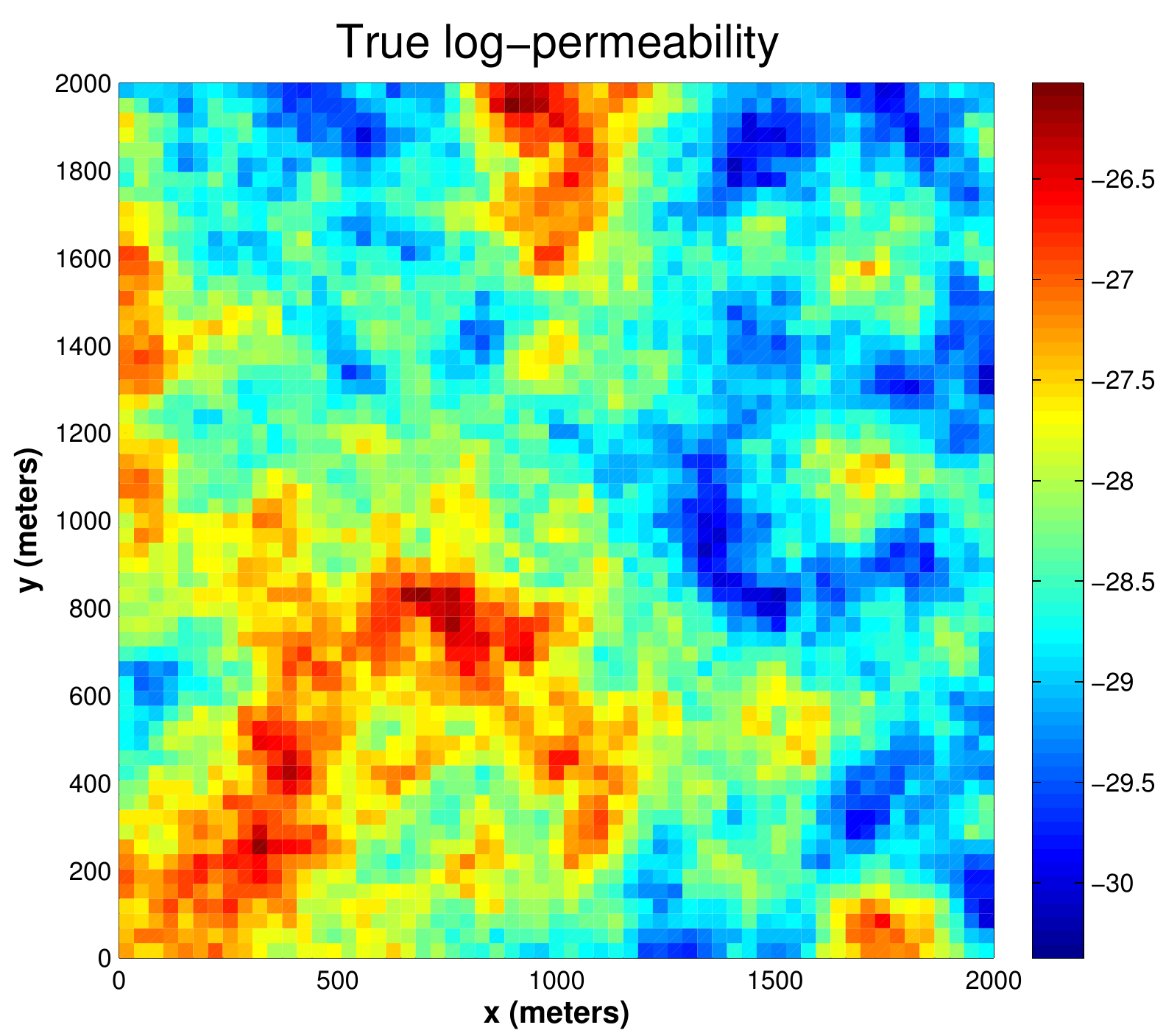}
\includegraphics[scale=0.25]{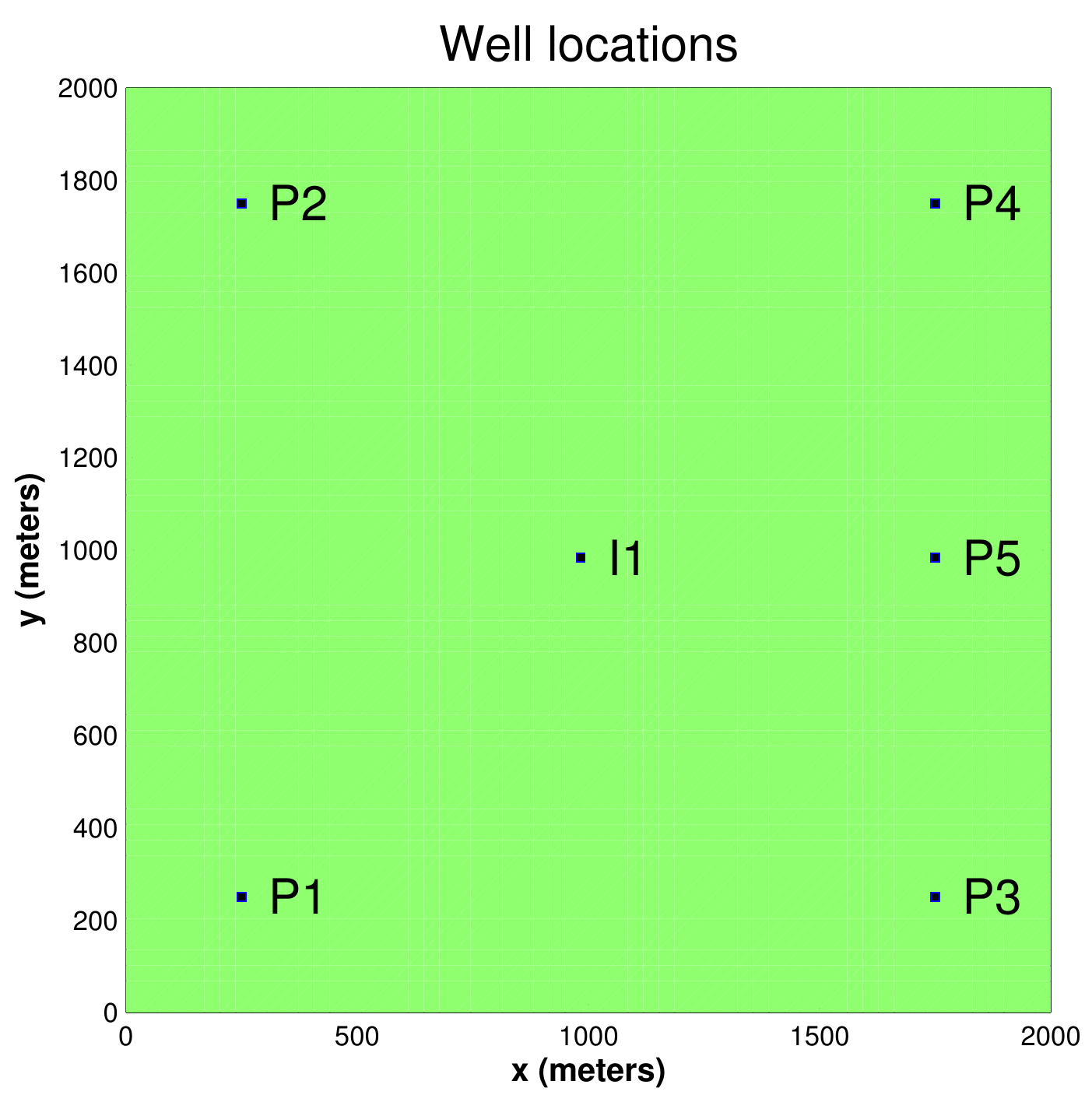}
\includegraphics[scale=0.25]{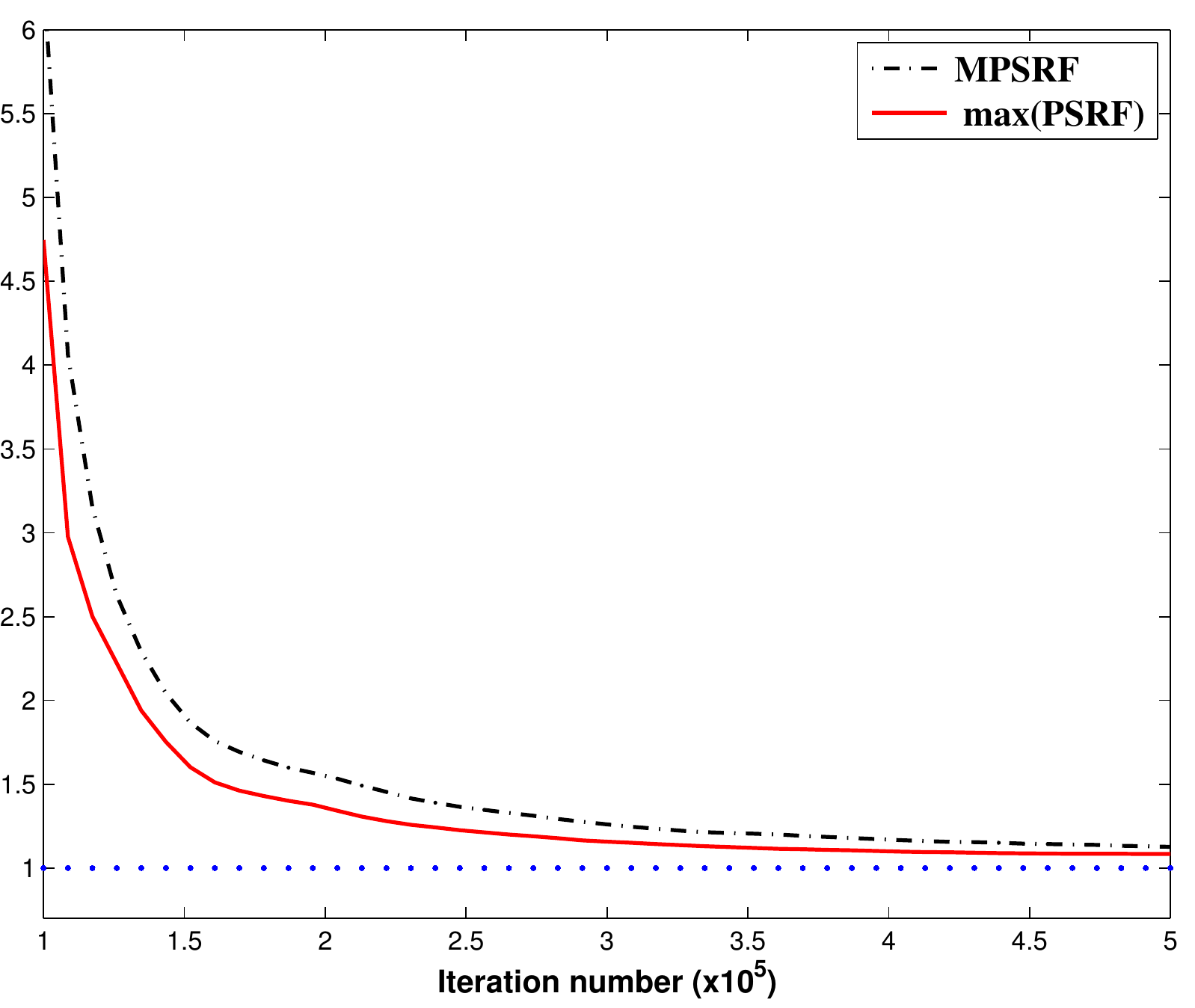}
\caption{Two-phase model (small number of wells). Left: True log-permeability [$\log{\textrm{m}^2}$]. Middle: Well configuration. Right: Gelman-rubin diagnostic}  
\label{Figure8}
\end{figure}

In this subsection we consider the oil-water reservoir model described in Section \ref{ReservoirModels}. The reservoir is a square $D=[0,L]\times [0,L]$ with a five-spot well configuration consisting on production wells $P_{1},\dots P_{4}$ and one injection well $I_{1}$ as displayed in Figure \ref{Figure8}, middle (well $P_5$ will play a role in later discussions). Relevant information concerning this model is displayed in Table \ref{Table1}. The prior distribution of log-permeability is defined by the Gaussian measure defined in (\ref{eq:5.1}) with same parameters except $\kappa$ which for this experiment is $\kappa=4.0$.  With these parameters, similar realizations to those of Figure \ref{Figure2} are obtained but with a range of variability with respect to the prior mean $\overline{u}$ increased by a factor of $\sqrt{2}$.  

Similarly to the previous example, for the generation of synthetic data we define the ``true log-permeability'' $u^{\dagger}$ displayed in Figure \ref{Figure8} (left). This ``true log-permeability'' is generated from the prior distribution described in the preceding paragraph. Note that $u^{\dagger}$ is the same as the one used for the previous experiment (see Figure \ref{Figure1} (left)) but with the magnitude of $u^{\dagger}-\overline{u}$ multiplied by $\sqrt{2}$. The generation of synthetic data is now conducted by computing $(p,s)$ from (\ref{eq:2.7})-(\ref{eq:2.7B}) with $u=u^{\dagger}$. Equation (\ref{eq:2.14}) is then used to compute $G(u^{\dagger})$ and zero mean Gaussian random error $\eta$ is added to generate data $y=G(u^{\dagger})+\eta$. The measurement times used in  (\ref{eq:2.11})-(\ref{eq:2.12}) are $t_{n}=0.67n~\textrm{years}$, $n=\{1,\dots,7\}$. According to the structure of (\ref{eq:2.13})-(\ref{eq:2.14}), $\eta$ has the following form
\begin{eqnarray}\label{eq:5.5}
\eta=(\eta_1, \dots, \eta_{N_{M}}),\qquad 
\eta_{n}=(\eta_{n}^{1,I}, \cdot, \eta_{n}^{N_{I},I},\eta_{n}^{1,P}, \dots, \eta_{n}^{N_{P},P})
\end{eqnarray}
We generate the components of $\eta$ as follows
\begin{eqnarray}\label{eq:5.6}
\eta_{n}^{1,I}\sim N(0,(3.2\times 10^{4}\textrm{Pa})^2)\qquad \eta_{n}^{1,P} \sim N(0,(0.25\textrm{m}^{3}/\textrm{day})^2) \nonumber \\
 \eta_{m}^{2,P},\eta_{n}^{3,P} ,\eta_{n}^{4,P}  \sim N(0,(0.02\textrm{m}^{3}/\textrm{day})^2).
\end{eqnarray}
for all $n\in \{1,\dots, 7\}$. In the previous definitions we consider larger measurement error at the production well $P_{1}$ since larger total flow rates are obtained. This is caused by the larger permeability region around $P_{1}$ and so the early water breakthrough at this well. 

\begin{figure}
\includegraphics[scale=0.55]{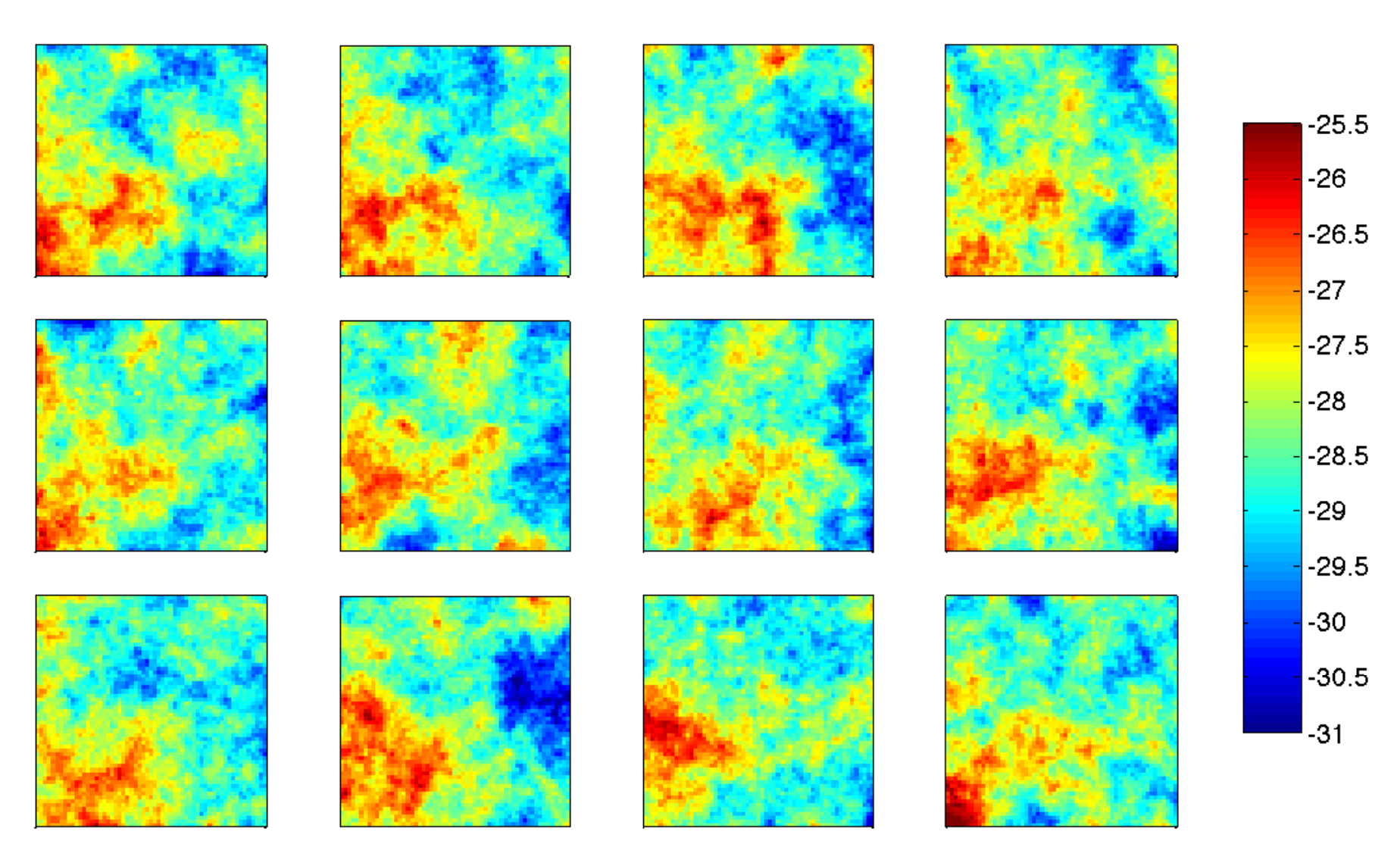}
\caption{Two-phase model (small number of wells). Samples from the posterior distribution (characterized with MCMC) [$\log{\textrm{m}^2}$]}  
\label{Figure9}
\end{figure}

The pCN-MCMC Algorithm \ref{al:MCMC} with $\beta=0.015$ is applied to this problem with the synthetic data previously described. 110 chains are generated starting from independent draws from the prior distribution. A burn-in period of $1.5\times 10^{4}$ was observed after which the chains were run $5\times 10^5$ iterations. By using the same  Gelman-Rubin indicated in the previous example, we determine convergence of our chains. This numerical evidence is presented in Figure \ref{Figure8} where, after $5\times 10^5$ iteration, both the max PRSF and MPRSF for the highest energy models converges to one. Uncorrelated samples (from independent chains) are shown in Figure \ref{Figure9}.

\begin{figure}
\includegraphics[scale=0.65]{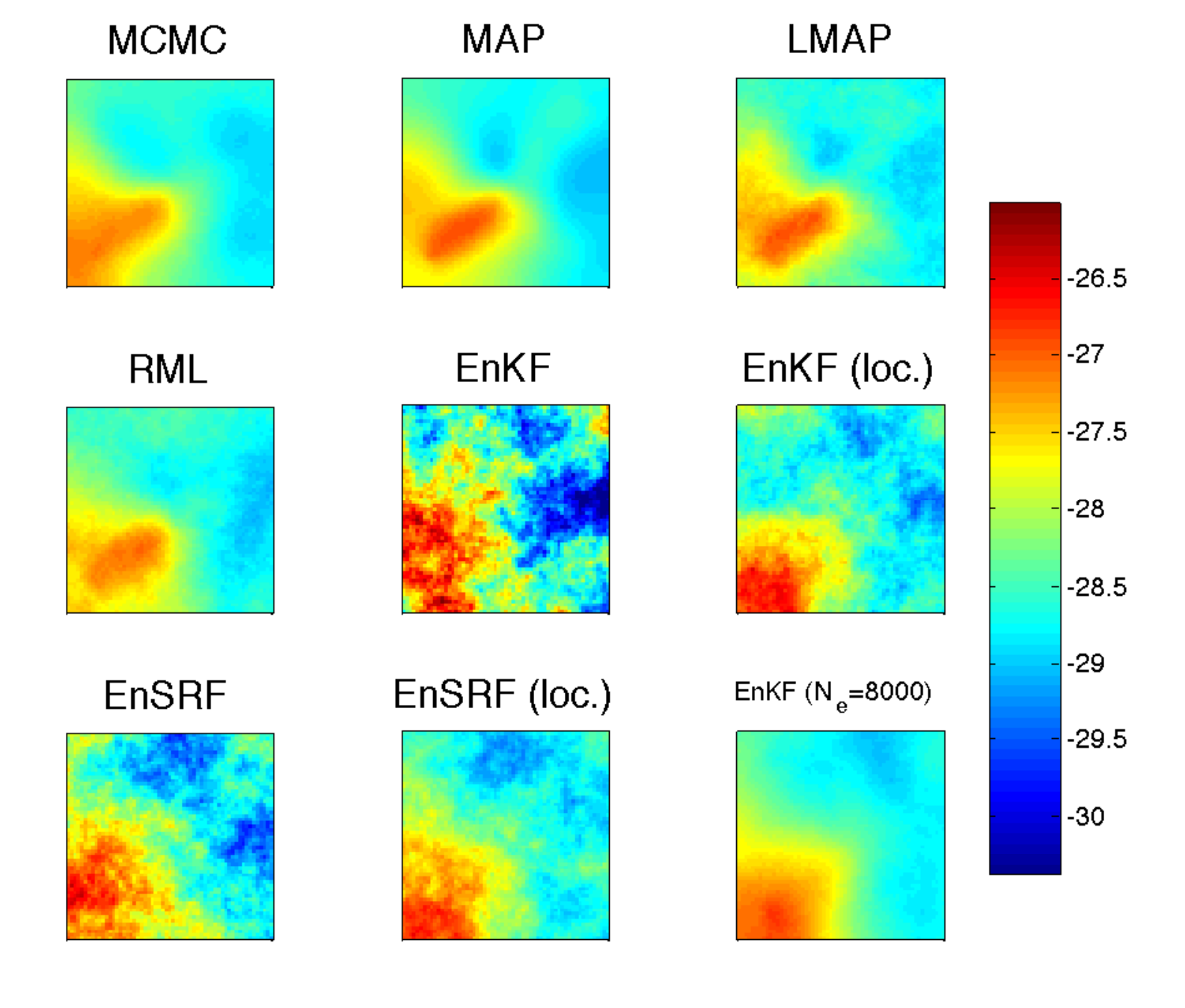}
\caption{Two-phase model (small number of wells). Mean of the posterior distribution (characterized with MCMC) and Gaussian approximations [$\log{\textrm{m}^2}$]}  
\label{Figure10}
\end{figure}

\begin{figure}
\includegraphics[scale=0.65]{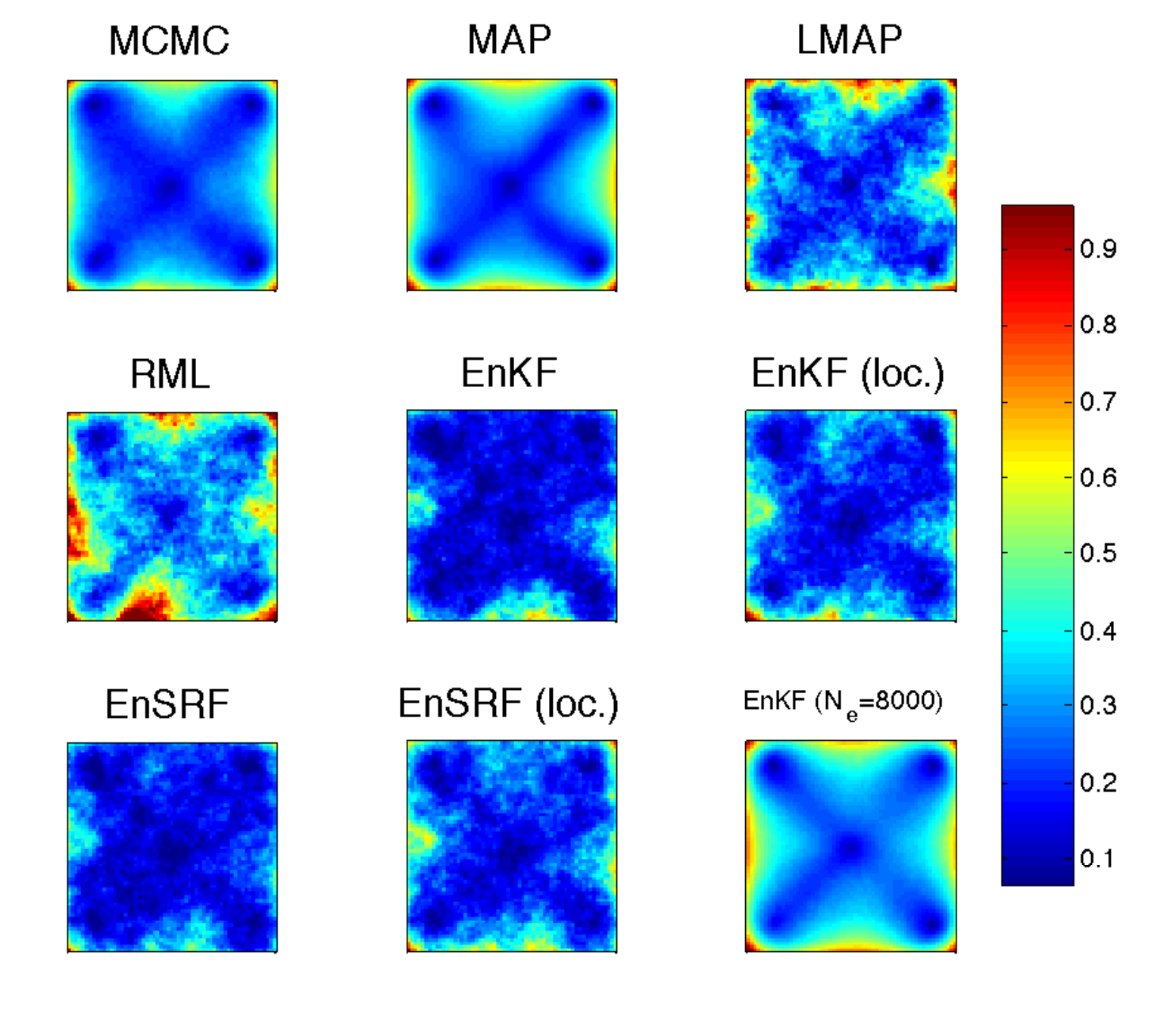}
\caption{Two-phase model (small number of wells). Variance of the posterior distribution (characterized with MCMC) and Gaussian approximations [$(\log{\textrm{m}^2})^2$]}  
\label{Figure11}
\end{figure}

Samples of the posterior from our converged chains define our gold standard. These are then used to compare the performance of  Gaussian approximations in terms of mean and variance.  Analogous to the previous example, we use an ensemble with of size $N_{e}=50$ and compute the mean and variance with LMAP, RML, EnKF, EnKF with localization, EnSRF and EnSRF with localization. Relative errors of the mean and variance (see expression (\ref{eq:5.3}) ) with respect to the posterior distribution are provided in Table \ref{Table3}. In Figure \ref{Figure10} and Figure \ref{Figure11} we present the mean and variance, respectively. 

In contrast to the previous experiment, the error in the approximation given by $N(u_{MAP},C_{MAP})$ is relatively large. The relative errors of mean and variance are $27\%$ and $14\%$ respectively. Similar to the previous experiment, RML provided the best approximation of the posterior in terms of the mean. However, the variance of the posterior was significantly overestimated by RML. The performance of the standard EnKF for $N_{en}=50$ was very poor. From Figure \ref{Figure10} we observe large values of the estimated field, which are typically found in standard EnKF applications with small ensemble sizes. Nevertheless, for the same size, covariance localization has a positive effect by reducing the error in the mean and variance with respect to the posterior. Similar to the previous experiment, EnKF and EnKF with localization were outperformed by the corresponding EnSRF implementations. As in the previous experiment, EnSRF with localization provided, among the techniques with $N_{e}=50$, the best approximation in terms of variance.

Unlike the preceding experiment, here we observe that increasing the size of the ensemble does not result in a decrease of the error with respect to the mean. This can be observed from at the end of Table \ref{Table3} where we report the results of EnKF implementations for $N_{en}=1000$, $N_{en}=3000$, $N_{en}=8000$. Both in terms of mean and variance, EnKF for large ensembles does not exhibit convergence to the posterior. Note that EnKF with $N_{en}=3000$ corresponds to the same computational cost of RML. Yet, the latter provides a better approximation in terms of the mean. Among all the techniques, LMAP provides reasonable approximations of both the mean and variance of the posterior. 

\begin{figure}
\includegraphics[scale=0.22]{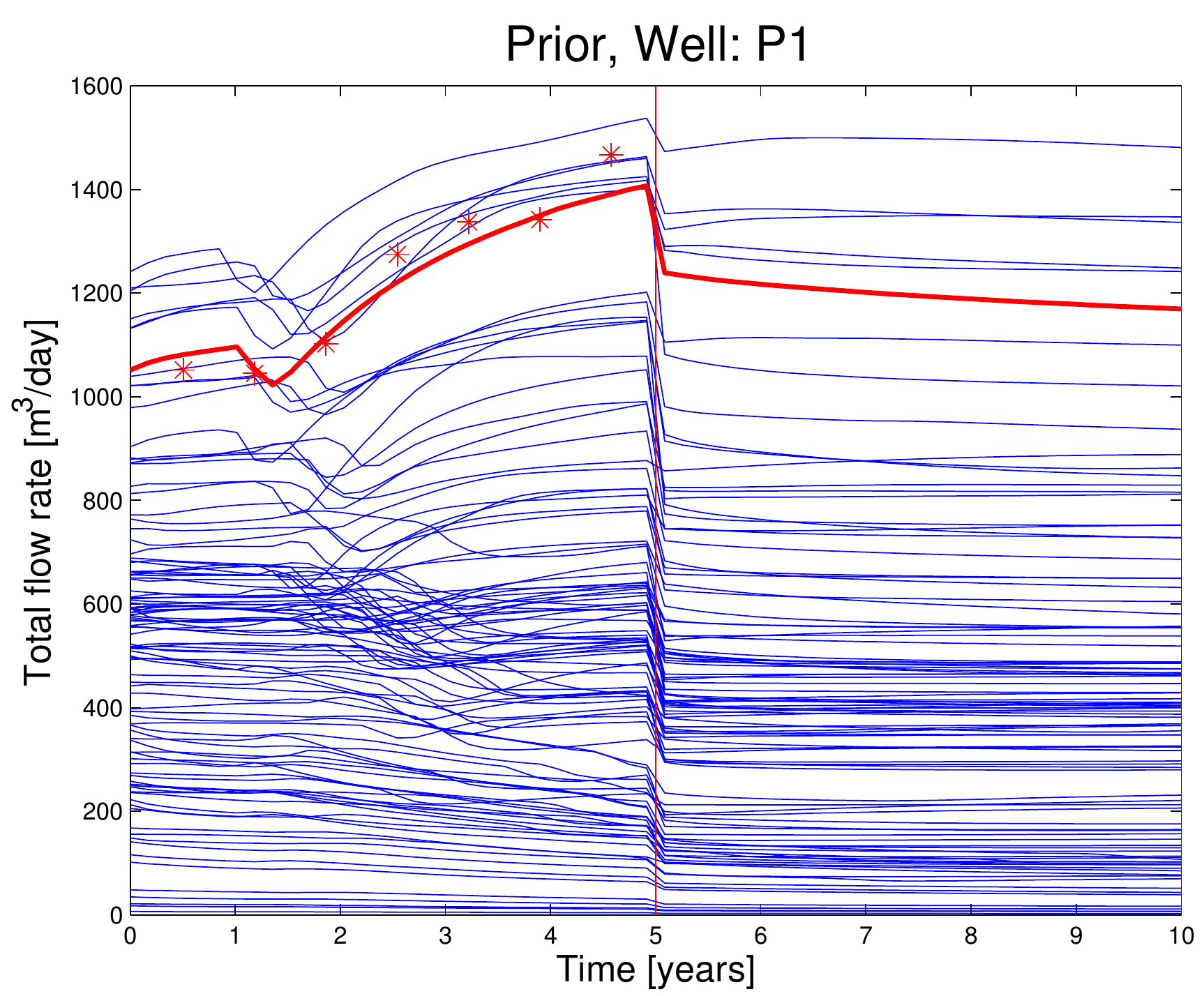}
\includegraphics[scale=0.22]{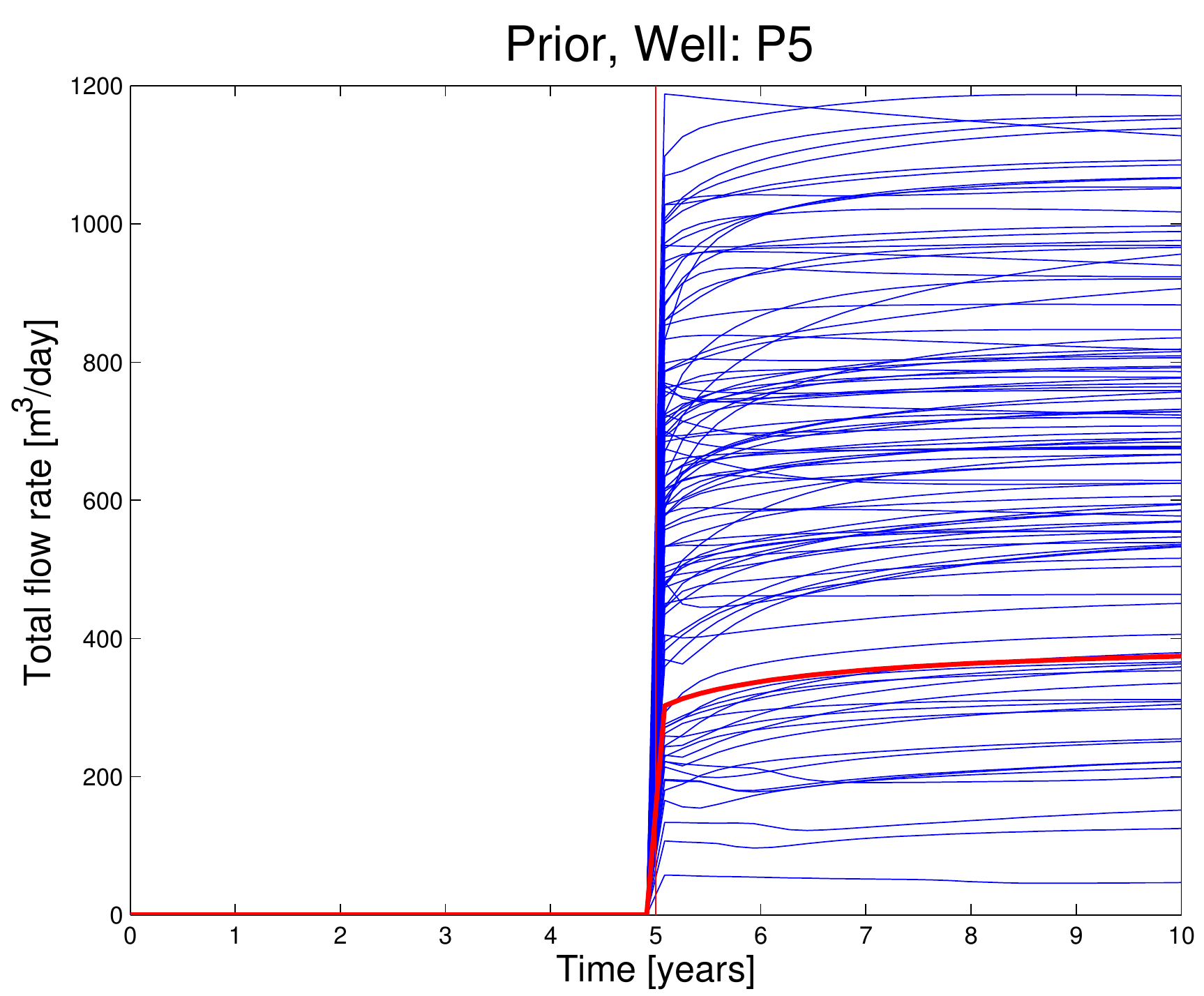}
\includegraphics[scale=0.22]{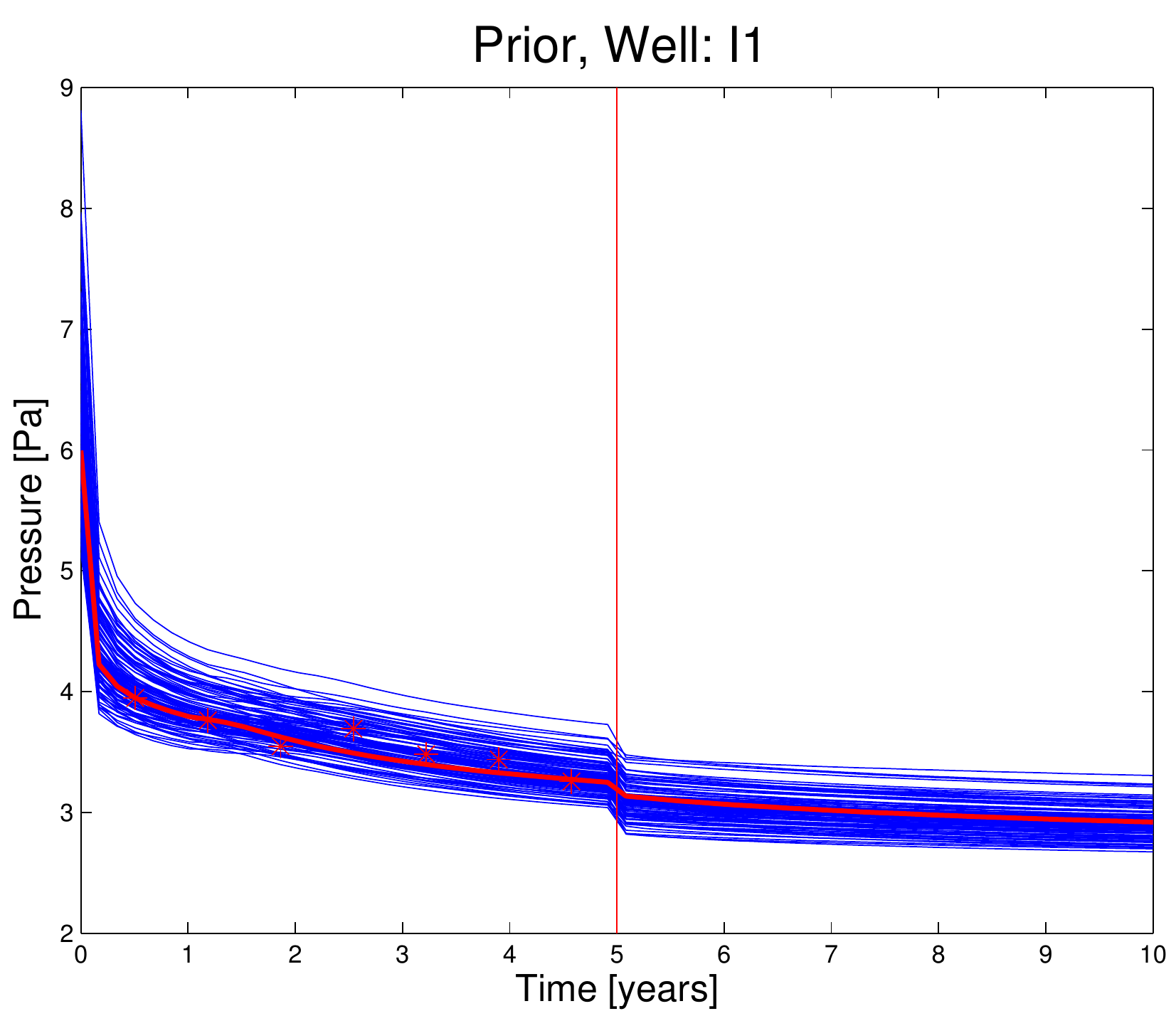}\\
\includegraphics[scale=0.22]{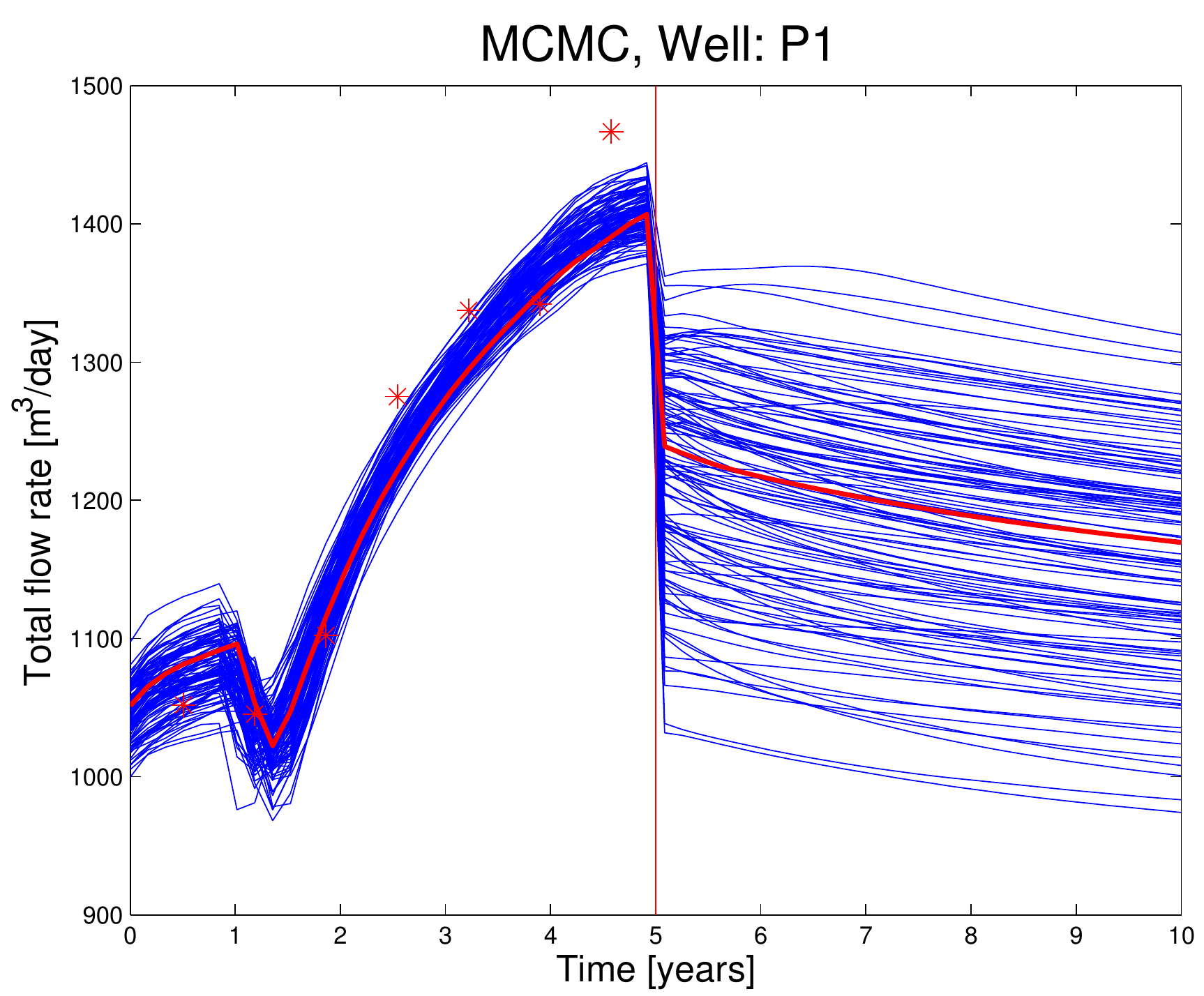}
\includegraphics[scale=0.22]{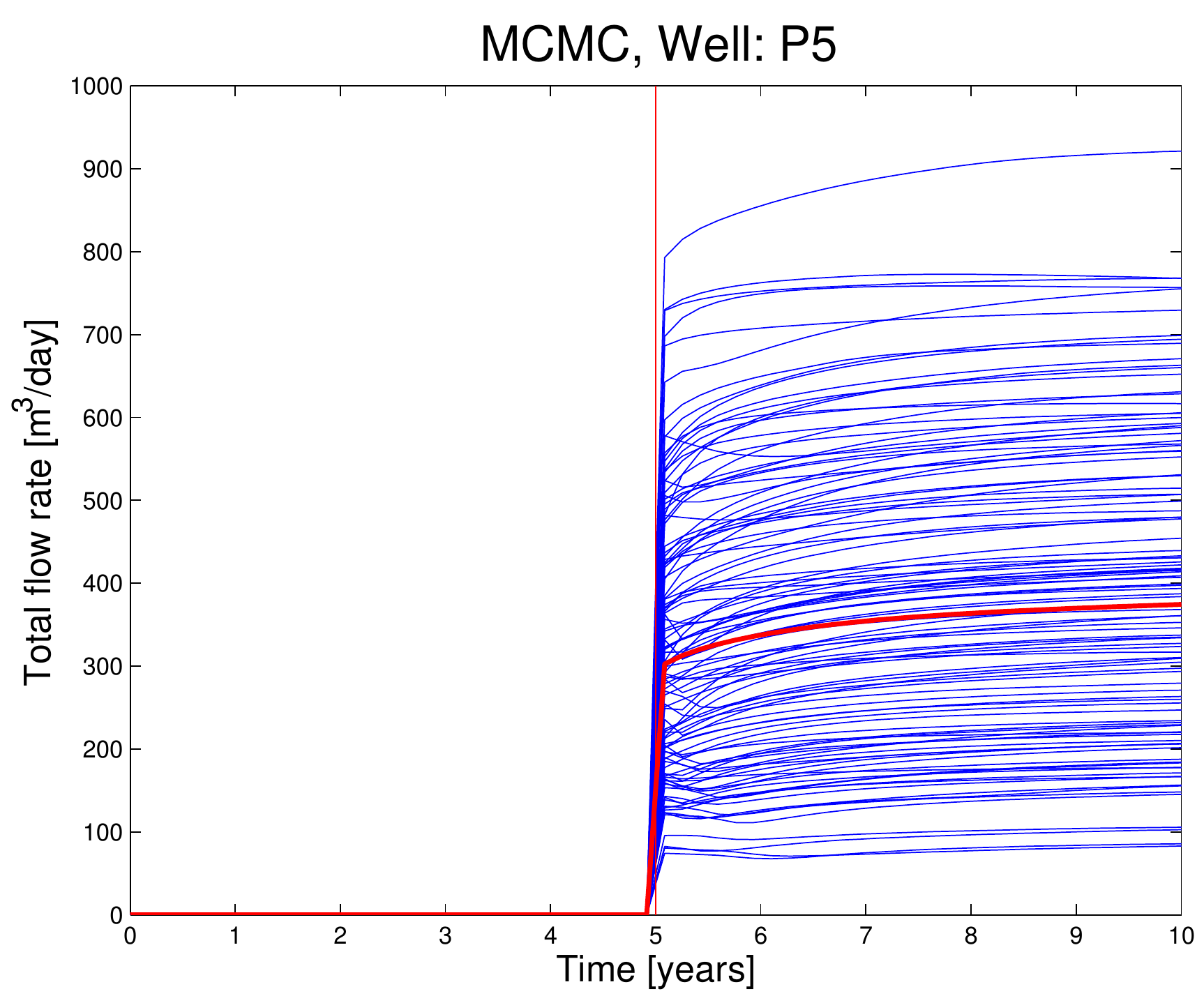}
\includegraphics[scale=0.22]{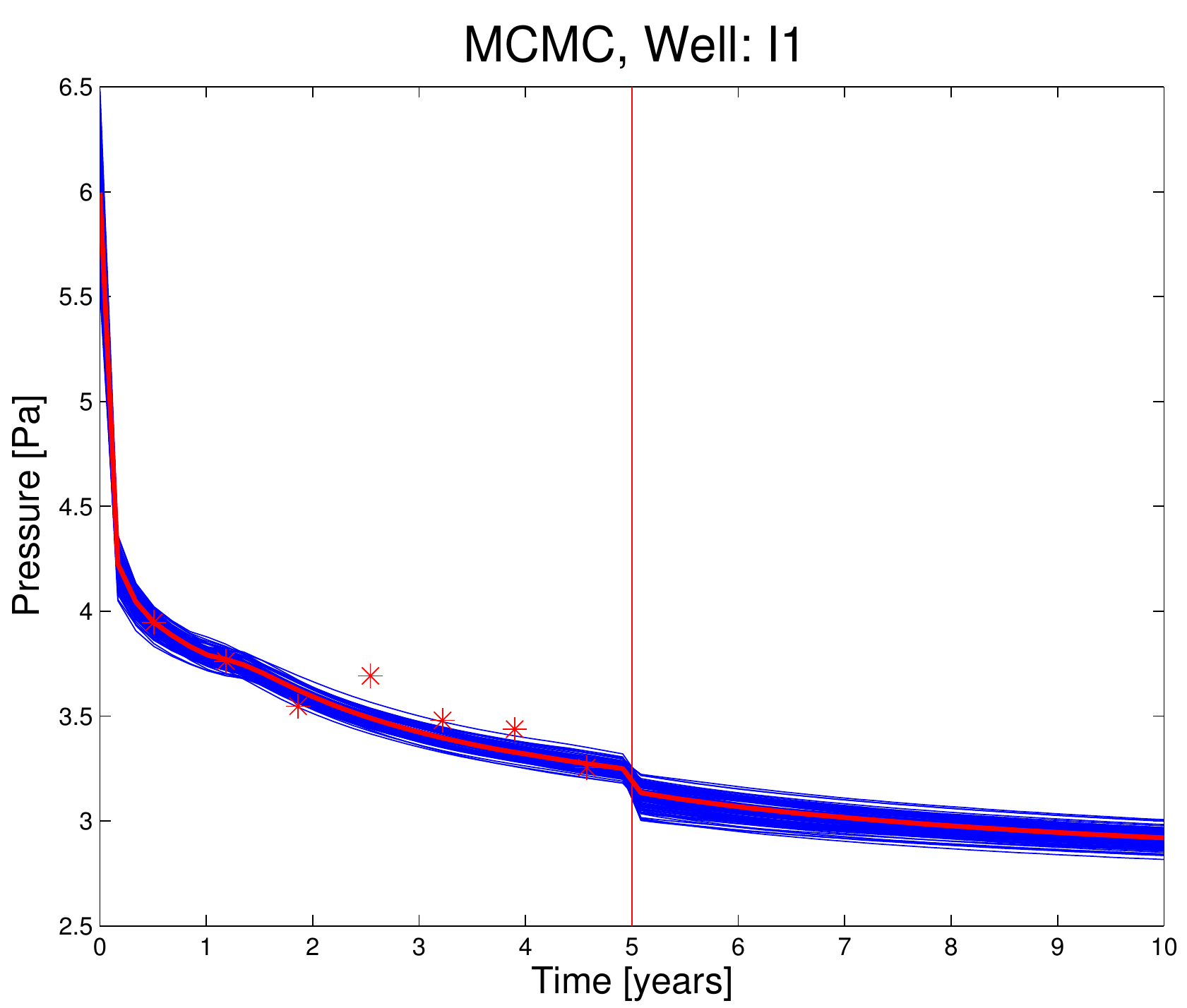}\\
\includegraphics[scale=0.22]{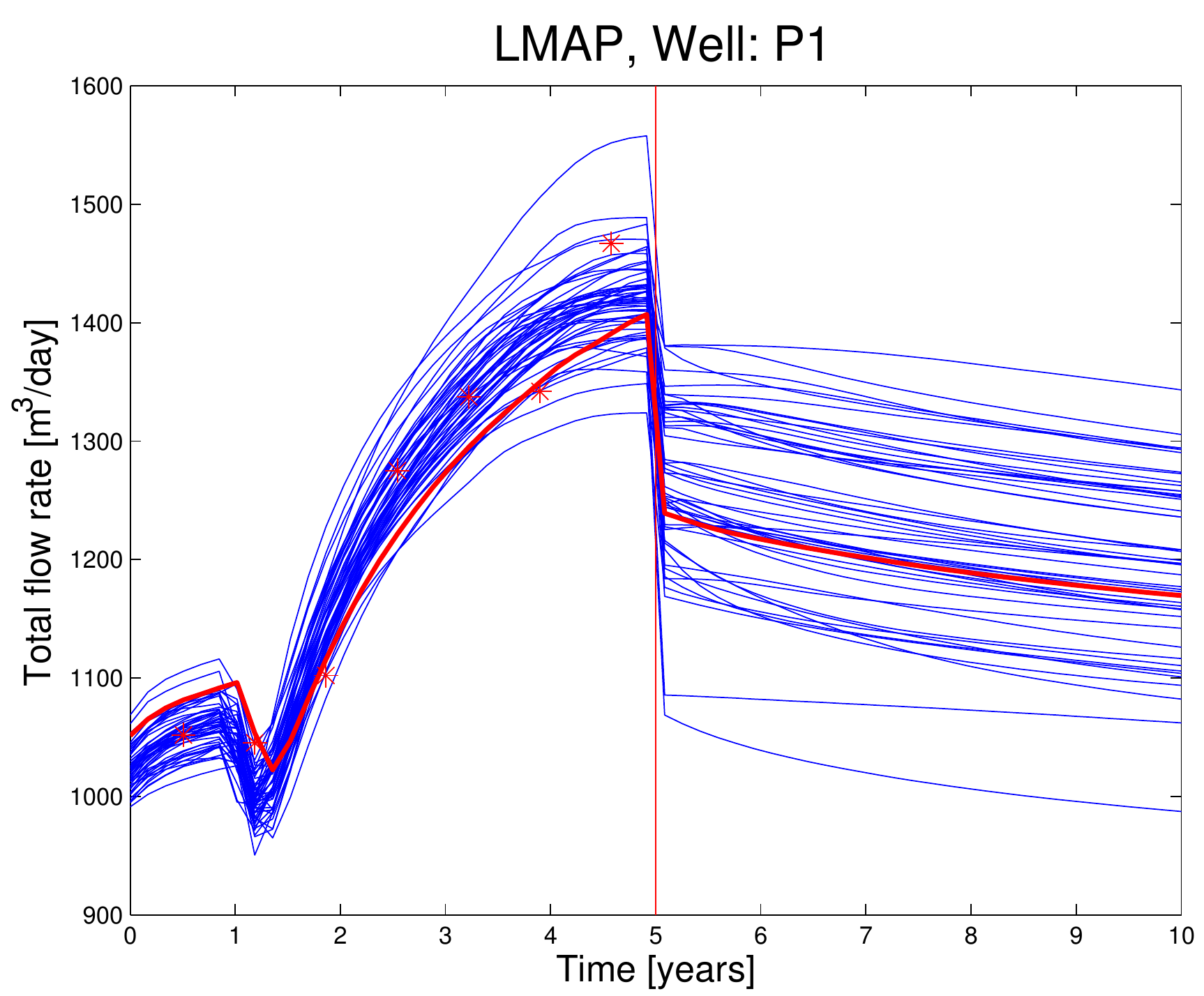}
\includegraphics[scale=0.22]{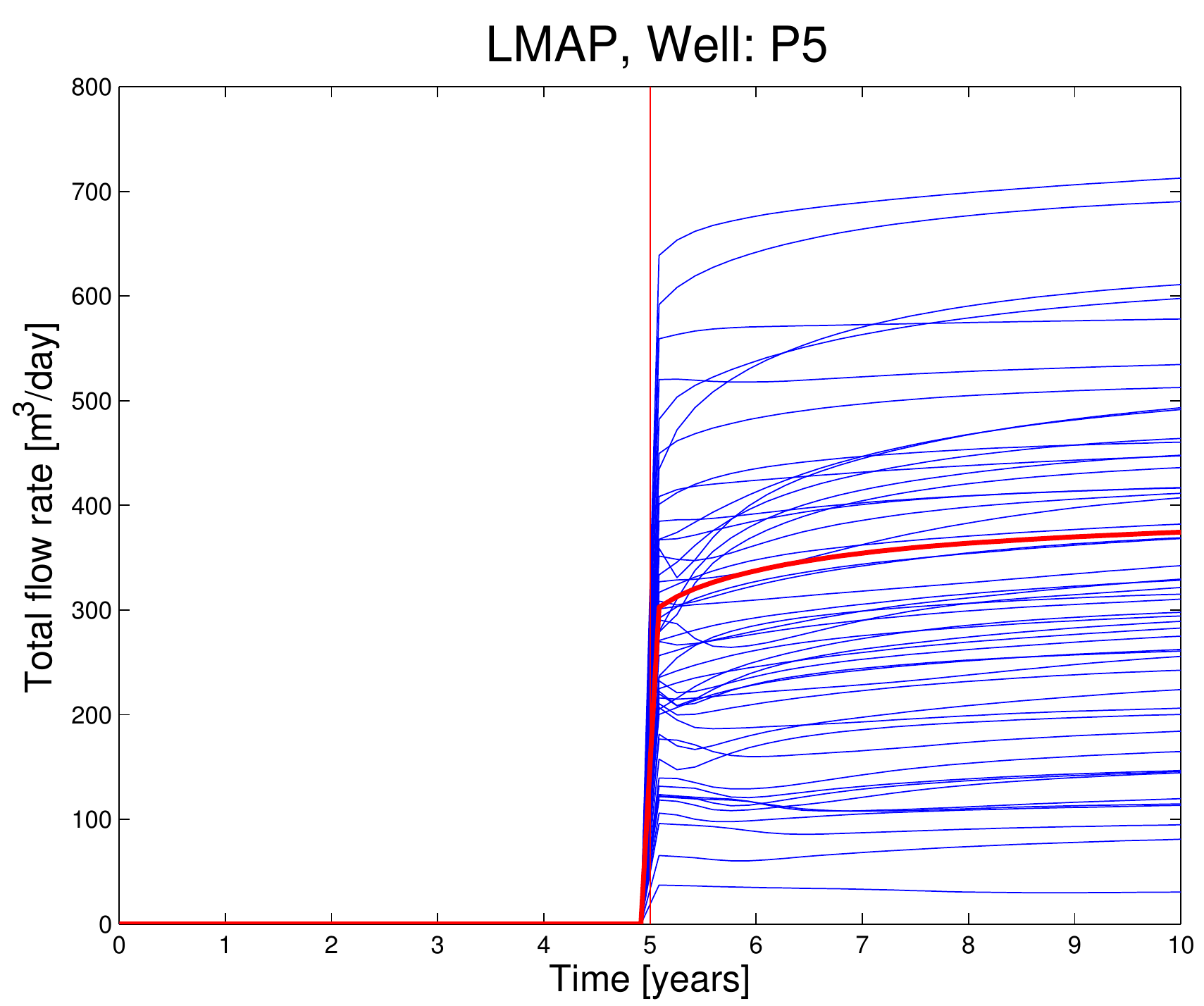}
\includegraphics[scale=0.22]{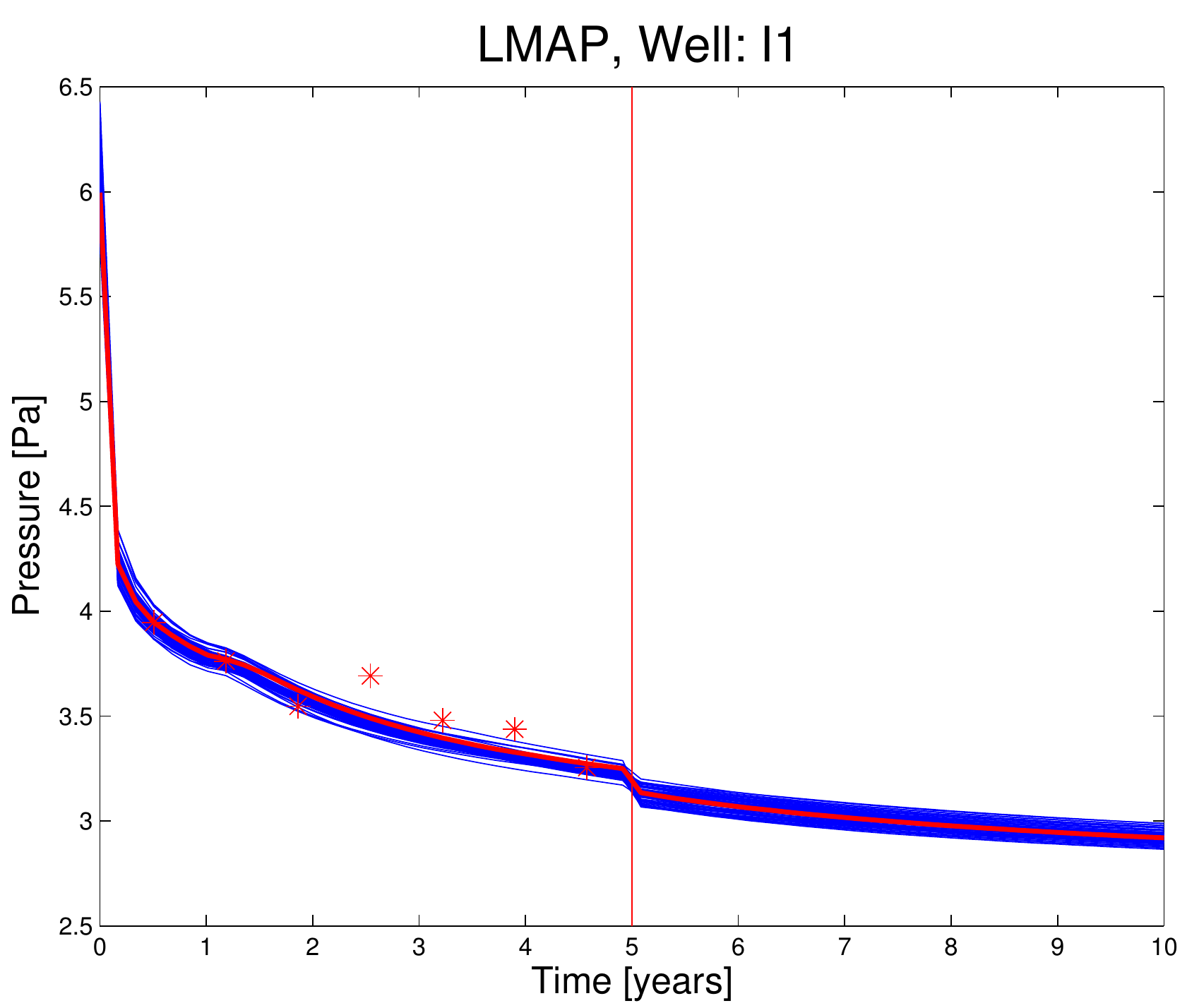}\\
\includegraphics[scale=0.22]{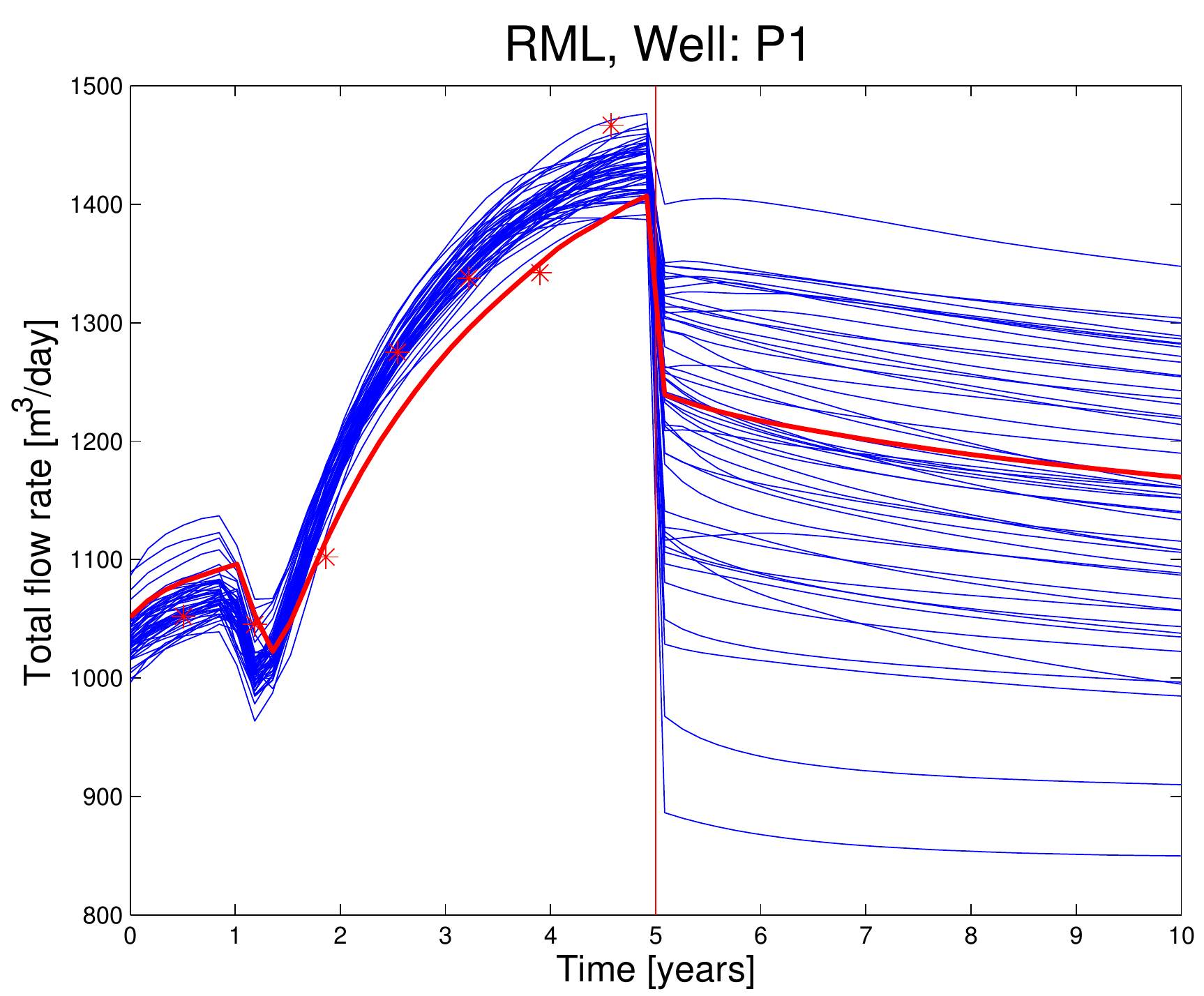}
\includegraphics[scale=0.22]{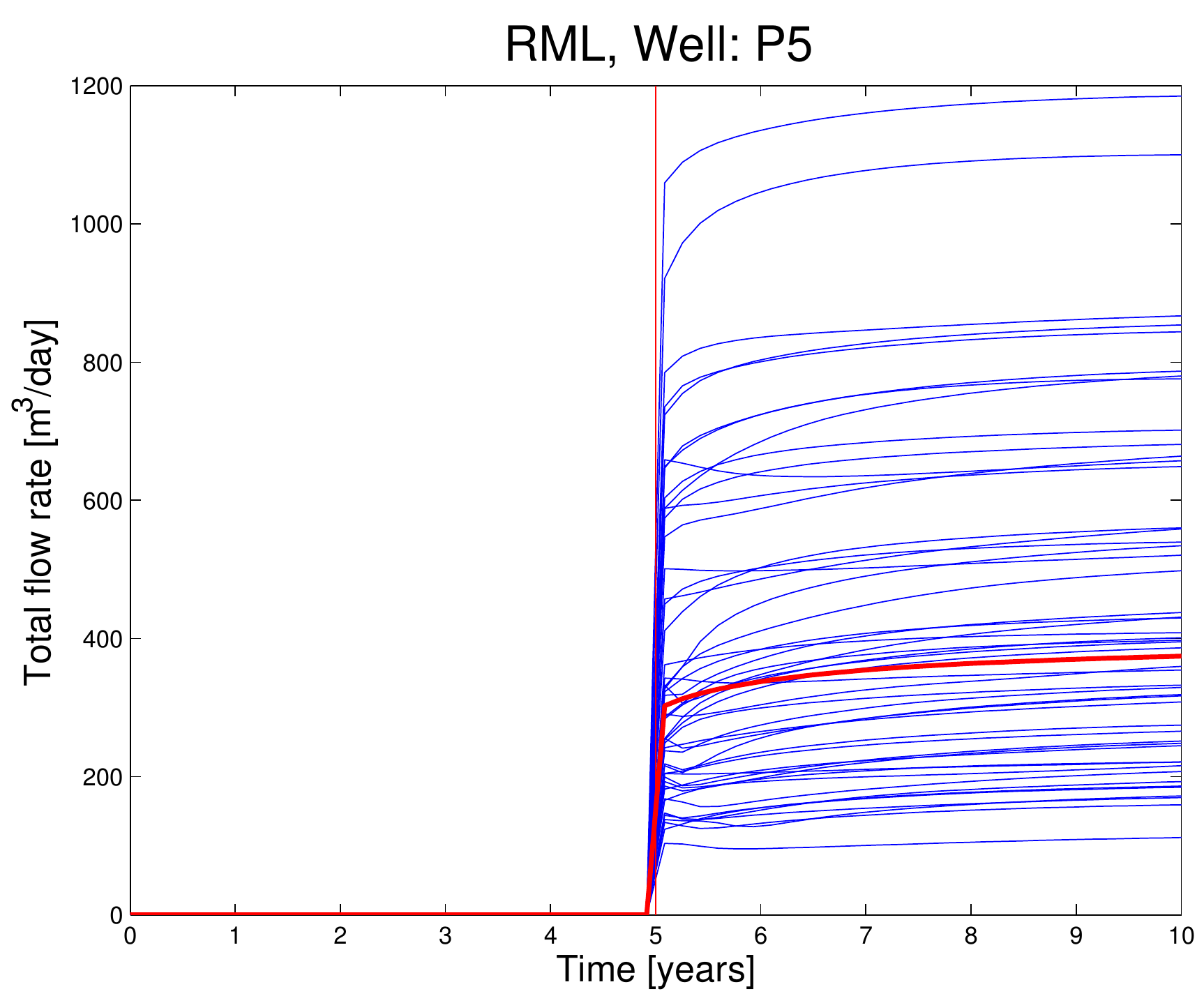}
\includegraphics[scale=0.22]{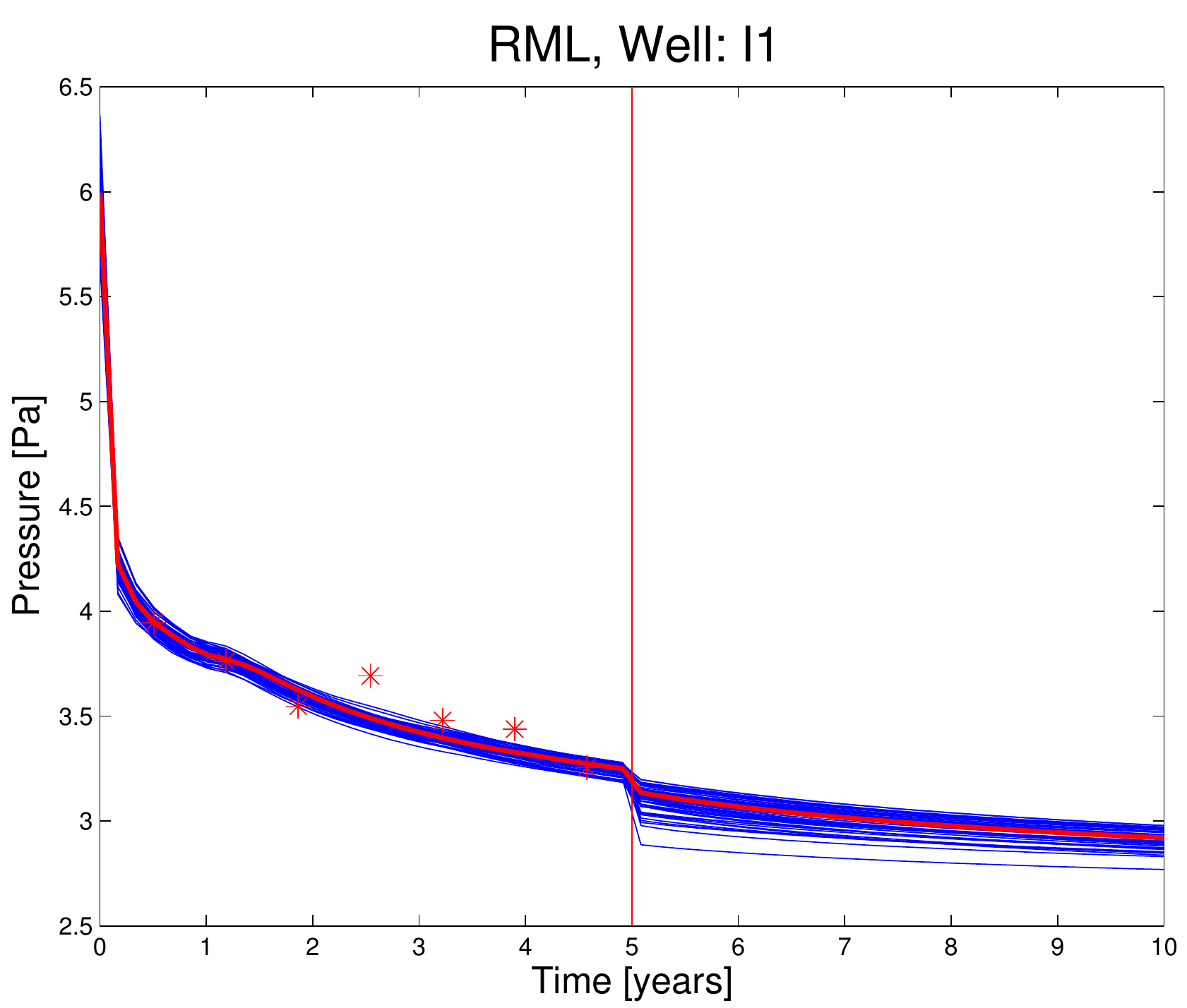}\\
\includegraphics[scale=0.22]{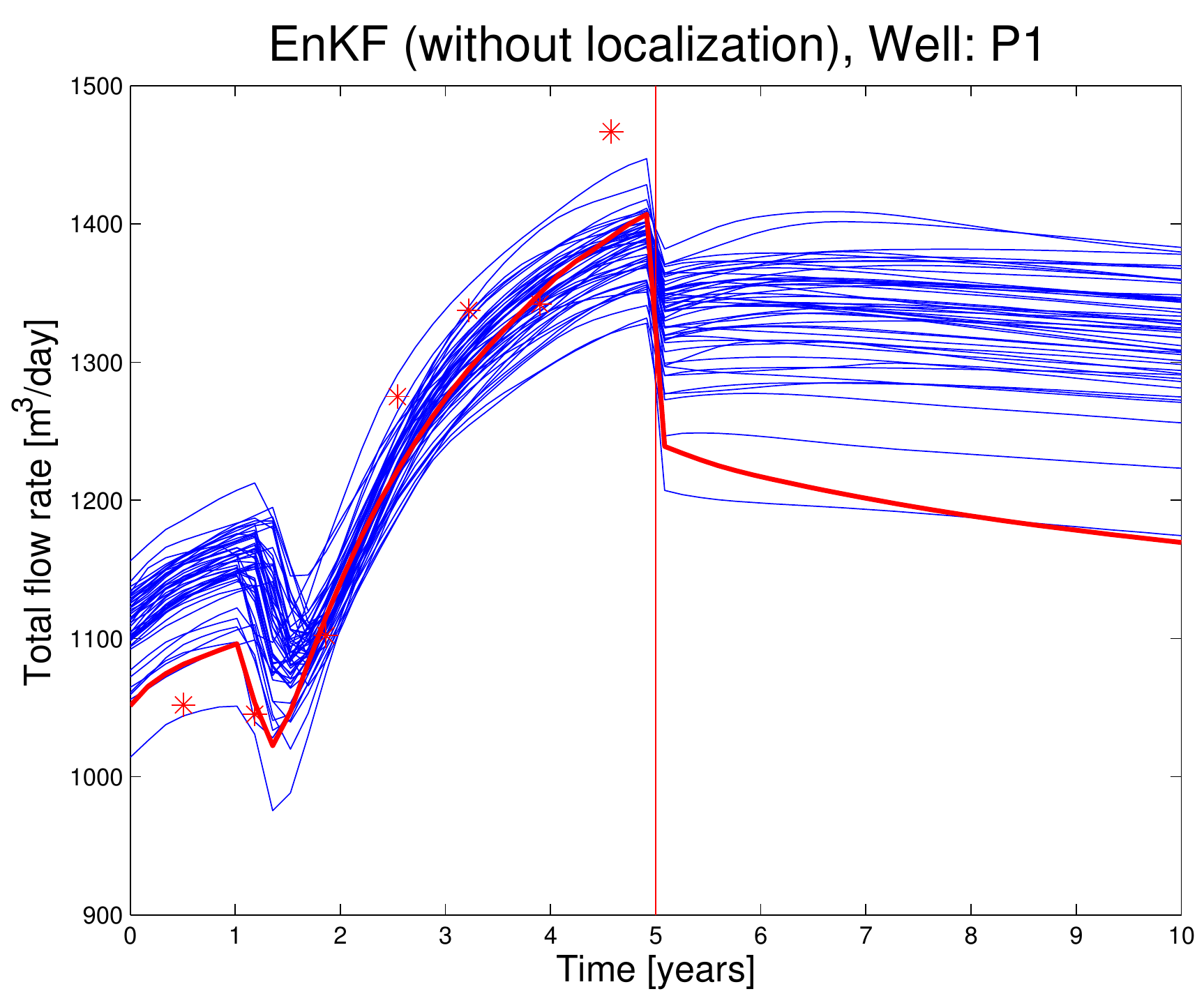}
\includegraphics[scale=0.22]{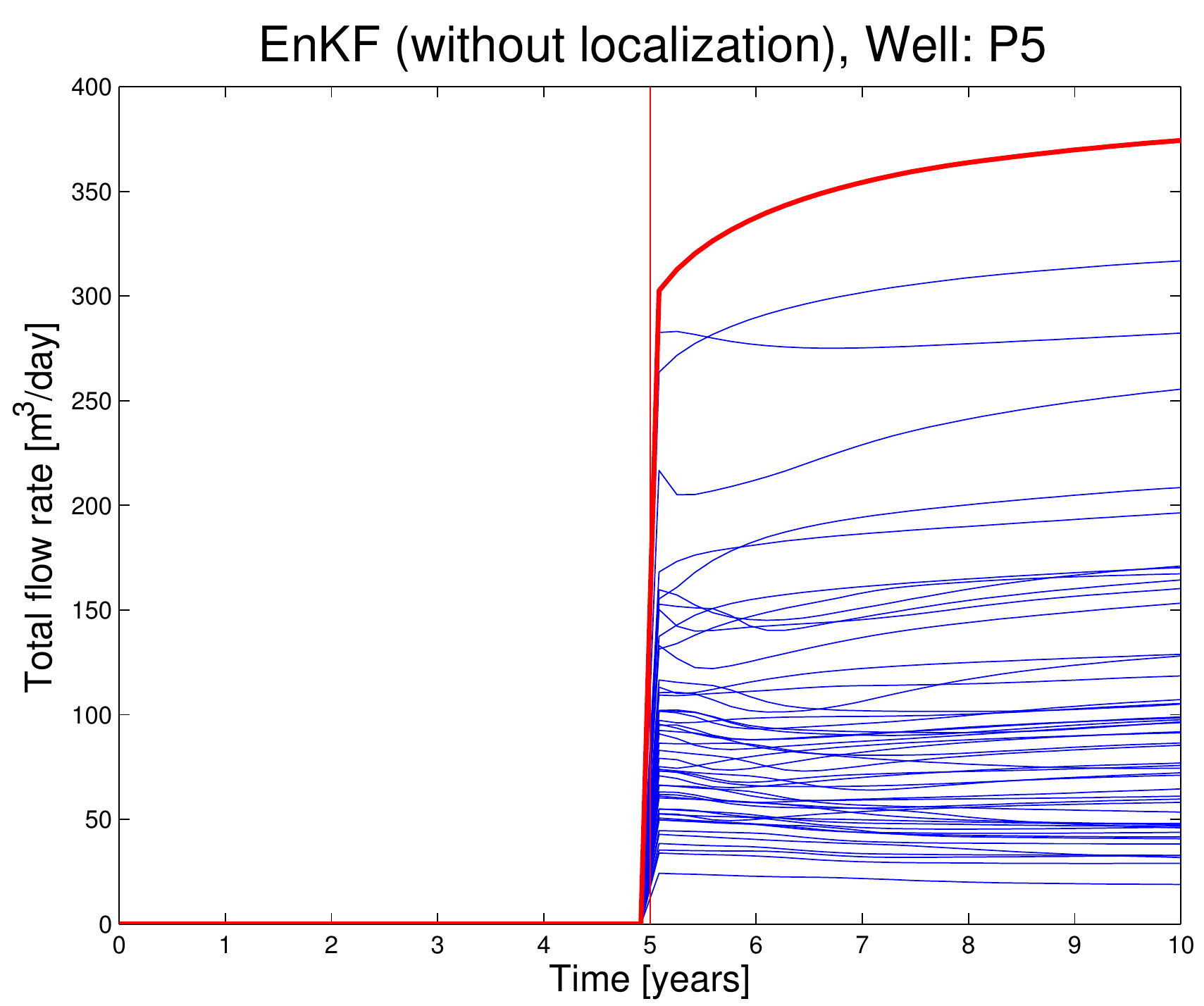}
\includegraphics[scale=0.22]{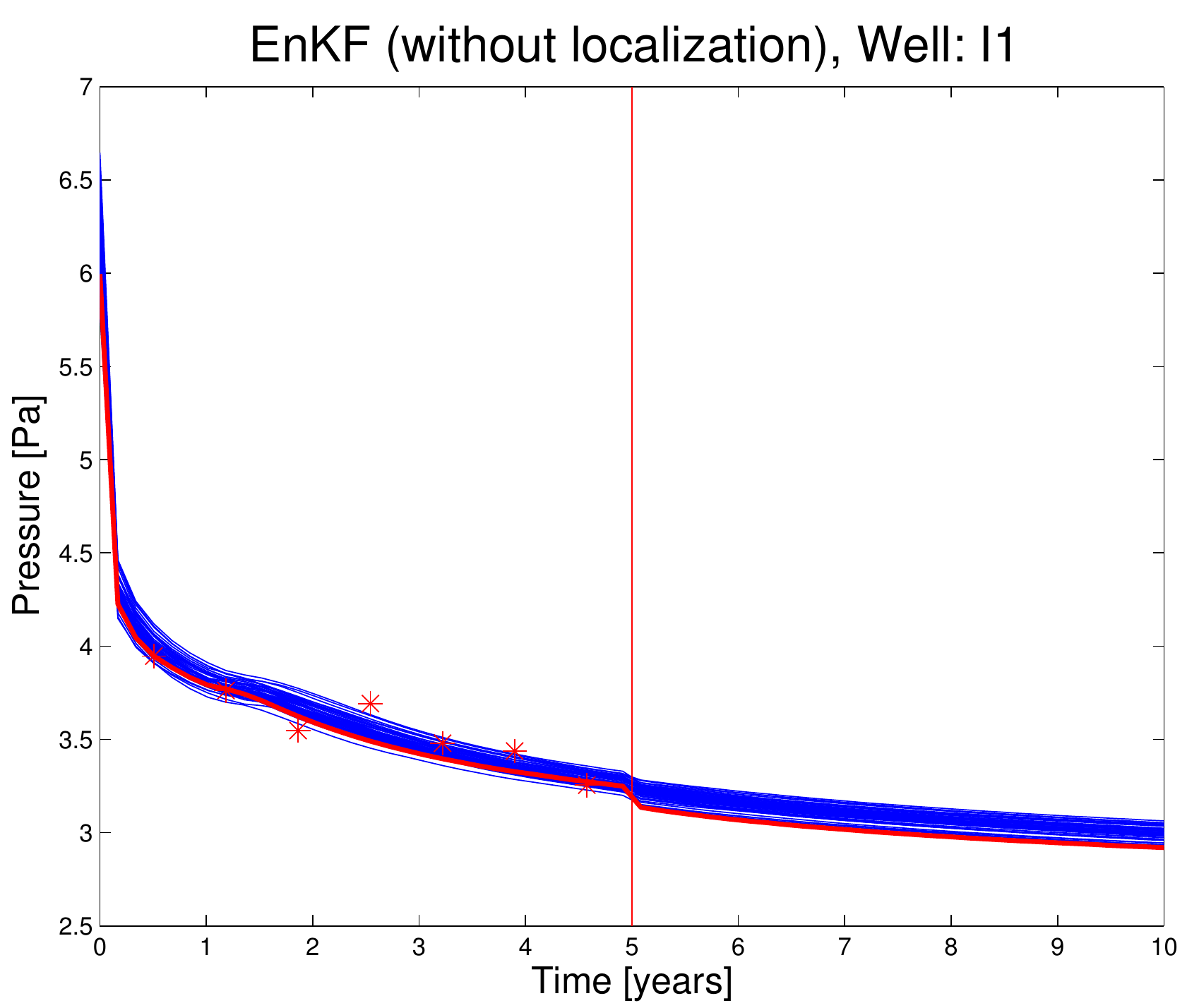}\\
\includegraphics[scale=0.22]{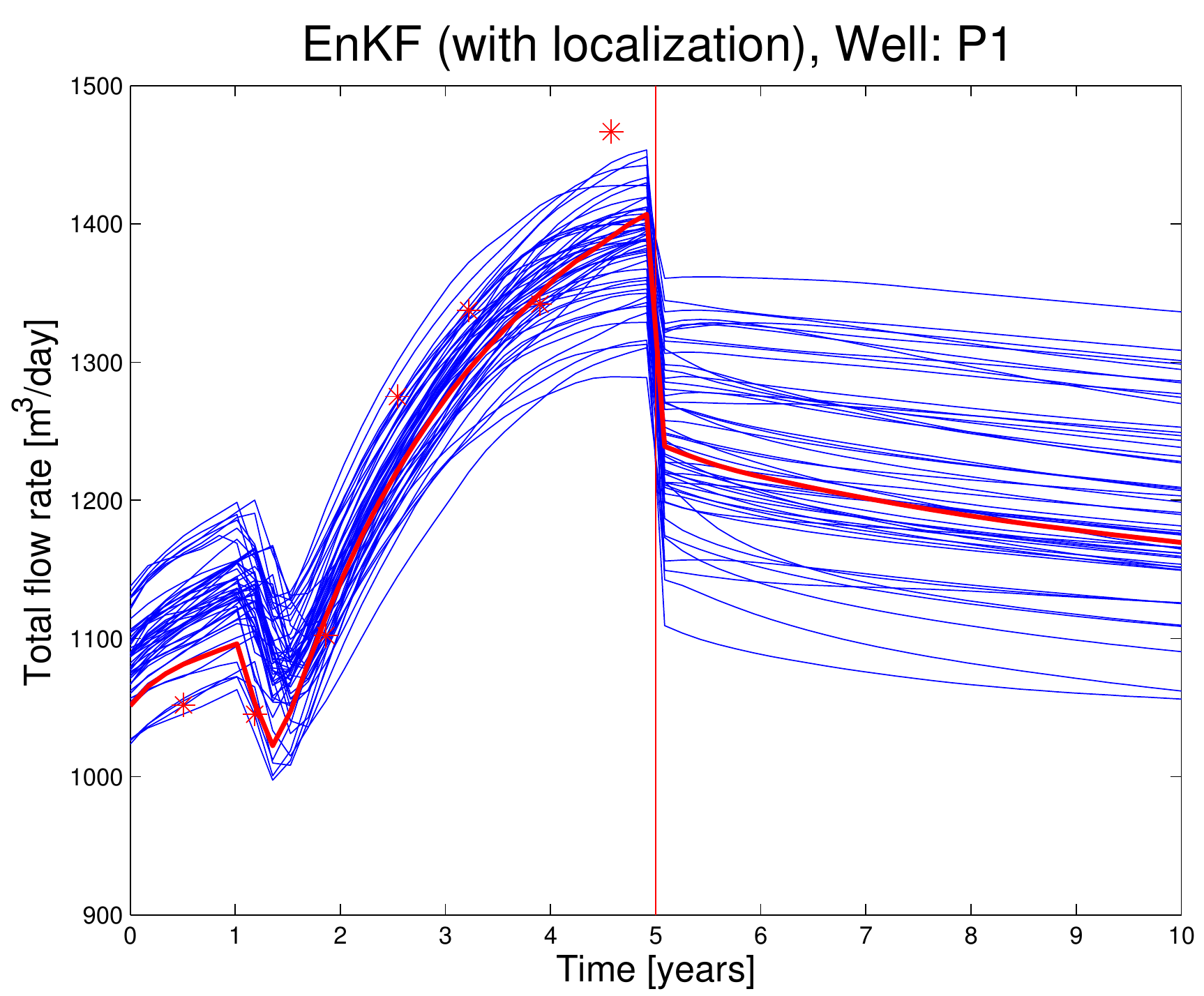}
\includegraphics[scale=0.22]{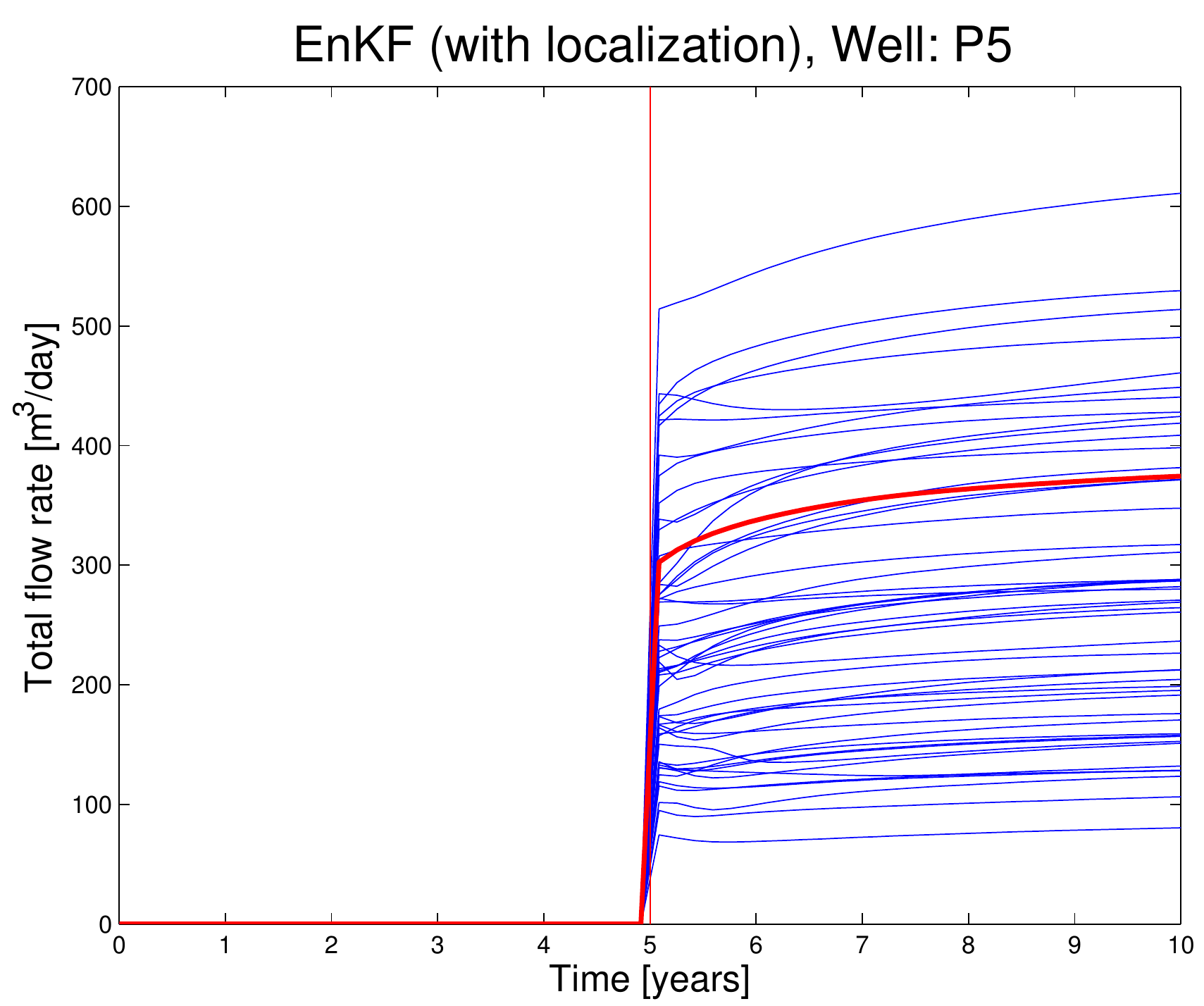}
\includegraphics[scale=0.22]{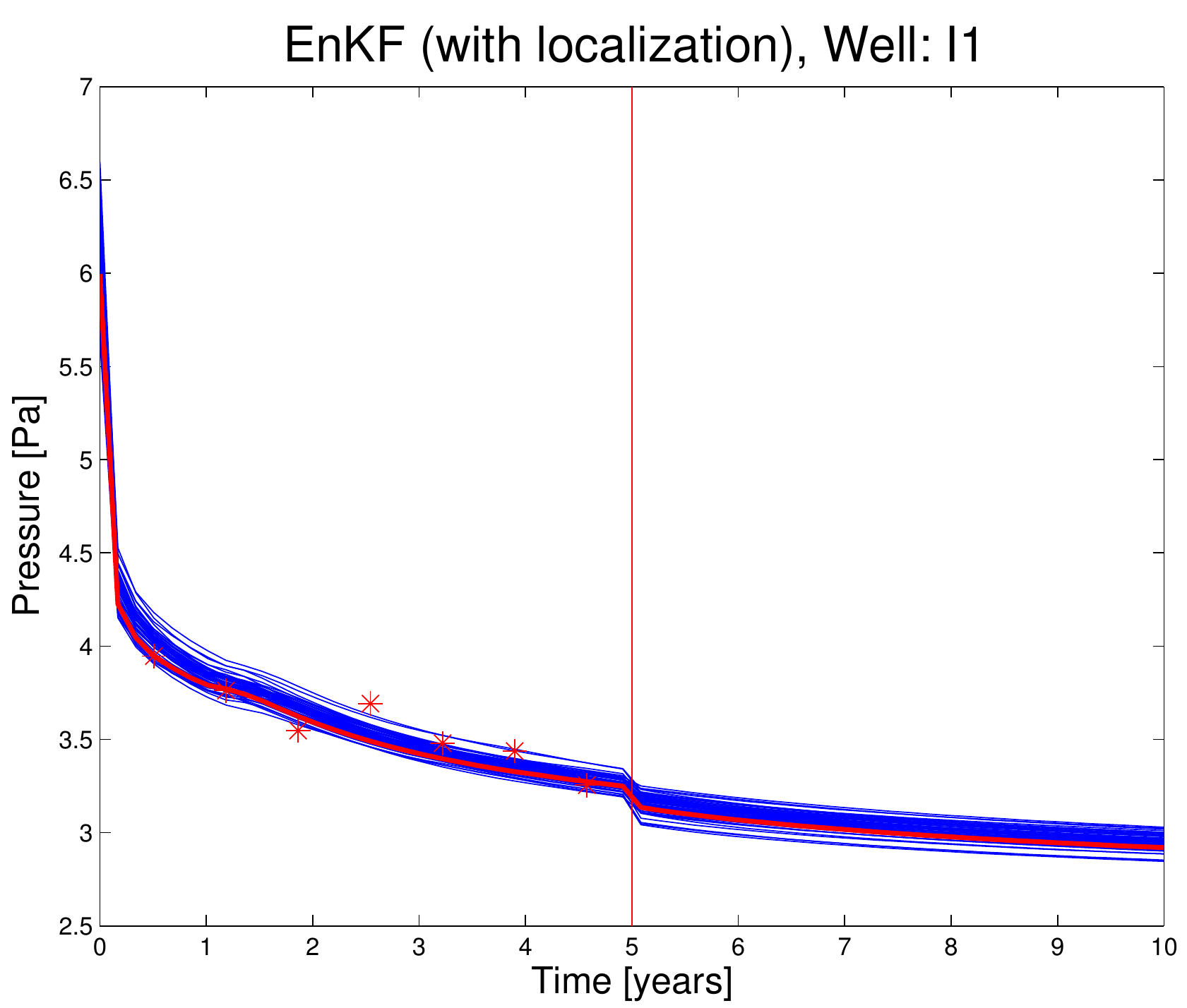}
\caption{Two-phase model (small number of wells). Total flow rates from $P_{1}$ (left column), $P_{5}$ (middle column) and bottom-hole pressure from $I_{1}$ (right column) simulated with permeabilities sampled from (top to bottom rows) the prior, the posterior, LMAP, RML, EnKF and EnKF with localization. }
\label{Figure12}
\end{figure}

In order to assess the performance of the approximate posterior samples at reproducing the predicting distribution, we consider an additional simulation period of $5$ years of forecast. For this additional $5$ years, a new well labeled as $P_{5}$ in Figure \ref{Figure8} (middle) is drilled and operated under constant fixed bottom-hole pressure $P_{bh}^{5}=2.7\times 10^{7}\textrm{Pa}$. In Figure \ref{Figure12} we present the total flow rates (from $P_{1}$ and $P_{5}$) and bottom-hole pressure (from $I_{1}$) simulated with permeabilities from the prior (first row), the posterior (second row), and some of the Gaussian approximations under analysis (third-sixth row). The vertical line divides the assimilation from the prediction. The red curve is computed from the posterior mean at the corresponding location.  We note that the poor performance of EnKF is reflected in the model predictions whose ensemble does not capture the prediction based on the mean of the posterior. These issues are alleviated by localization as we observe from Figure \ref{Figure12} (last row).

\begin{figure}
\includegraphics[scale=0.35]{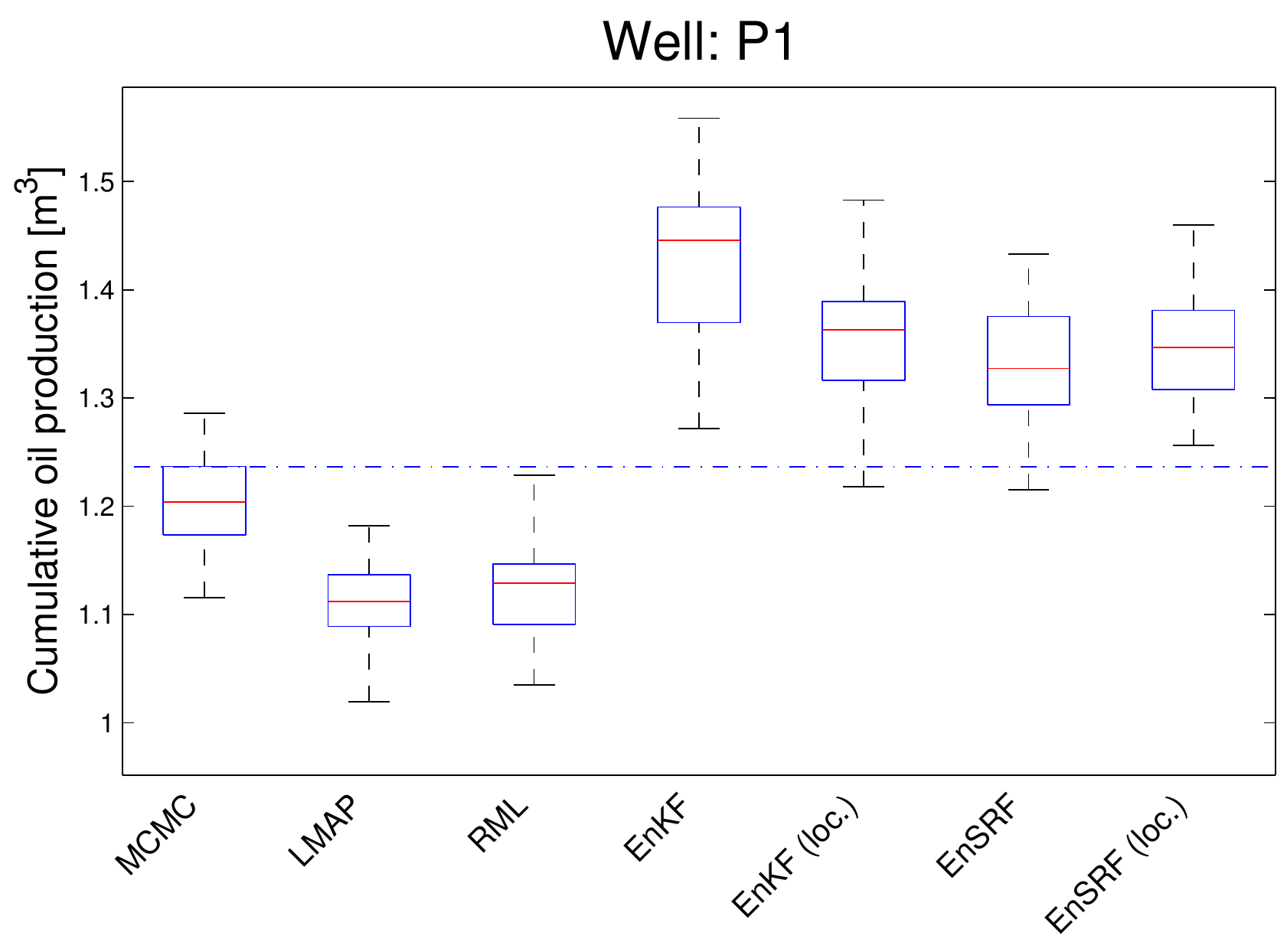}
\includegraphics[scale=0.35]{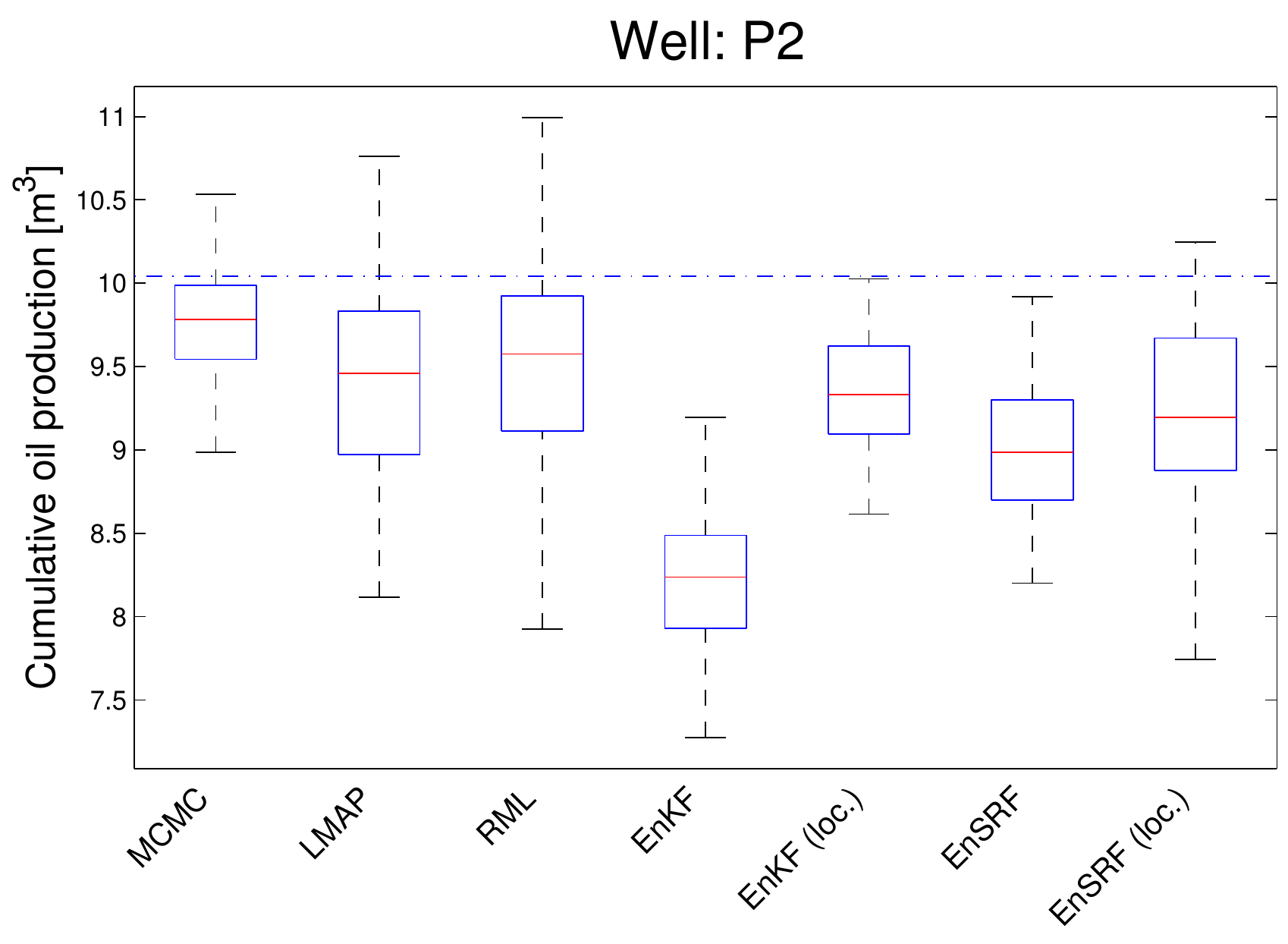}\\
\includegraphics[scale=0.35]{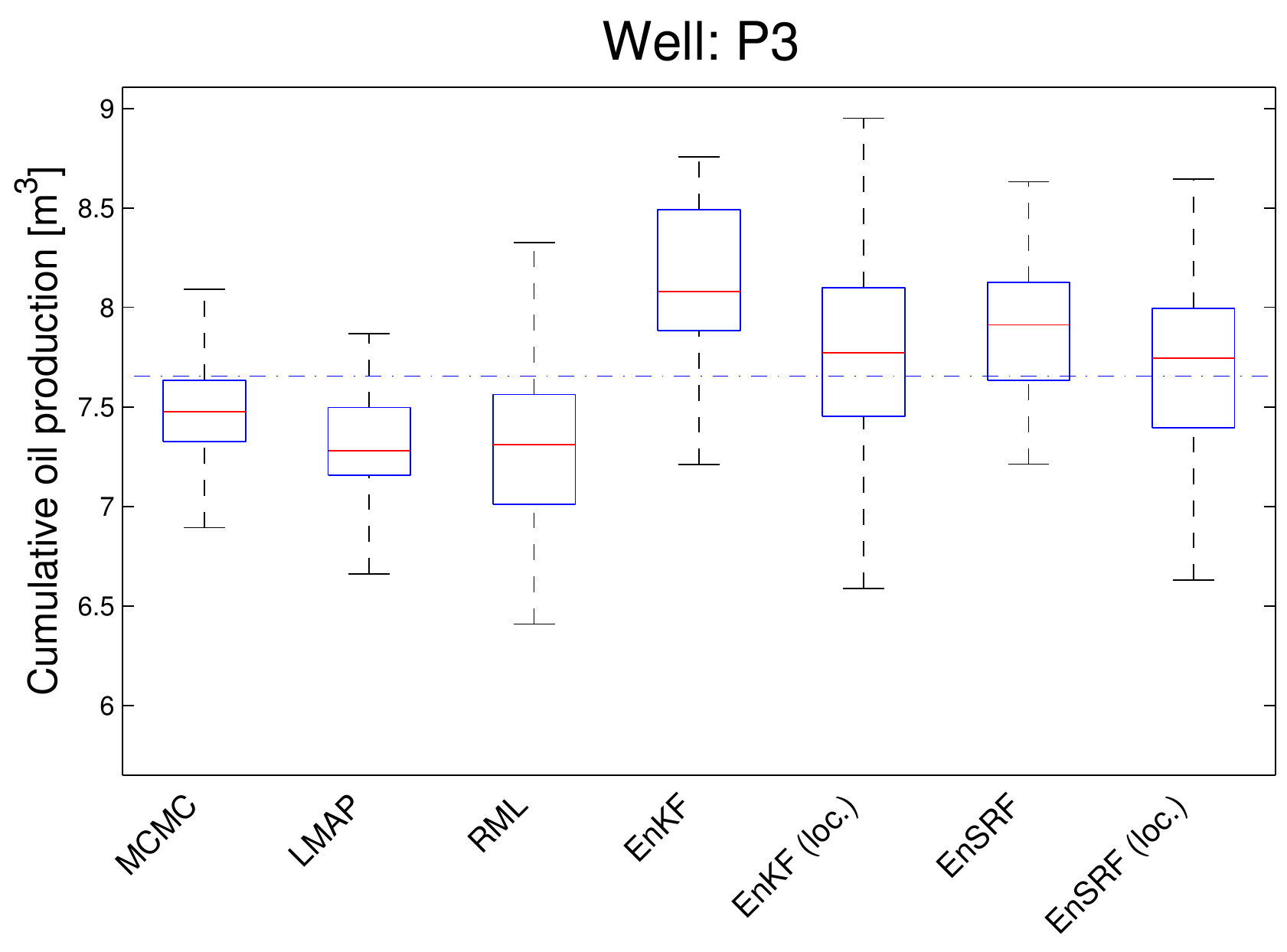}
\includegraphics[scale=0.35]{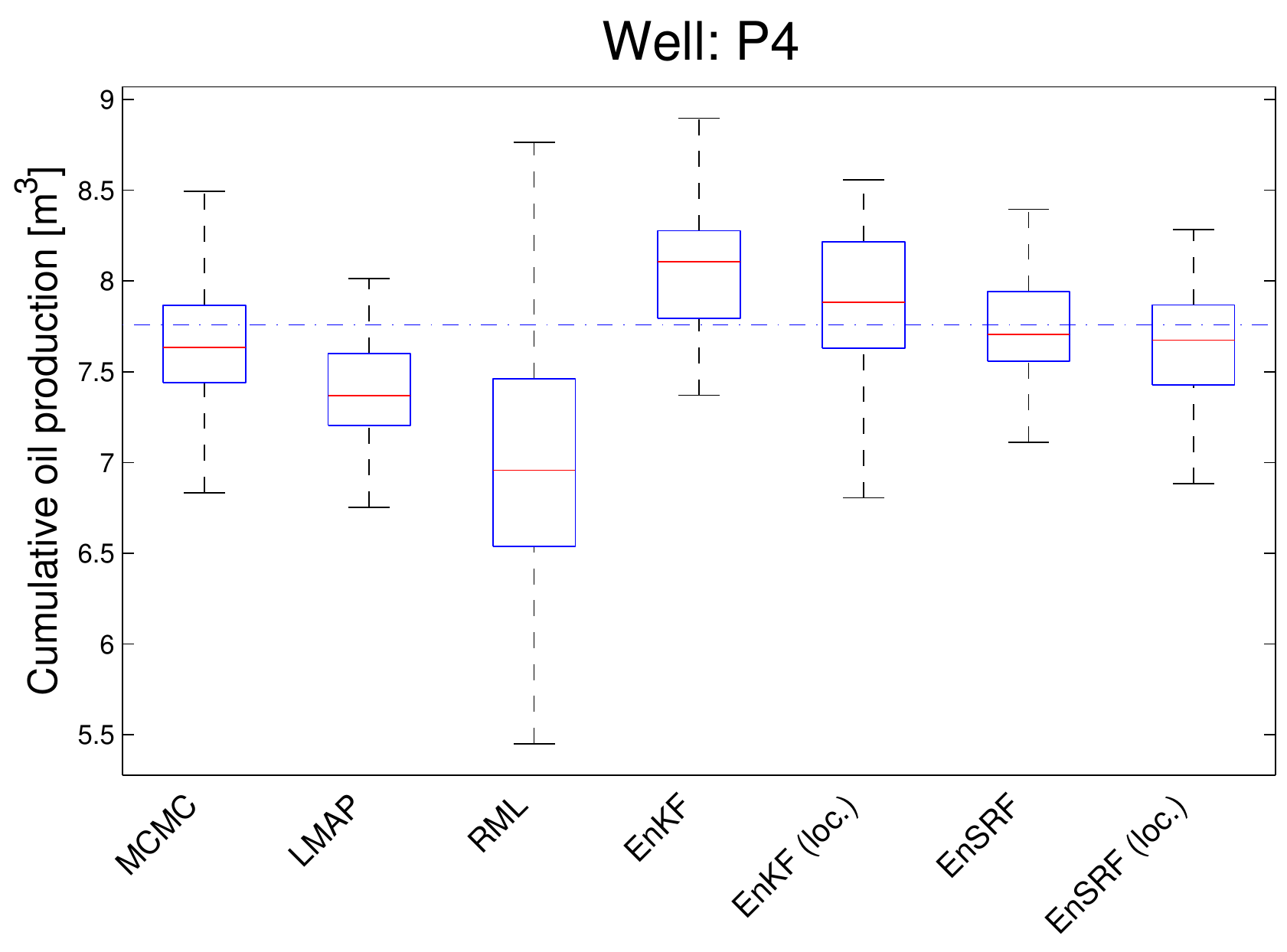}\\
\includegraphics[scale=0.35]{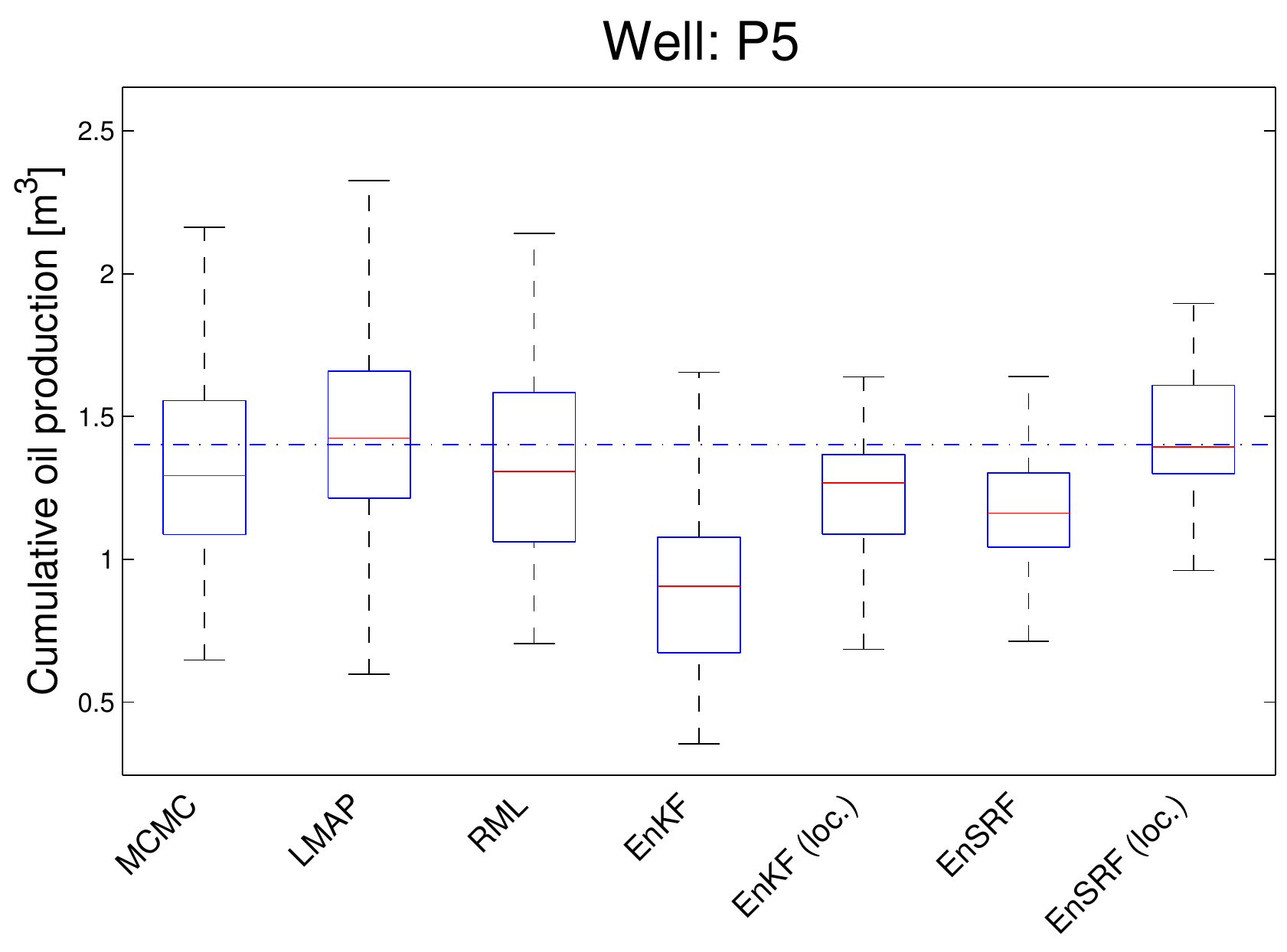}
\caption{Two-phase model (small number of wells). Distribution of cumulative oil production at wells $P_{1},\dots, P_{5}$ at the final time of simulation $t=10~ \textrm{years}$}  
\label{Figure13}
\end{figure}

In Figure \ref{Figure13} we display the distribution of the final time cumulative oil production simulated from the posterior and the Gaussian approximation. Note that even though LMAP provides a reasonable approximation in terms of both mean and variance, the approximation provides a deficient characterization of the predicting distribution. In general all the Gaussian approximations exhibit poor performance at reproducing the predicting distribution.

\begin{table}
\caption{Evaluation of Gaussian approximations for the two-phase model. Case with small number of wells.}
\label{Table3}       
\begin{tabular}{lccc}
\hline\noalign{\smallskip}
Method & Relative error &  Relative error & Computational cost\\
 &  in the mean $\epsilon_{u}$& in the variance $\epsilon_{\sigma}$ &  [Forward model runs]\\
\noalign{\smallskip}\hline\noalign{\smallskip}
MCMC  &0.000 & 0.000 &  $5.5\times 10^{7}$\\
MAP 				&0.277 & 0.143 &  $6.0\times 10^{1}$\\
LMAP  ($N_{e}=50$) &0.284 & 0.286 &  $6.0\times 10^{1}$\\
RML  ($N_{e}=50$)   & 0.253& 0.475 &  $3.0\times 10^{3}$\\
EnKF  ($N_{e}=50$)      &1.159& 0.424 &  $5.0\times 10^{1}$\\
EnKF (localization, $N_{e}=50$)  & 0.600 &0.263 &$5.0\times 10^{1}$\\
EnSRF  ($N_{e}=50$)		&0.715 & 0.397&  $5.0\times 10^{1}$\\
EnSRF (localization,  $N_{e}=50$)  & 0.483 &0.259 &$5.0\times 10^{1}$\\ \hline
EnKF  ($N_{e}=1000$) & 0.353 & 0.209&$1.0\times 10^{3}$\\
EnKF  ($N_{e}=3000$) & 0.301& 0.222 &$3.0\times 10^{3}$\\
EnKF  ($N_{e}=8000$) & 0.337 & 0.216&$8.0\times 10^{3}$\\
\noalign{\smallskip}\hline
\end{tabular}
\end{table}

\subsection{Oil-water reservoir: Large number of wells}

\begin{figure}
\includegraphics[scale=0.25]{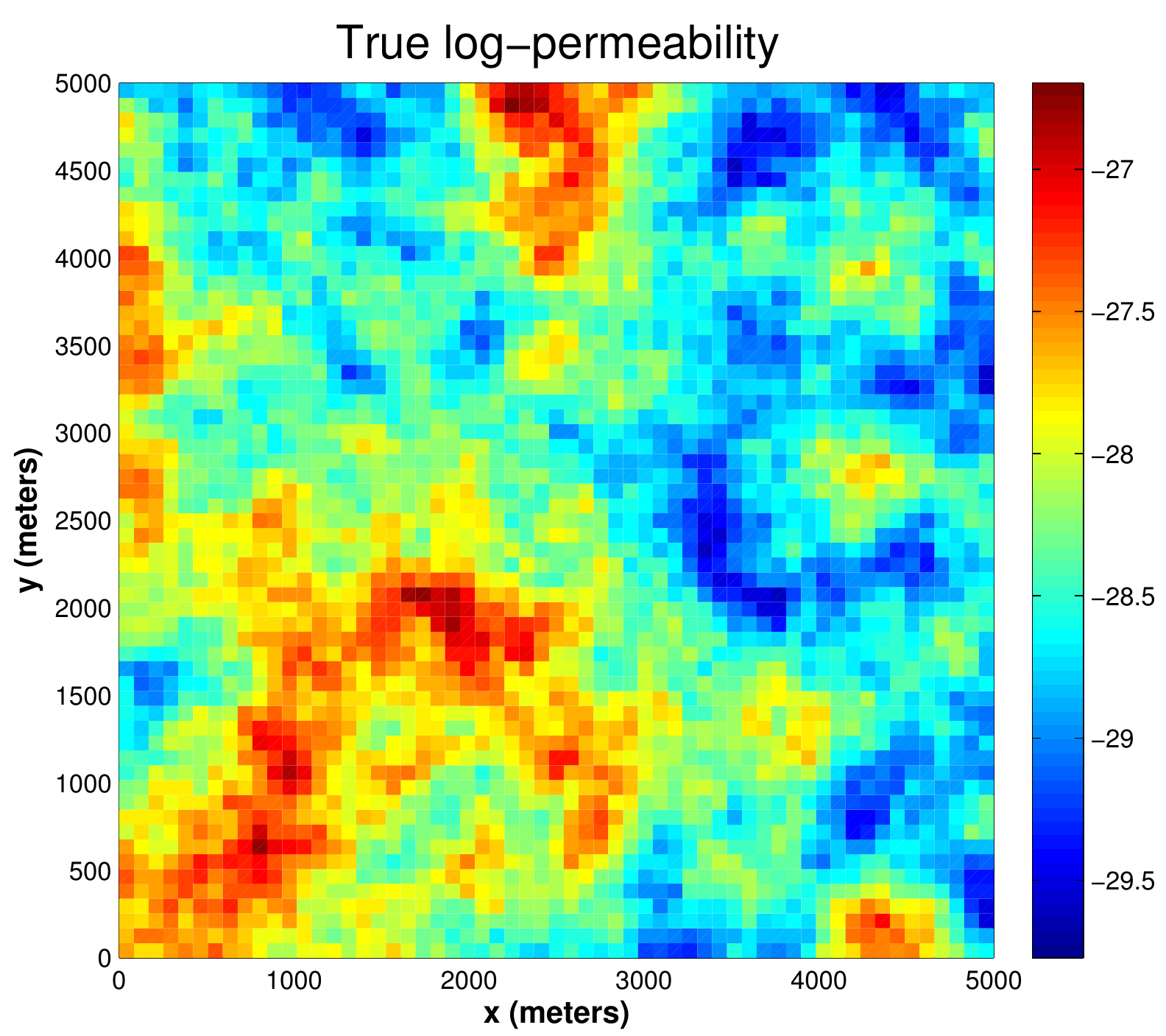}
\includegraphics[scale=0.25]{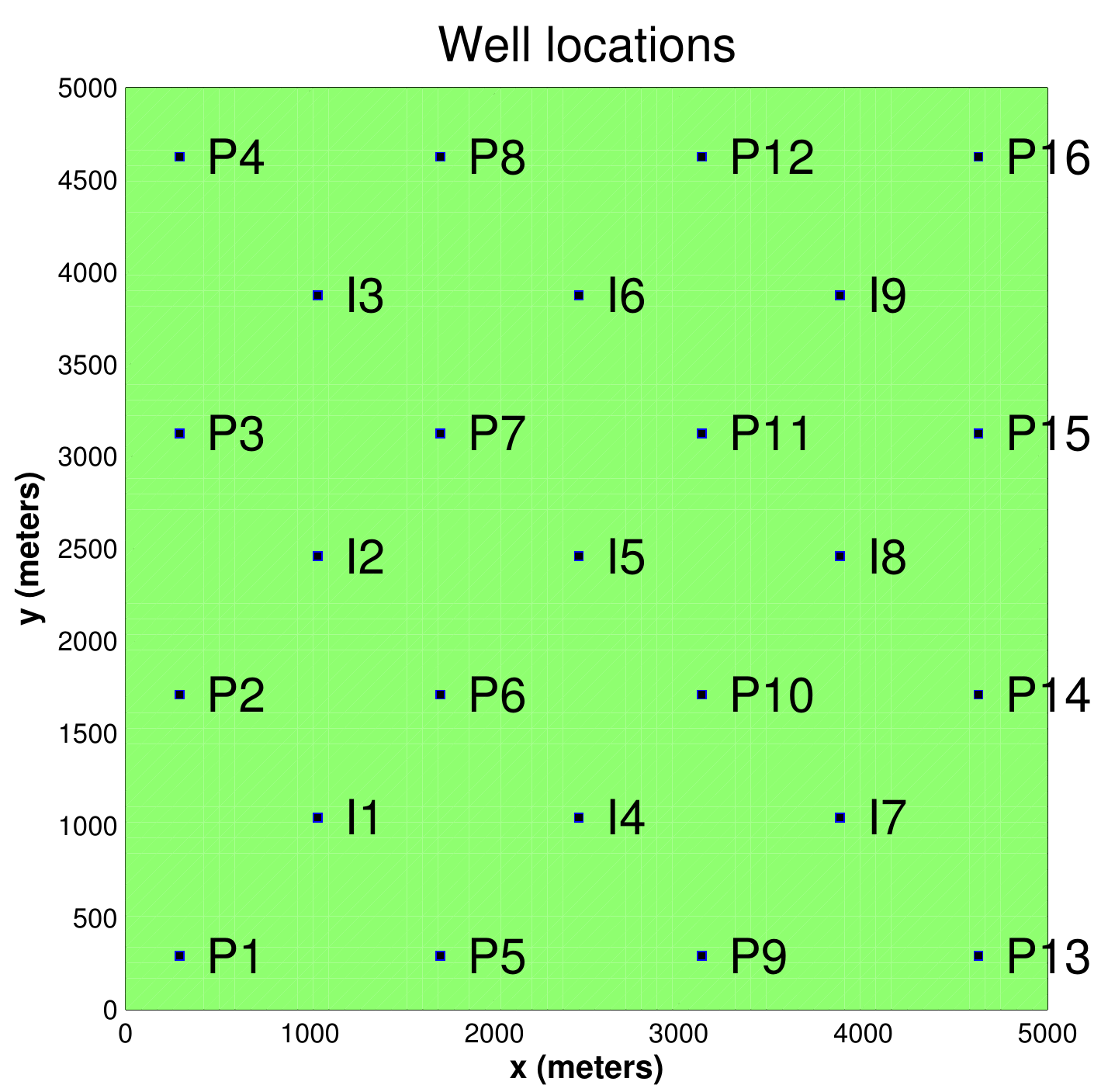}
\includegraphics[scale=0.25]{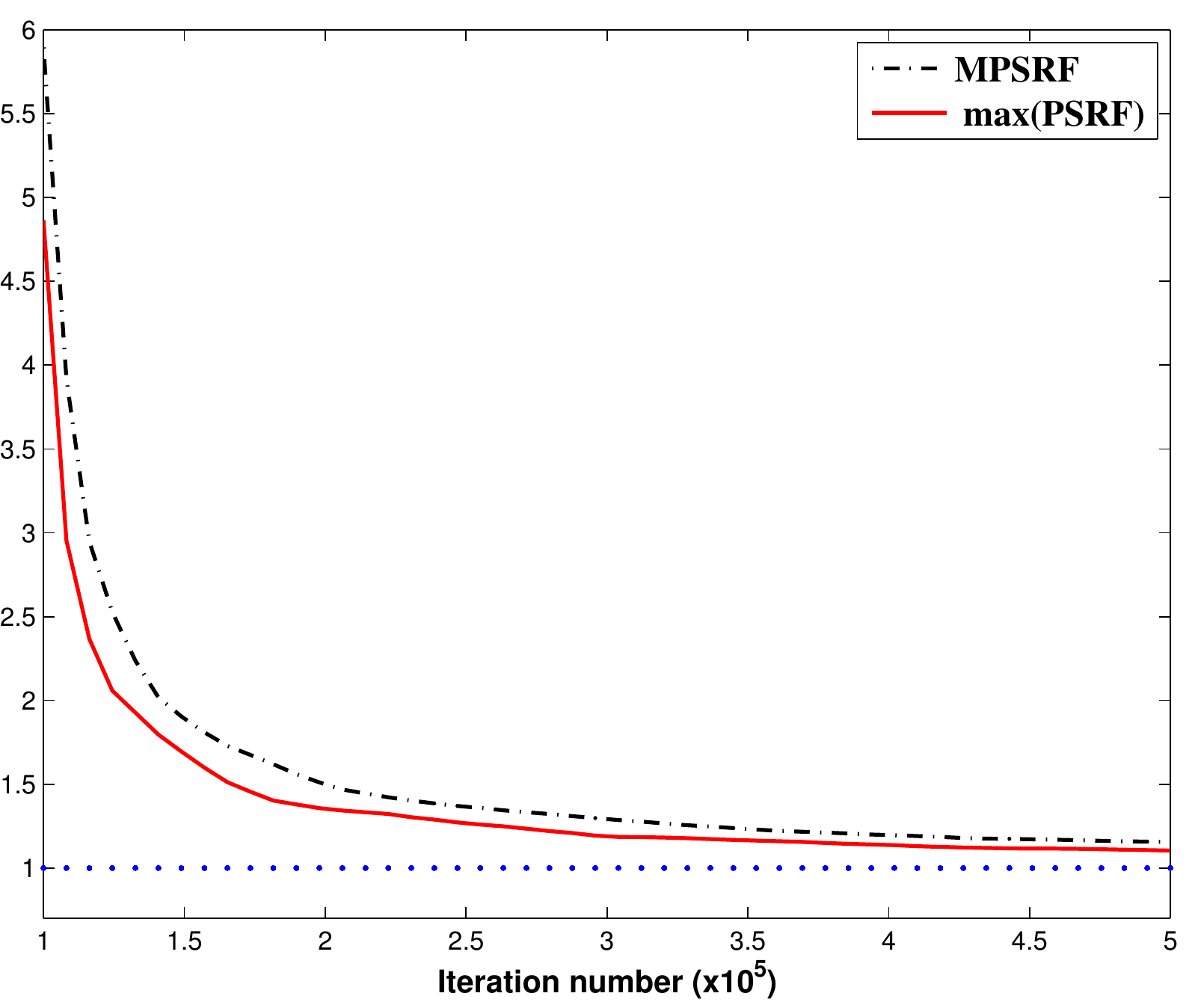}
\caption{Two-phase model (large number of wells). Left: True log-permeability [$\log{\textrm{m}^2}$]. Middle: Well configuration. Right: Gelman-rubin diagnostic}  
\label{Figure14}
\end{figure}

In this subsection we consider again the oil-water reservoir model described in Section \ref{ReservoirModels}. The well configuration for this case is displayed in \ref{Figure14} (middle) and relevant information can be found in Table \ref{Table1}. The aim of this experiment is to evaluate the performance of Gaussian approximations when measurements from many wells are available. The prior distribution of log-permeability is defined by the Gaussian measure defined in (\ref{eq:5.1}) with the same tunable parameters used in the experiment of subsection \ref{singleResults}.

\begin{figure}
\includegraphics[scale=0.65]{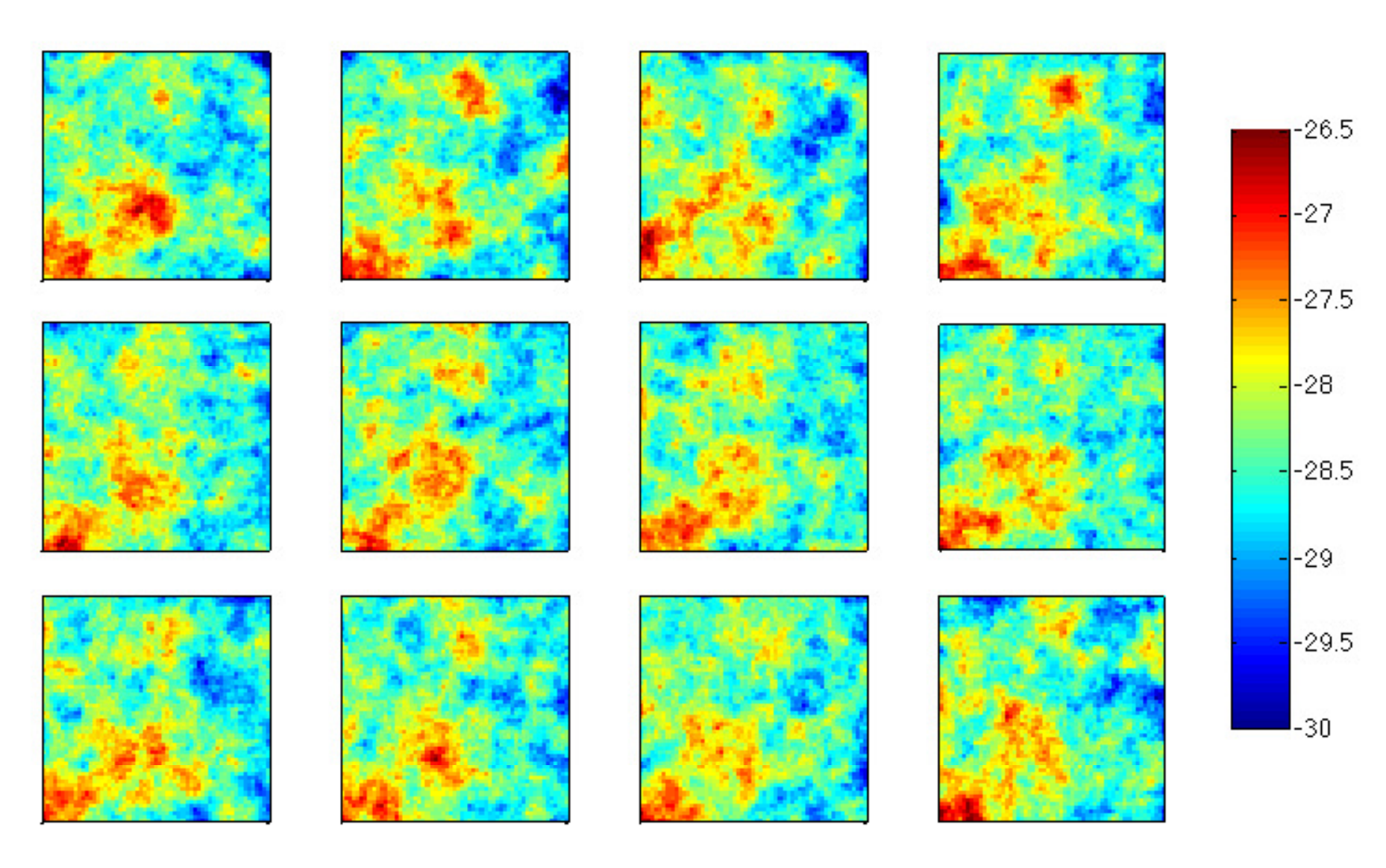}
\caption{Two-phase model (small number of wells). Samples from the posterior distribution (characterized with MCMC) [$\log{\textrm{m}^2}$]}  
\label{Figure15}
\end{figure}

Generation of synthetic data was conducted with the same procedure described before. The ``true log-permeability'' $u^{\dagger}$ is displayed in Figure \ref{Figure14} (left). The random error $\eta$ added to $G(u^{\dagger})$ has the form of (\ref{eq:5.5}) with $N_{I}=9$ and $N_{P}=16$ and
 \begin{eqnarray}\label{eq:5.7}
\eta_{n}^{j,I}\sim N(0,2.7\times 10^{4}\textrm{Pa}),\qquad j\in\{1,\dots,N_{I}\},\nonumber \\
\eta_{n}^{j,P}  \sim N(0,0.06\textrm{m}^{3}/\textrm{day}),\qquad j\in\{1,\dots,N_{P}\}.
\end{eqnarray}
for all $n\in \{1,\dots, 7\}$. Measurement times are $t_{n}=0.467n~\textrm{years}$, $n=\{1,\dots,7\}$
The pCN-MCMC Algorithm \ref{al:MCMC} is applied to generate 110 chains starting from independent draws from the prior distribution. After a burn-in period of $1.5\times 10^{4}$ the chains are run for $5\times 10^5$ iterations. The Gelman-Rubin diagnostic is conducted as describe in the preceding sections. Figure \ref{Figure14} (right) show the PRSF and MPRSF as defined previously. Uncorrelated samples (from independent chains) are shown in Figure \ref{Figure15}.

\begin{figure}
\includegraphics[scale=0.75]{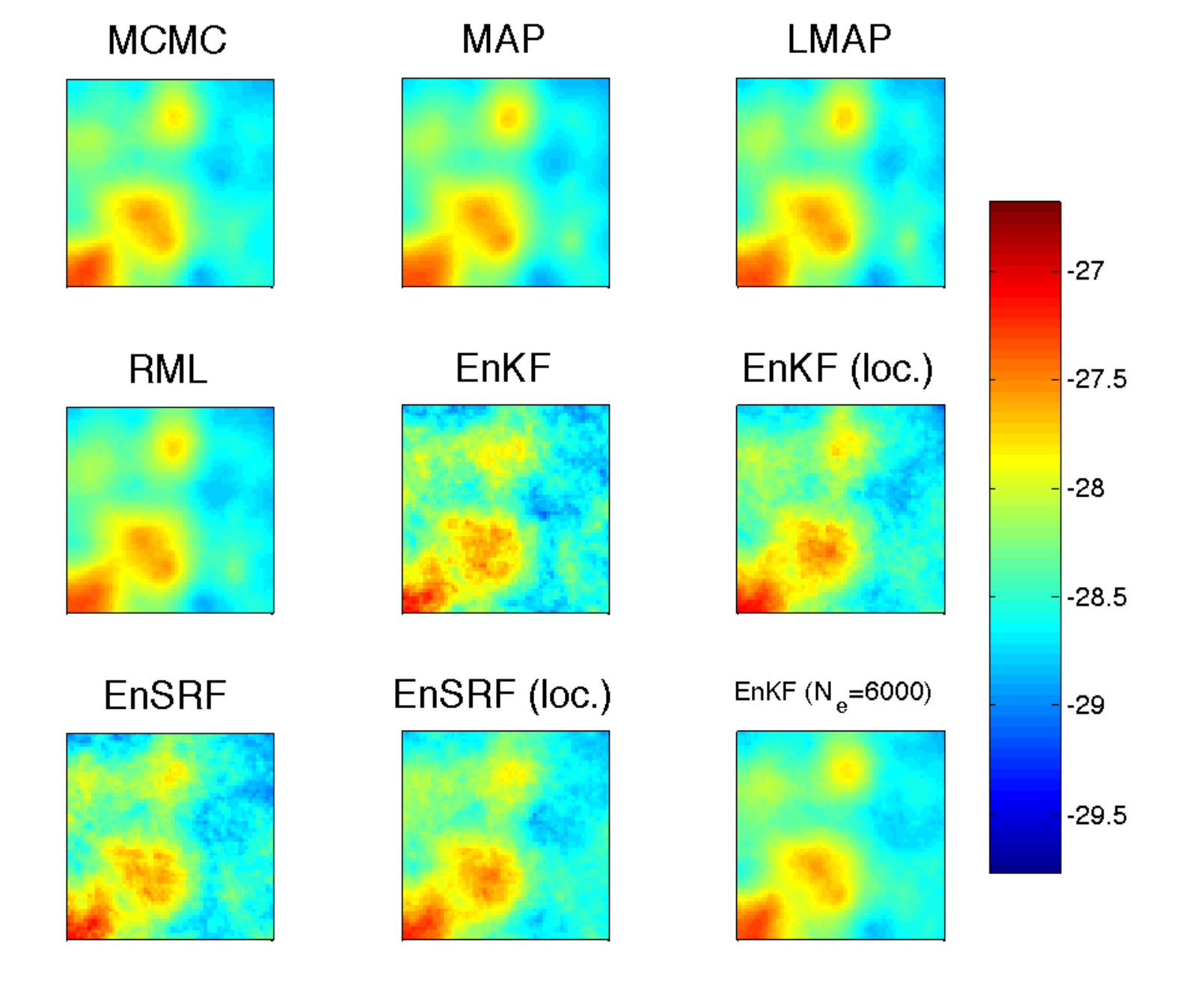}
\caption{Two-phase model (large number of wells). Mean of the posterior distribution (characterized with MCMC) and Gaussian approximations [$\log{\textrm{m}^2}$]}   
\label{Figure16}
\end{figure}

\begin{figure}
\includegraphics[scale=0.75]{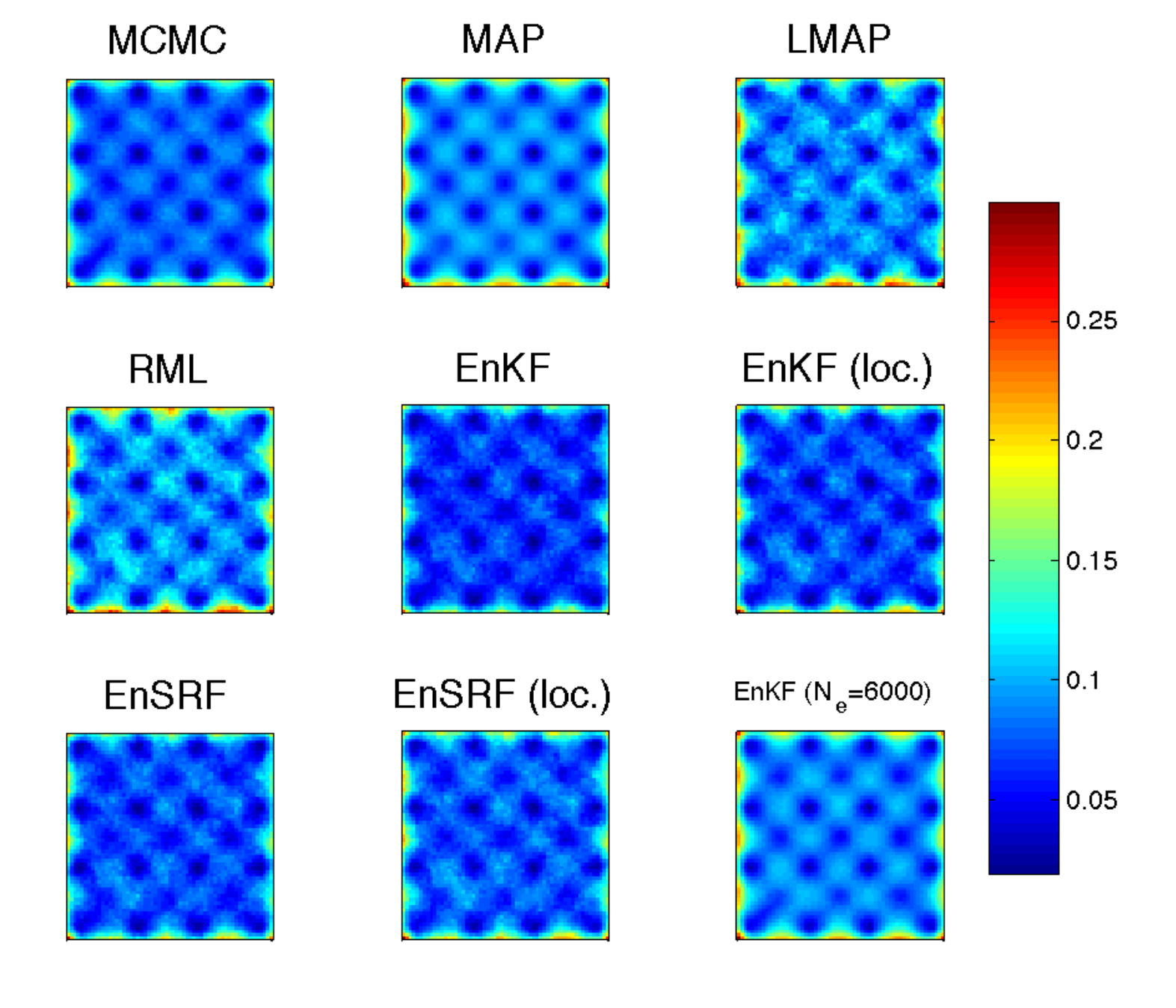}
\caption{Two-phase model (large number of wells). Variance of the posterior distribution (characterized with MCMC) and Gaussian approximations [$(\log{\textrm{m}^2})^2$]}  
\label{Figure17}
\end{figure}

Similar to the previous section, converged chains provide the posterior against which to compare the performance of  Gaussian approximations in terms of mean and variance. The first part of Table \ref{Table4} provides the results when an ensemble with of size $N_{e}=50$ is used. In Figure \ref{Figure16} and Figure \ref{Figure17} we display mean and variance, respectively.

\begin{figure}
\includegraphics[scale=0.22]{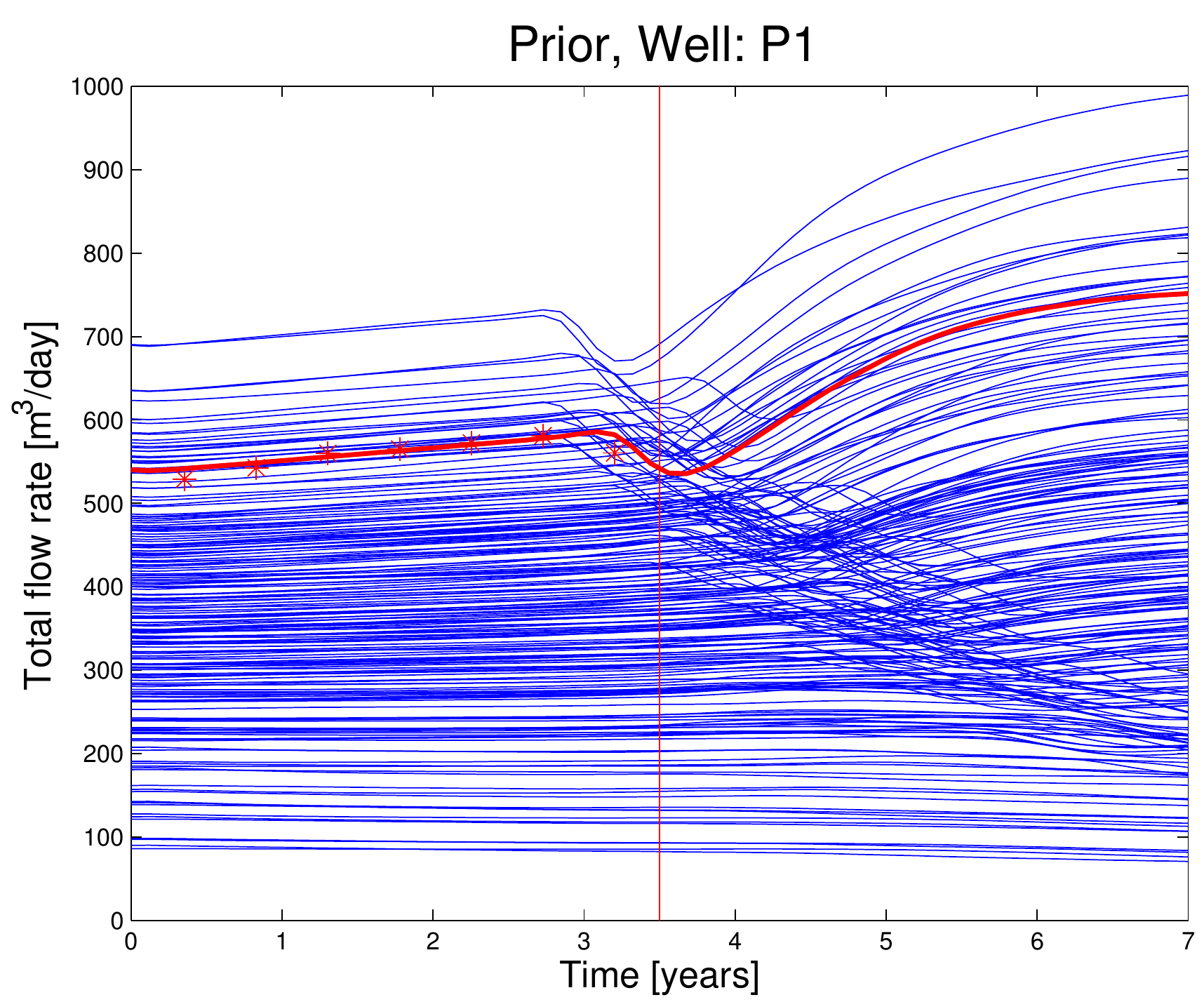}
\includegraphics[scale=0.22]{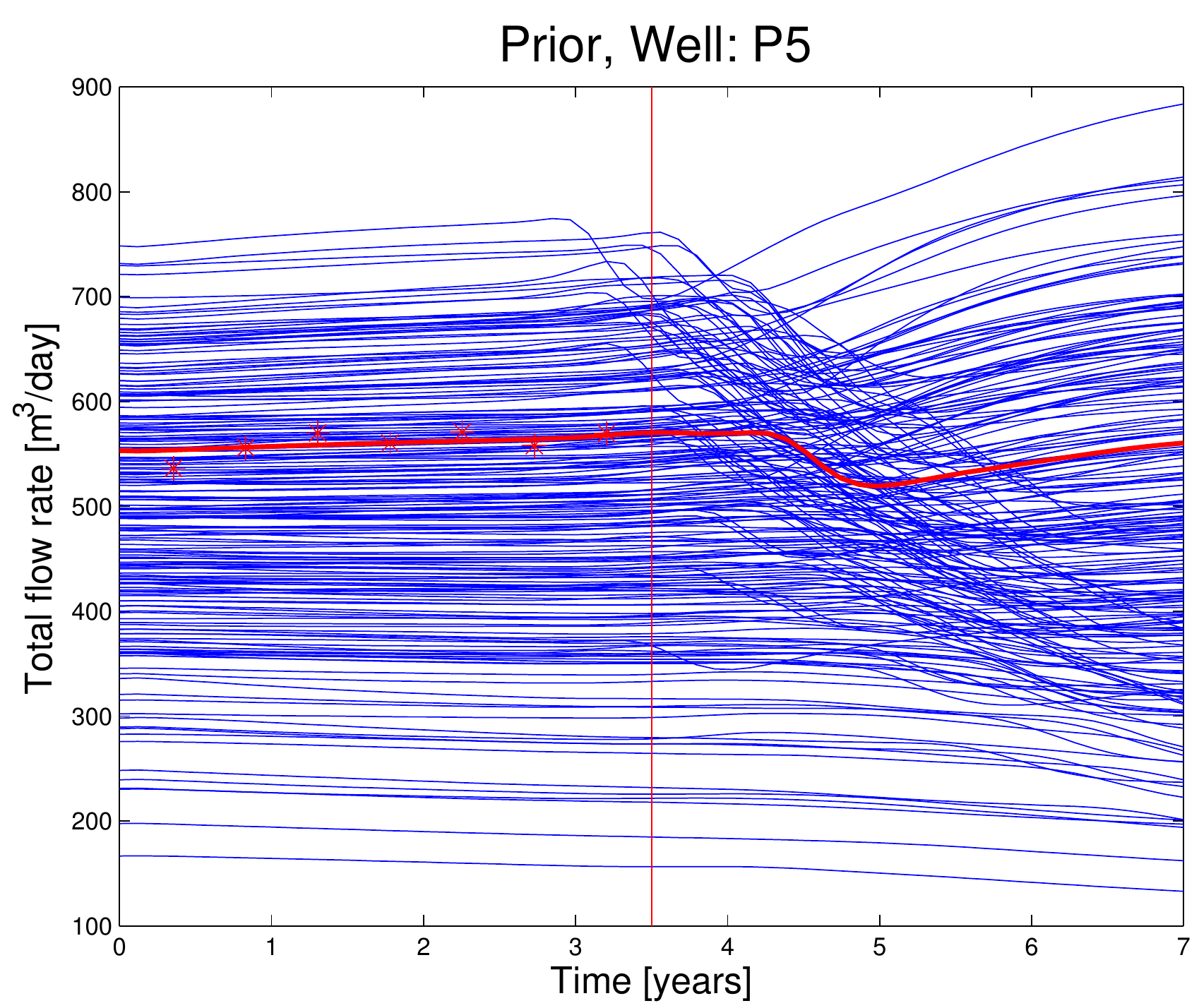}
\includegraphics[scale=0.22]{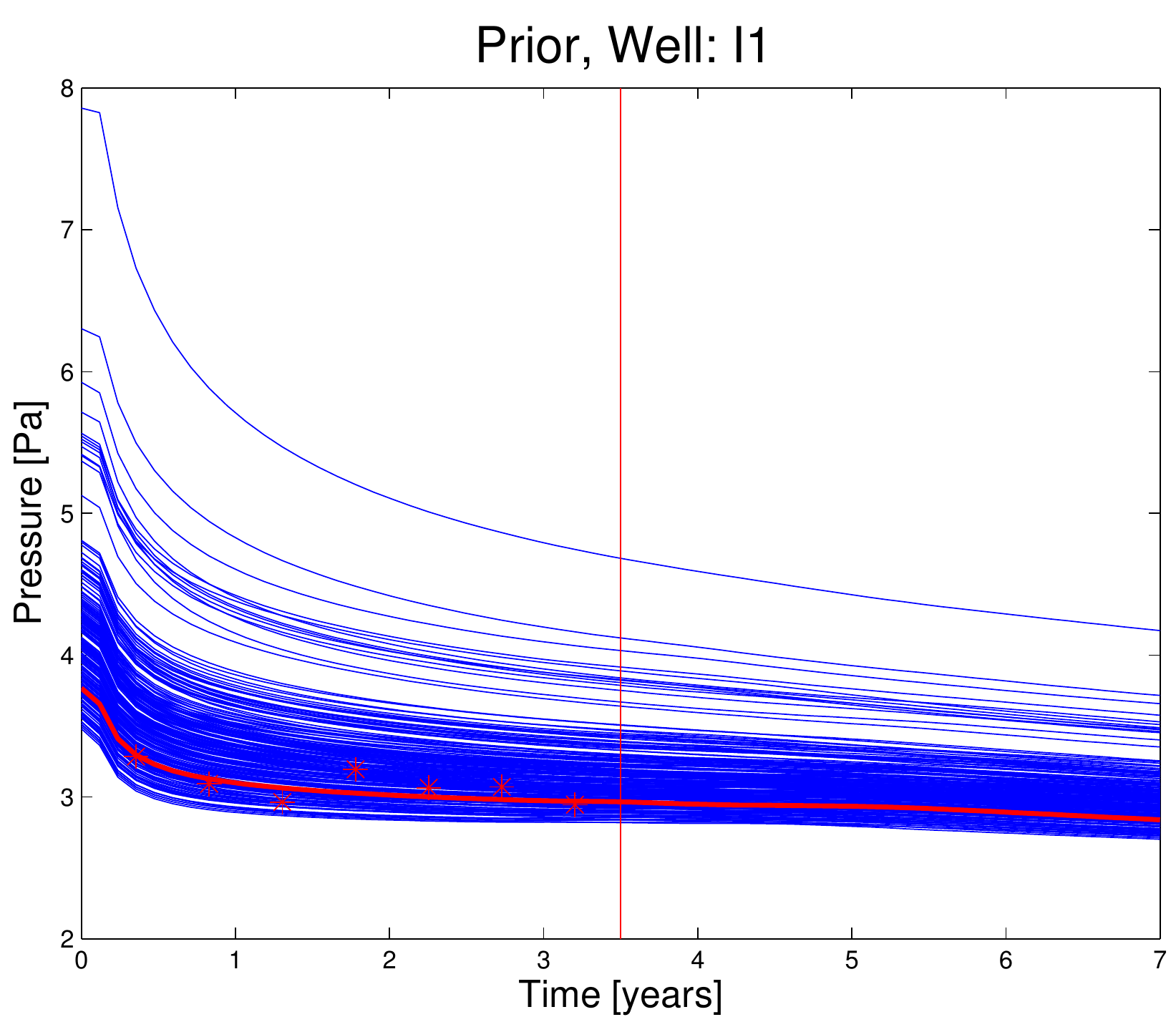}\\
\includegraphics[scale=0.22]{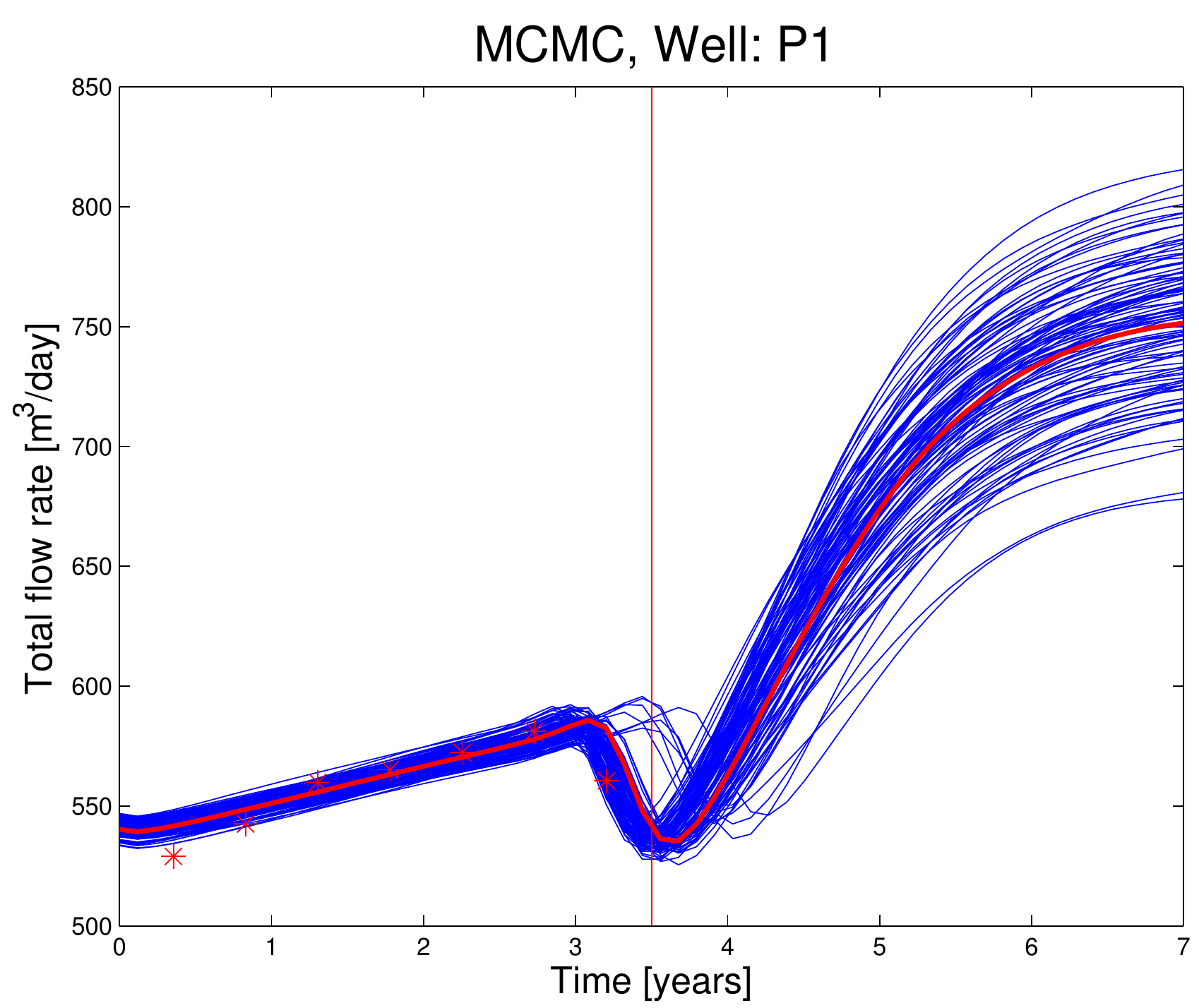}
\includegraphics[scale=0.22]{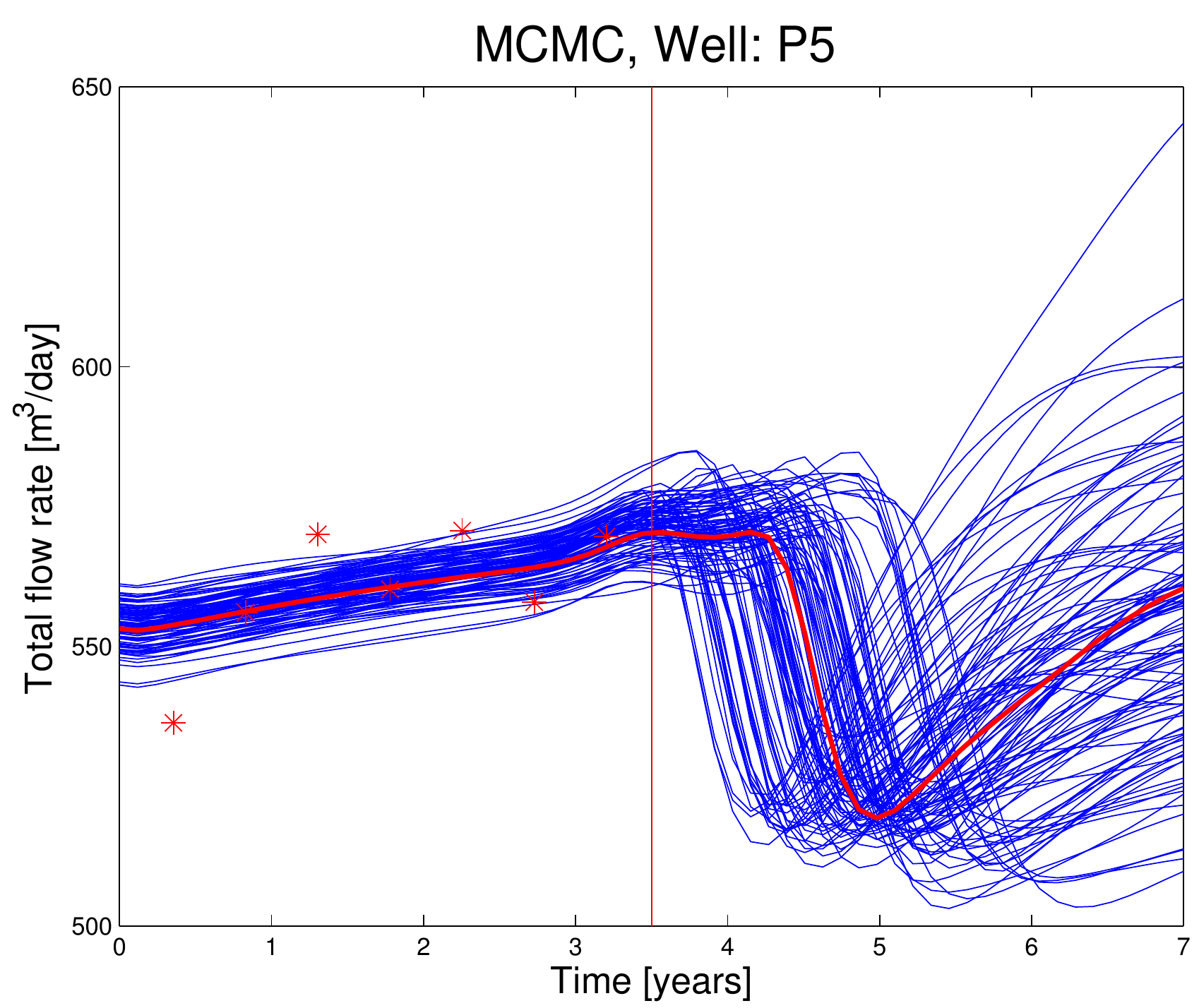}
\includegraphics[scale=0.22]{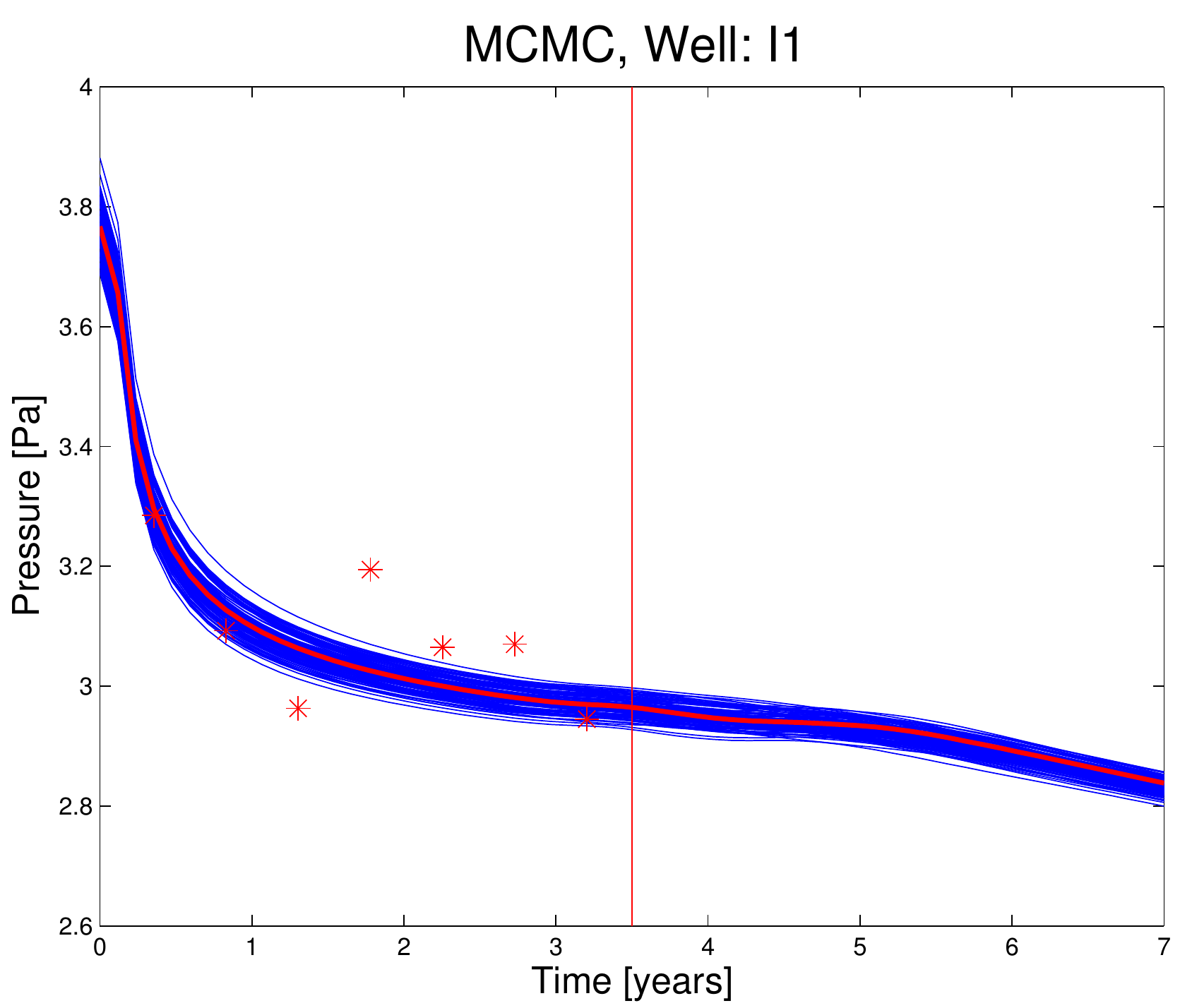}\\
\includegraphics[scale=0.22]{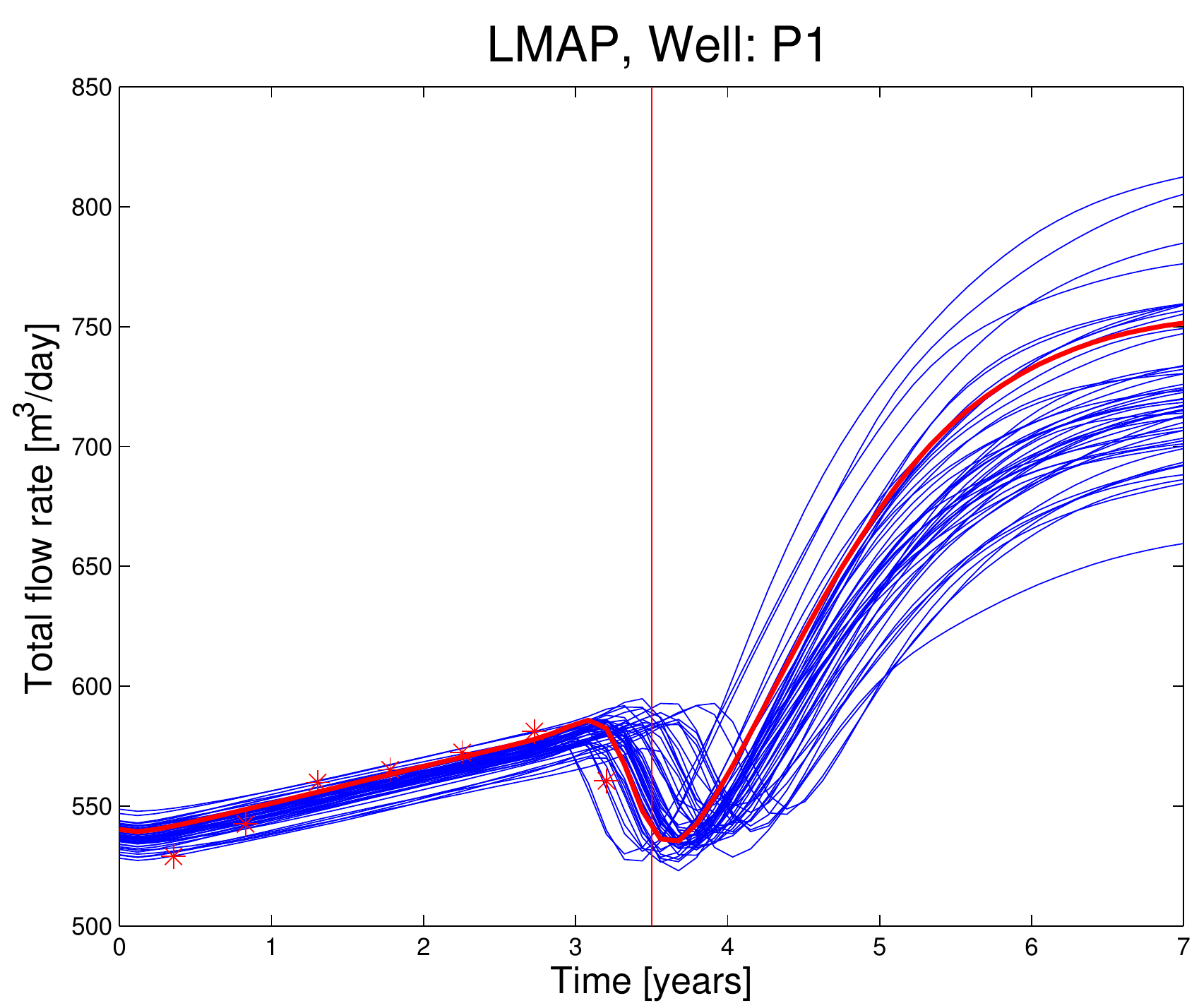}
\includegraphics[scale=0.22]{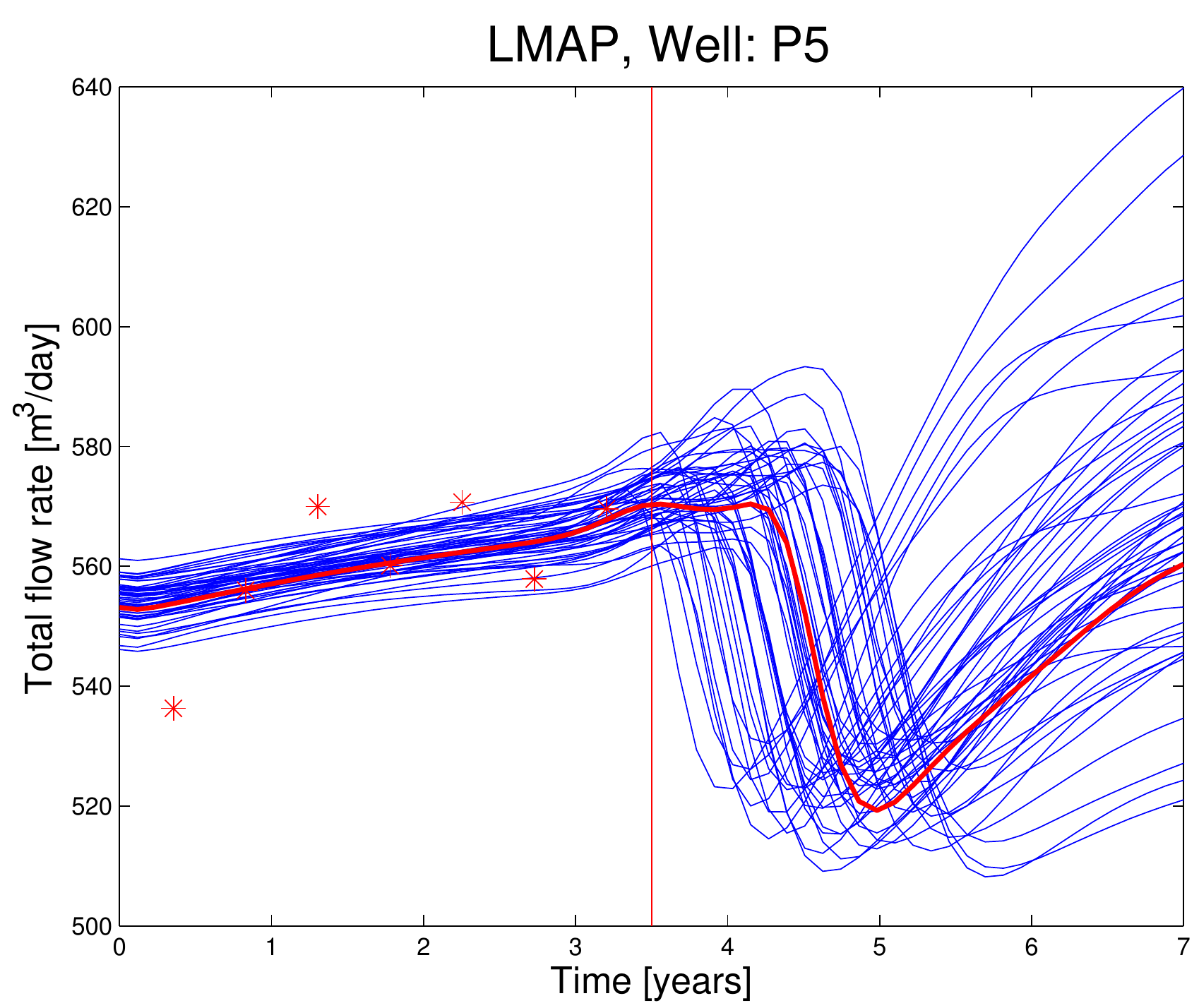}
\includegraphics[scale=0.22]{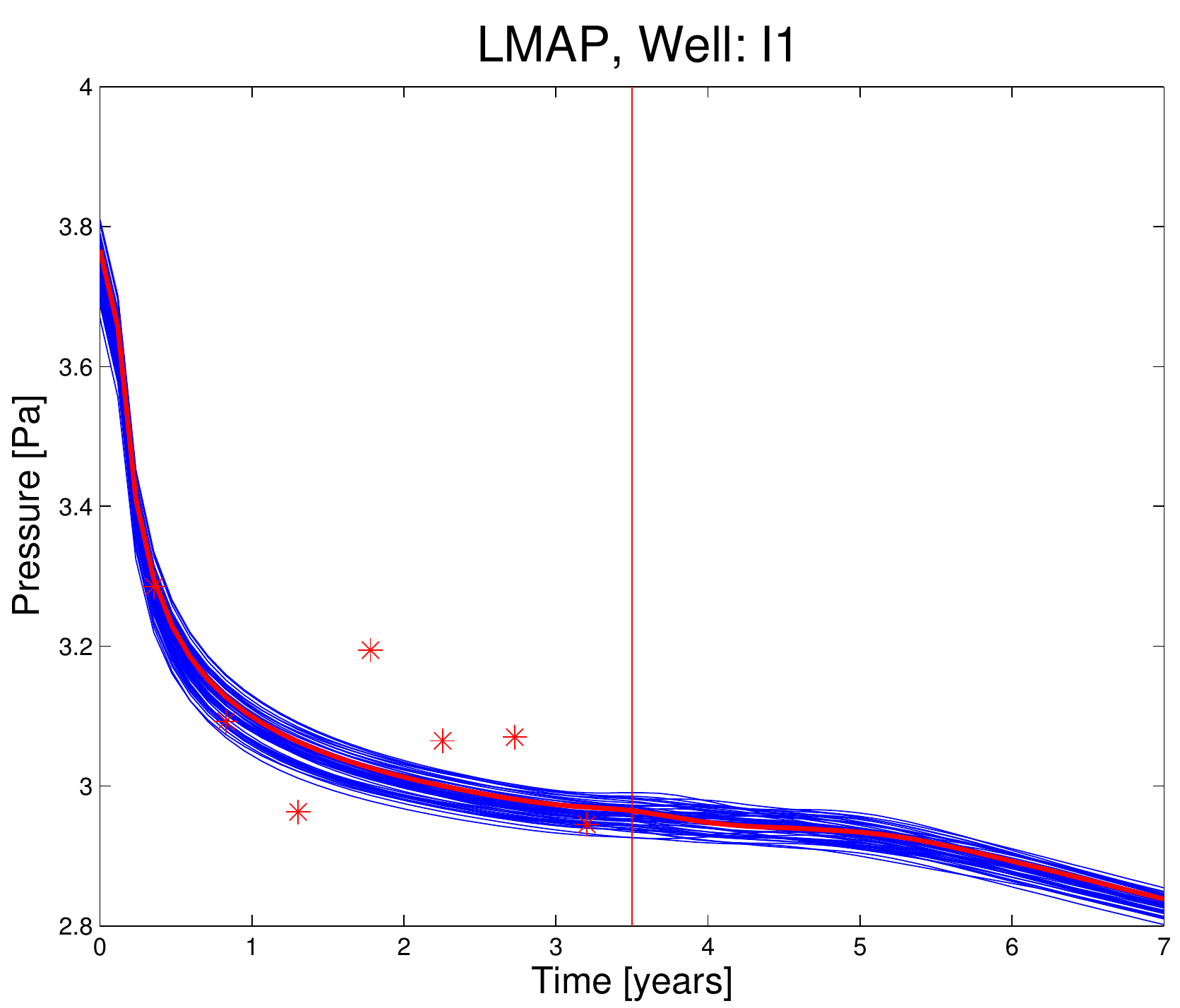}\\
\includegraphics[scale=0.22]{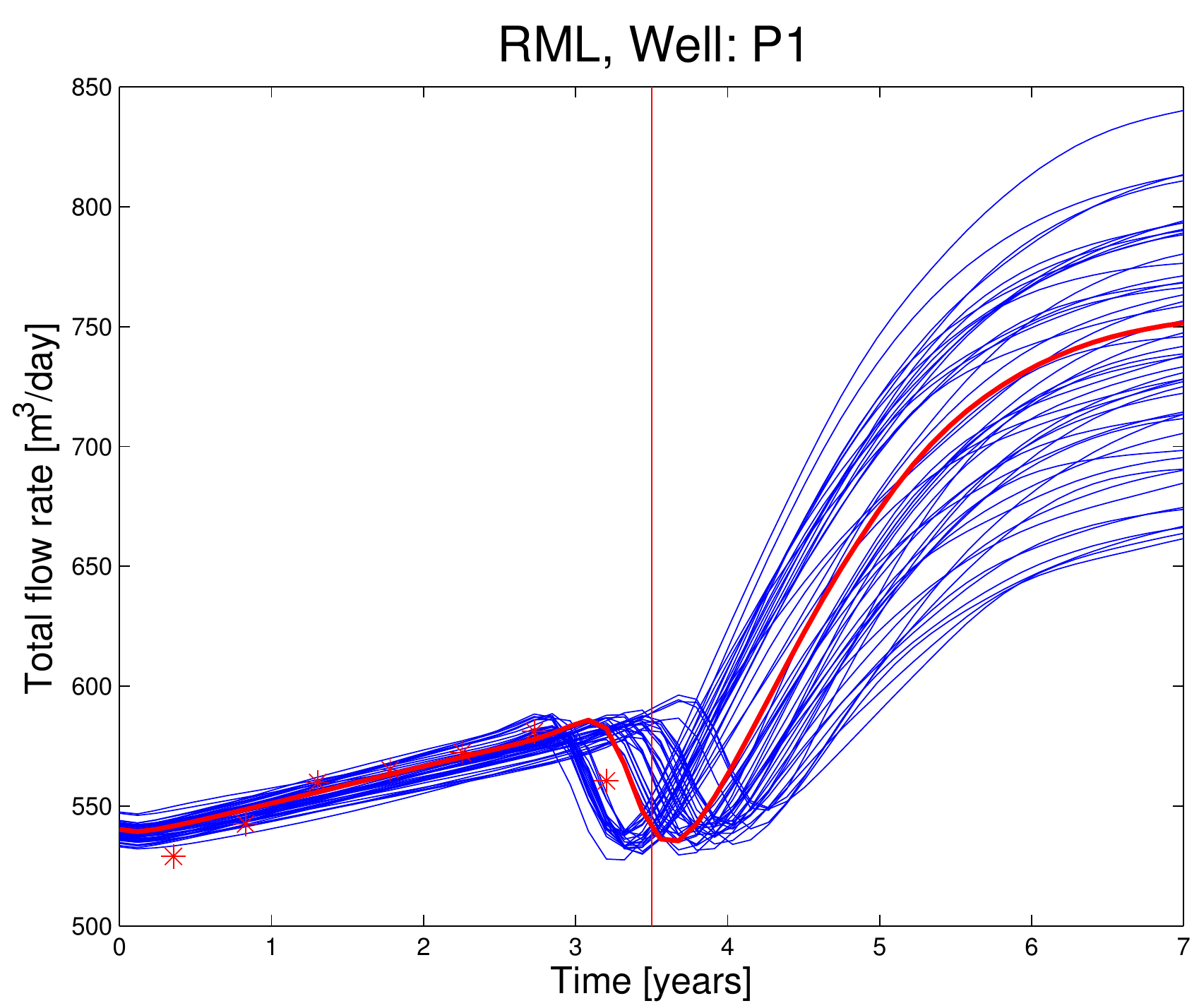}
\includegraphics[scale=0.22]{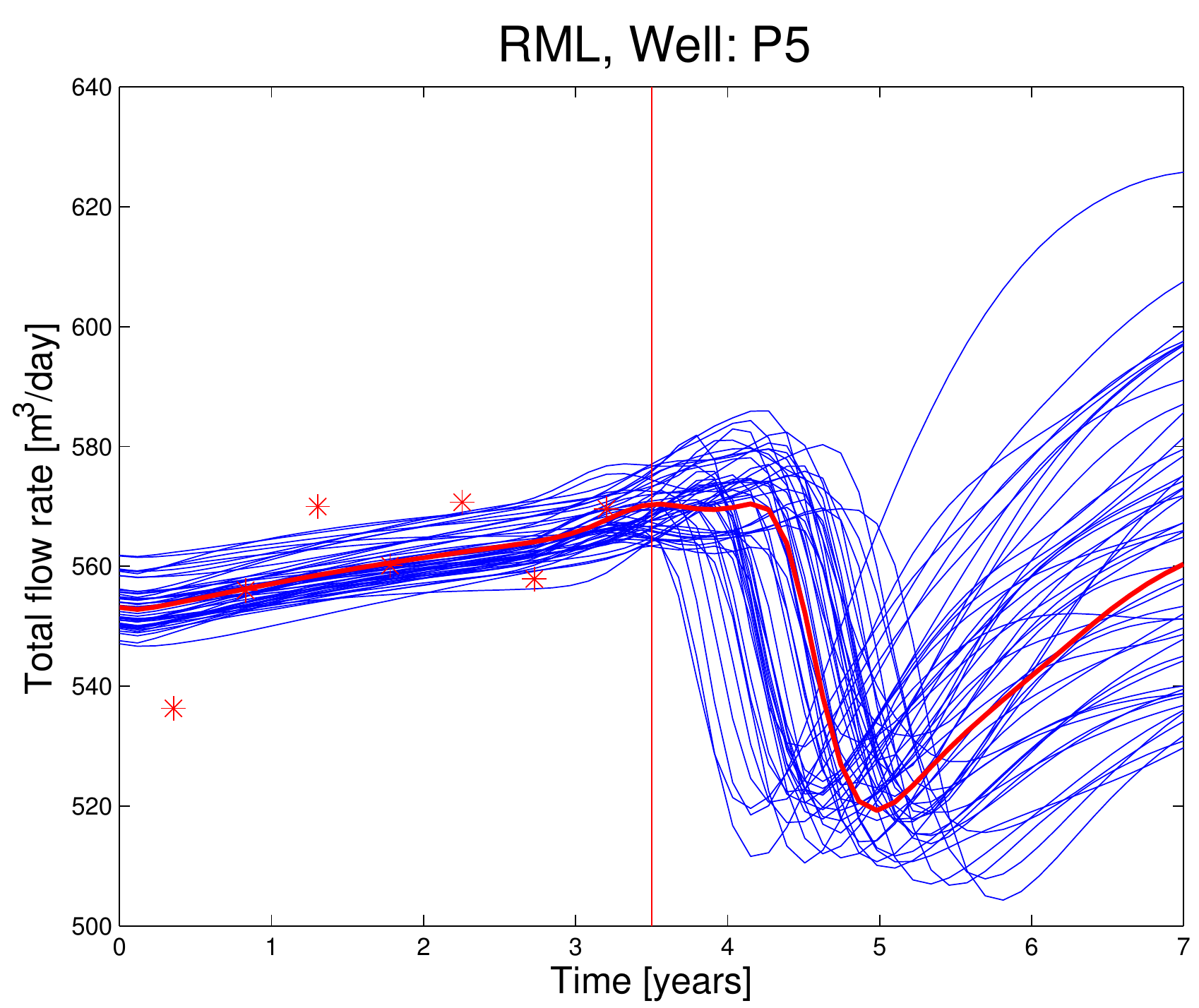}
\includegraphics[scale=0.22]{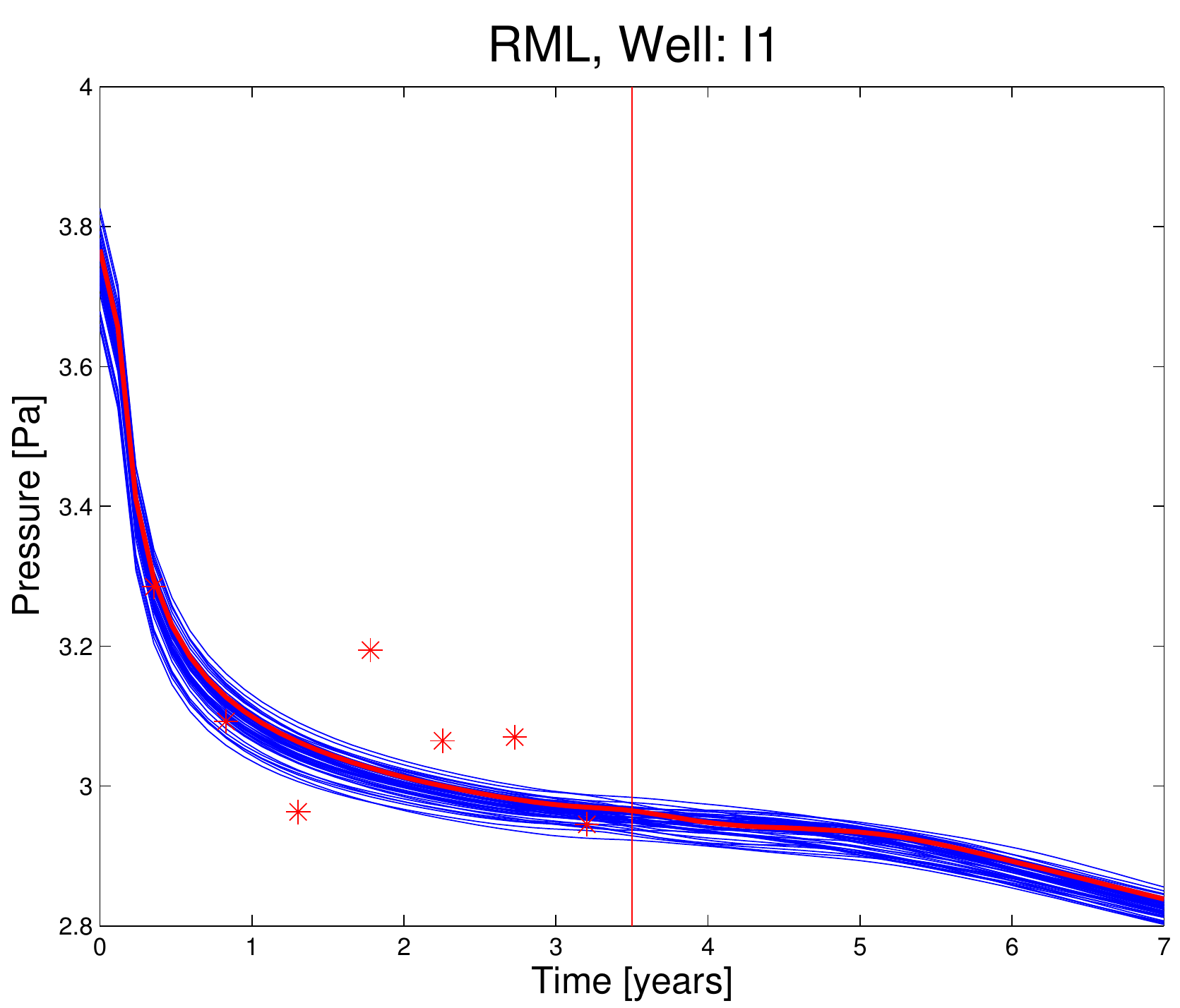}\\
\includegraphics[scale=0.22]{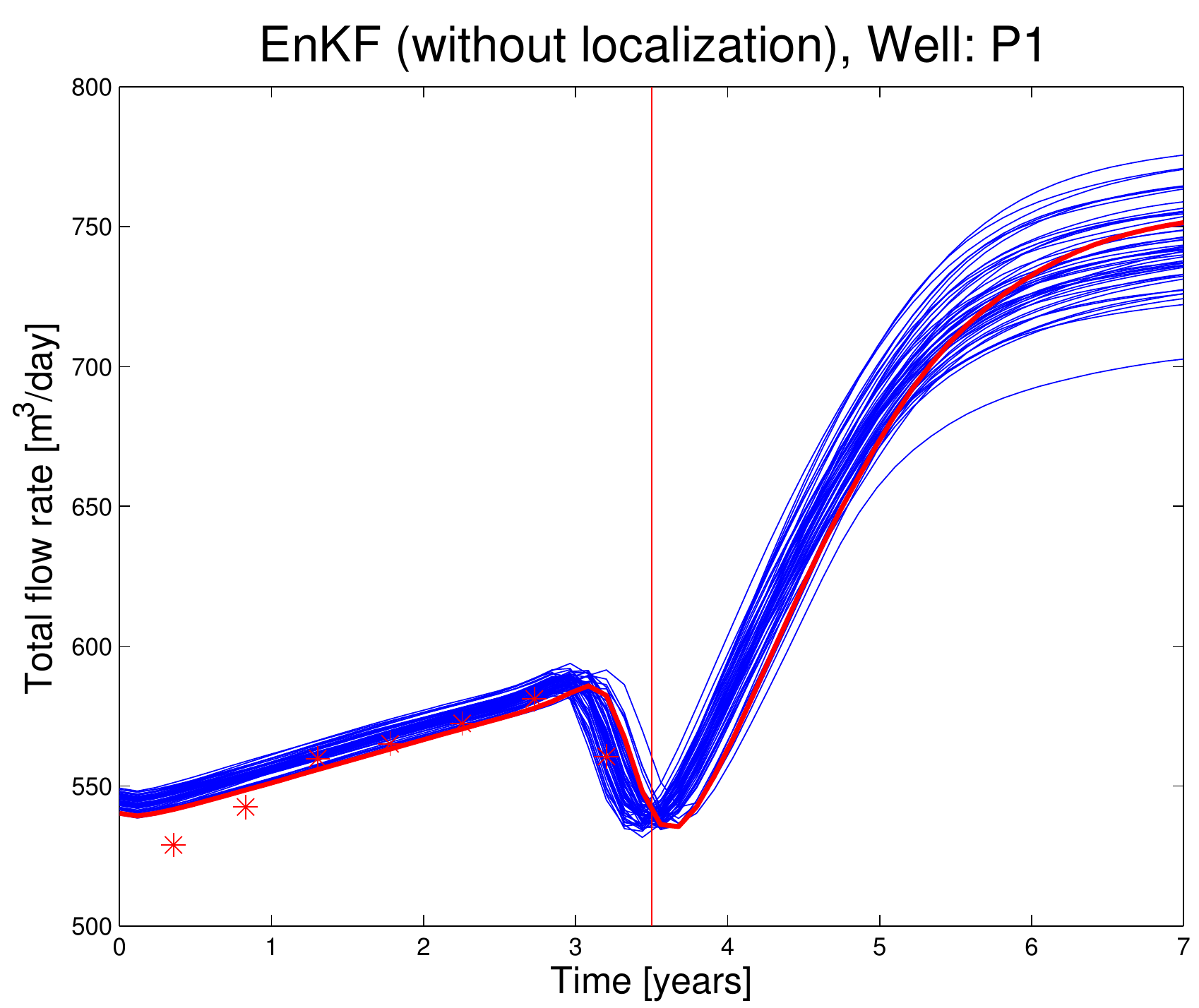}
\includegraphics[scale=0.22]{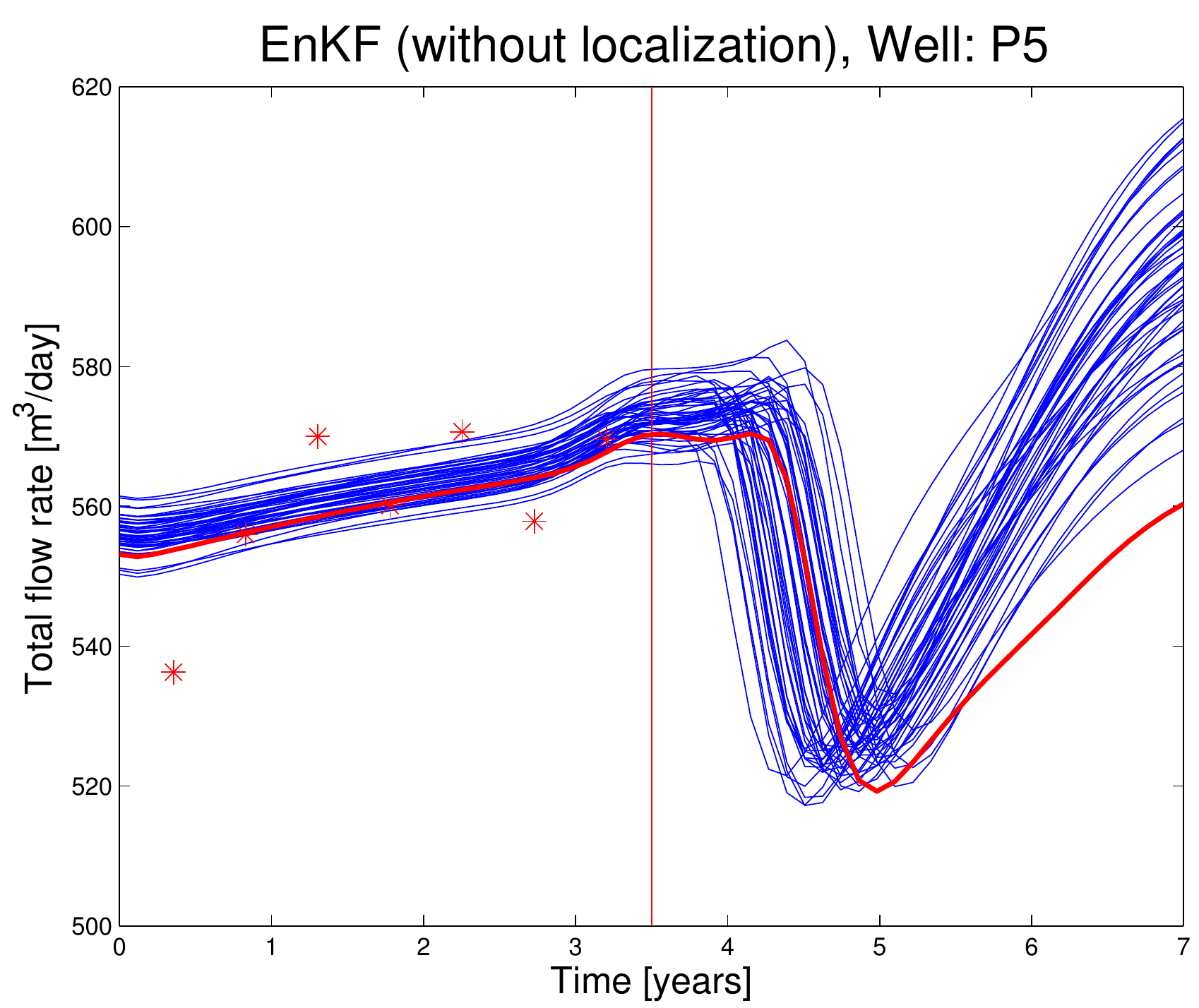}
\includegraphics[scale=0.22]{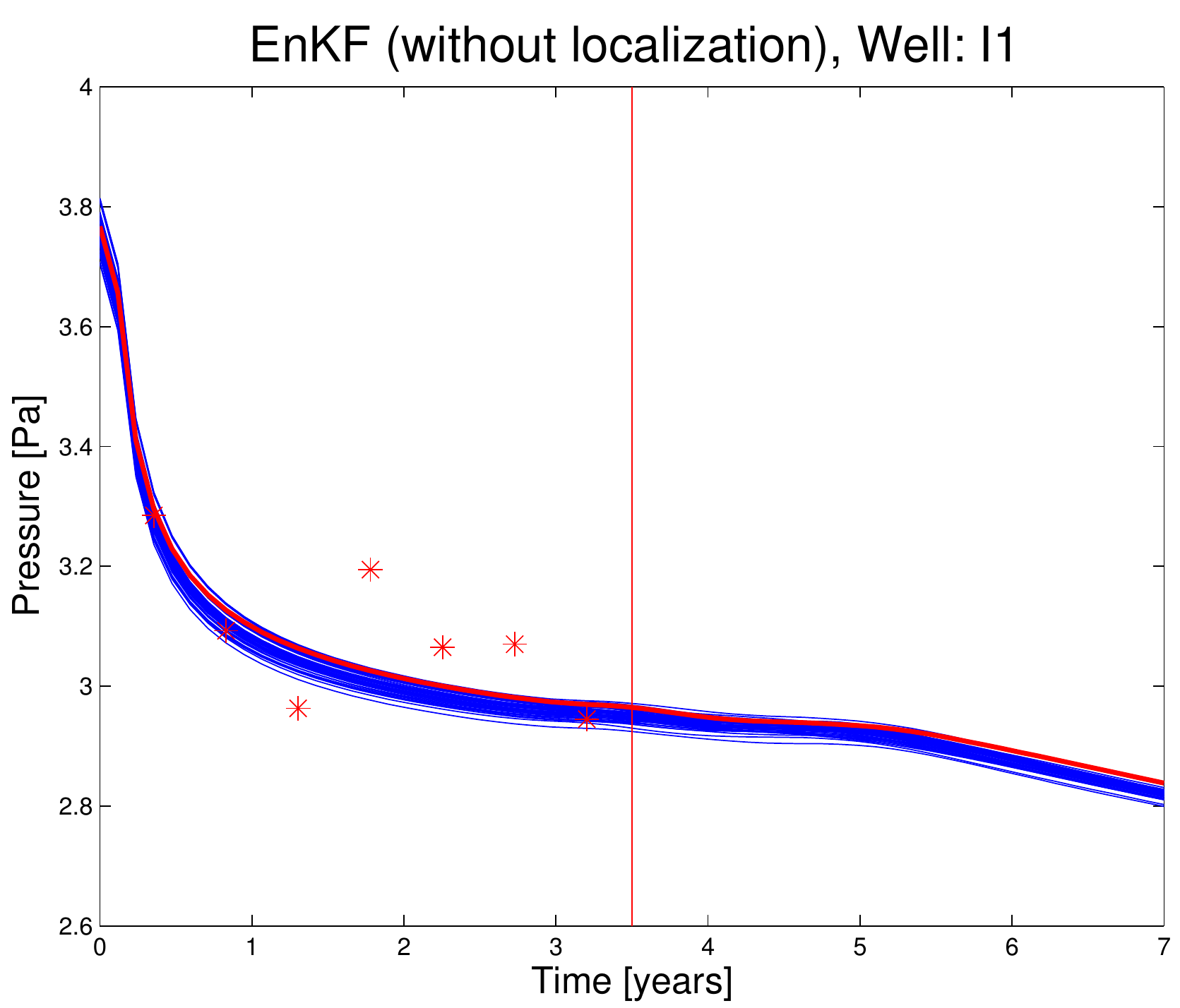}\\
\includegraphics[scale=0.22]{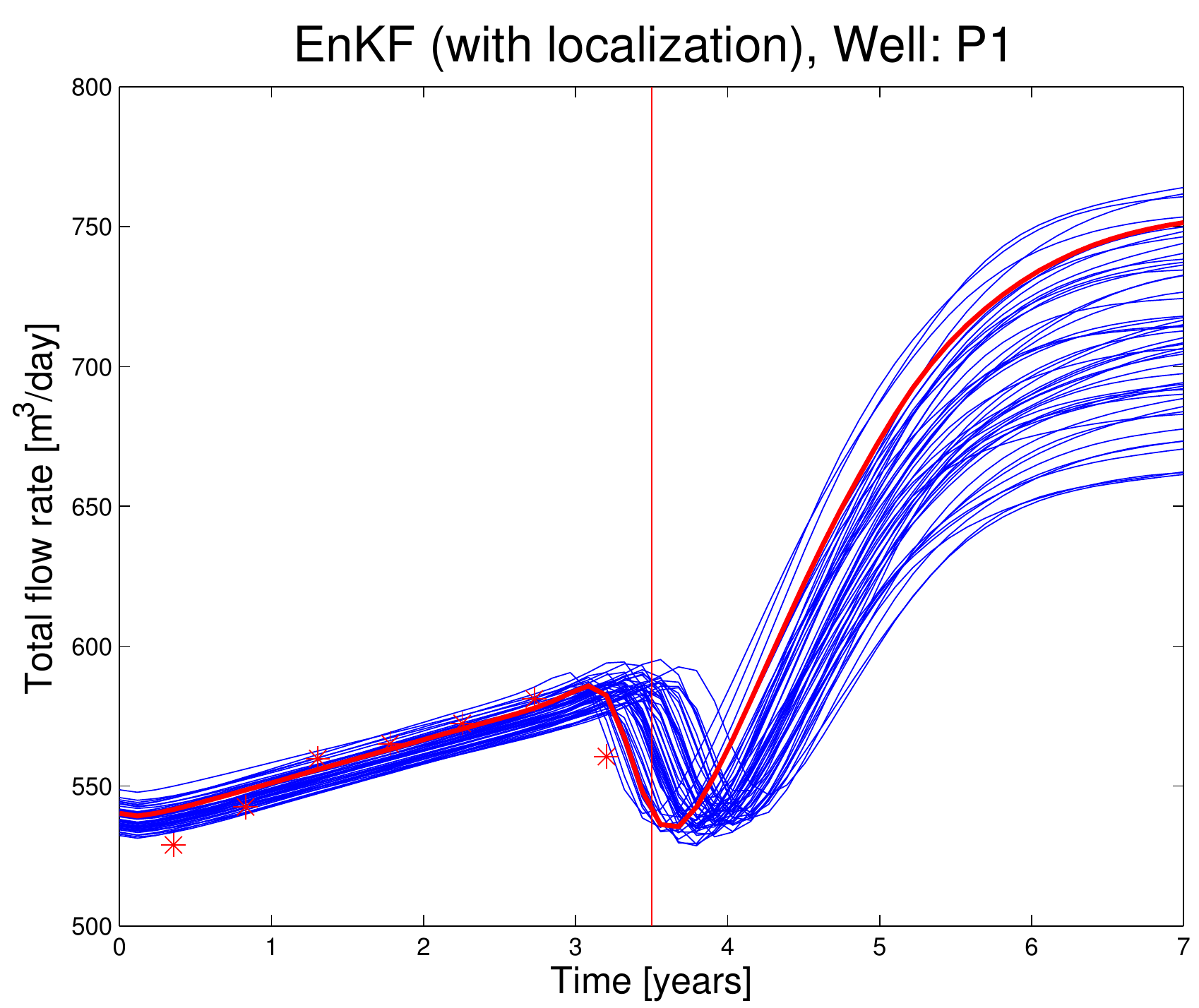}
\includegraphics[scale=0.22]{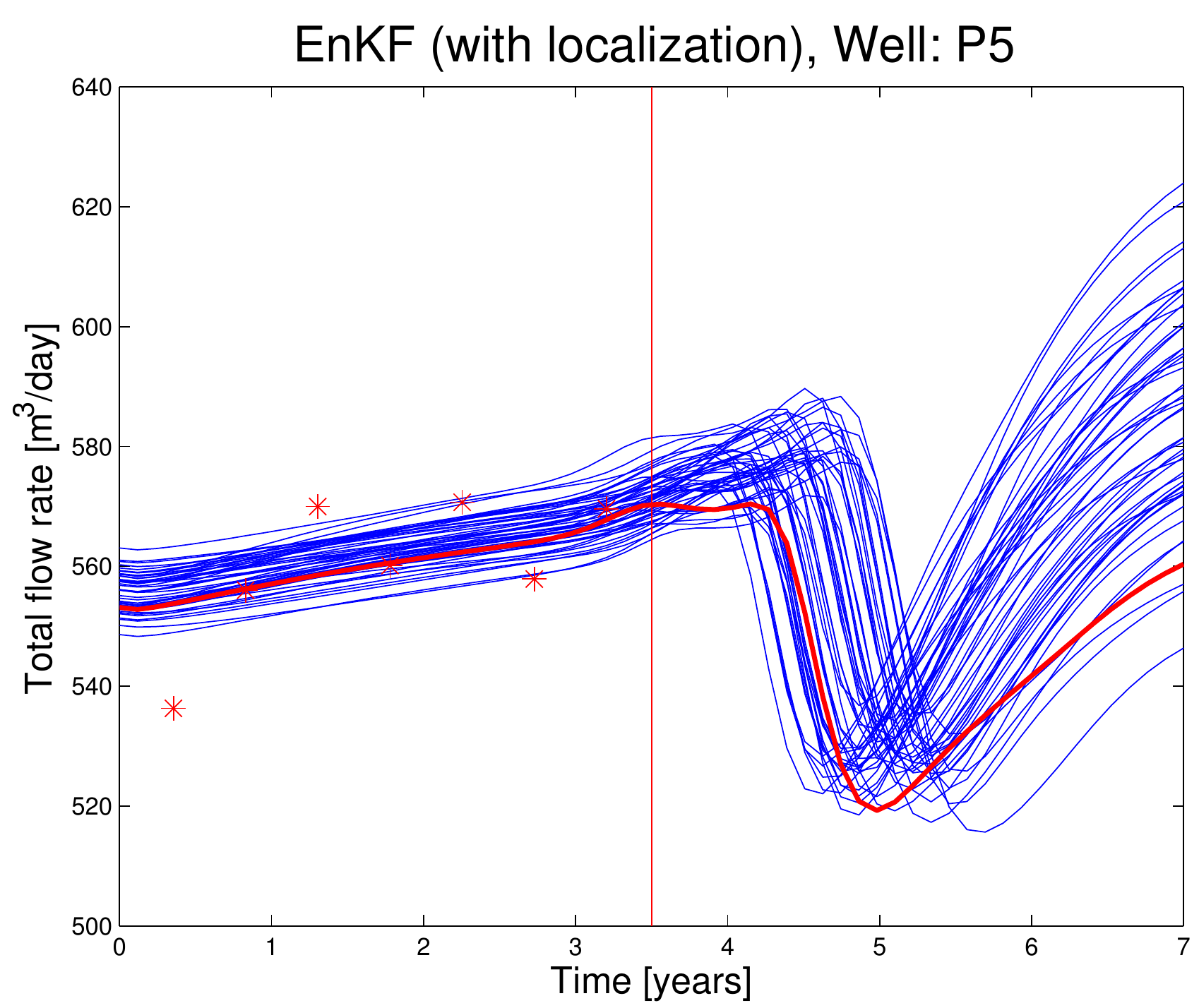}
\includegraphics[scale=0.22]{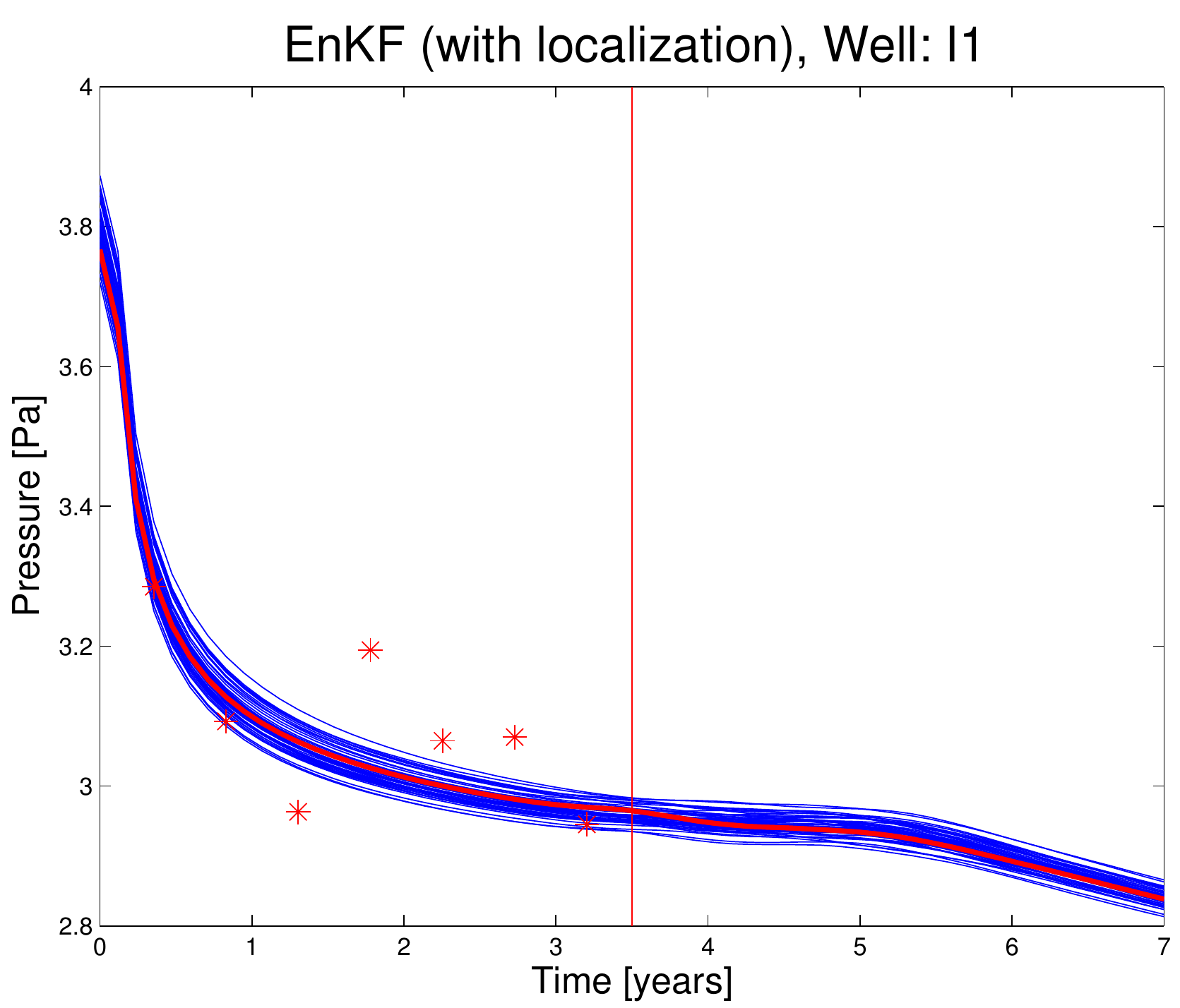}
\caption{Two-phase model (large number of wells). Total flow rates from $P_{1}$ (left column), $P_{5}$ (middle column) and bottom-hole pressure from $I_{1}$ (right column) simulated with permeabilities sampled from (top to bottom rows) the prior, the posterior, LMAP, RML, EnKF and EnKF with localization. }
\label{Figure18}
\end{figure}

\begin{figure}
\includegraphics[scale=0.35]{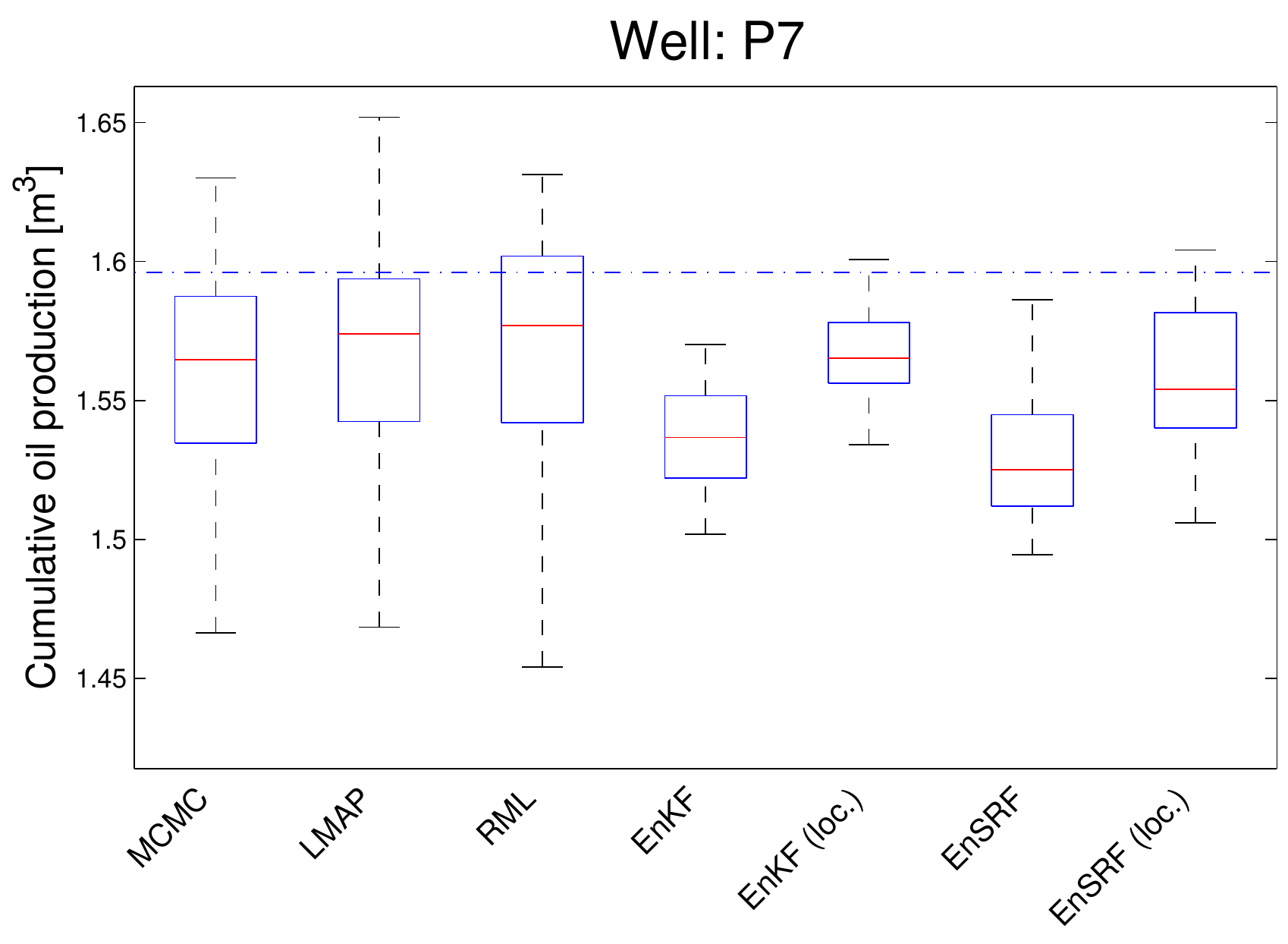}
\includegraphics[scale=0.35]{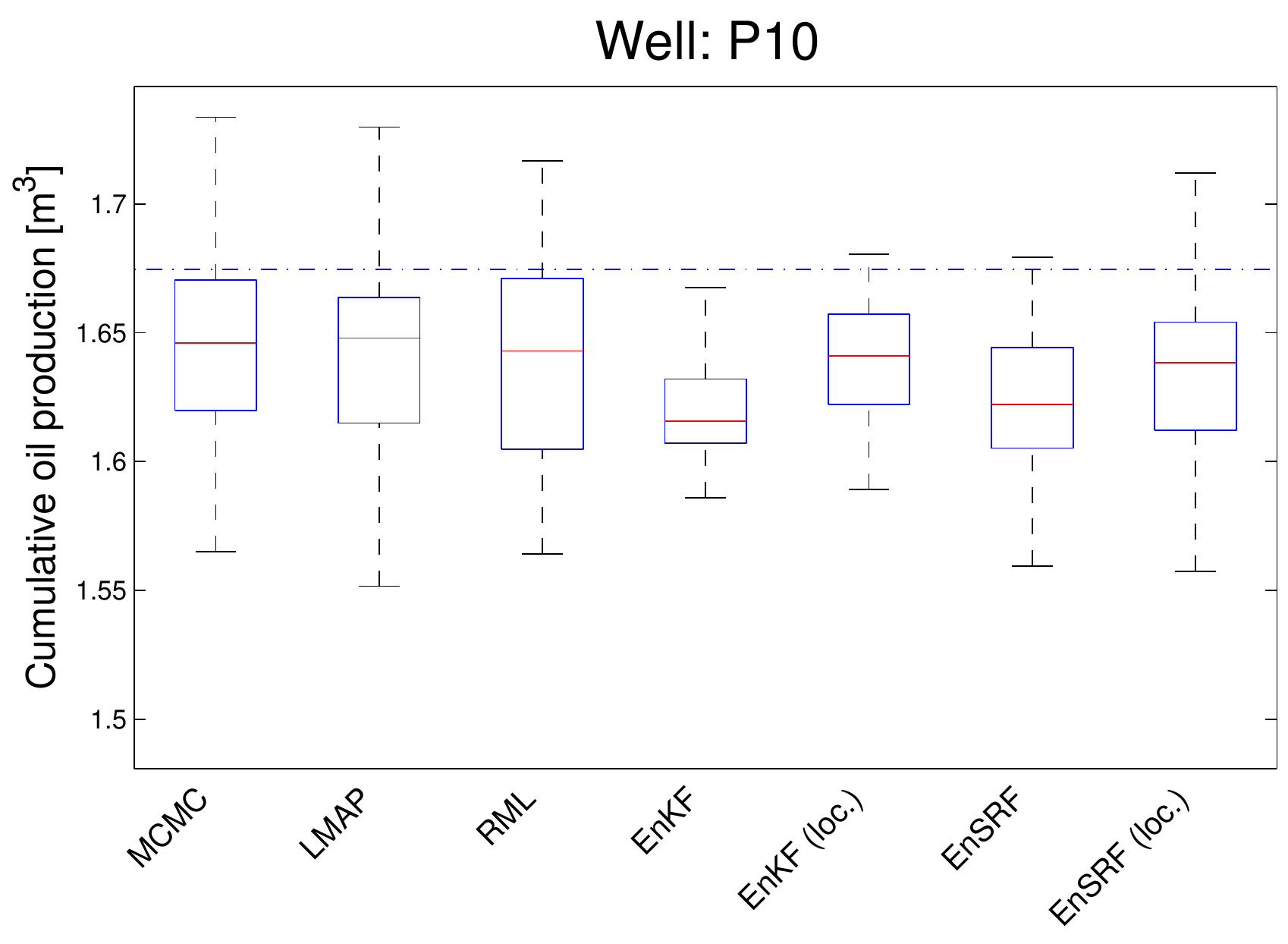}\\
\includegraphics[scale=0.35]{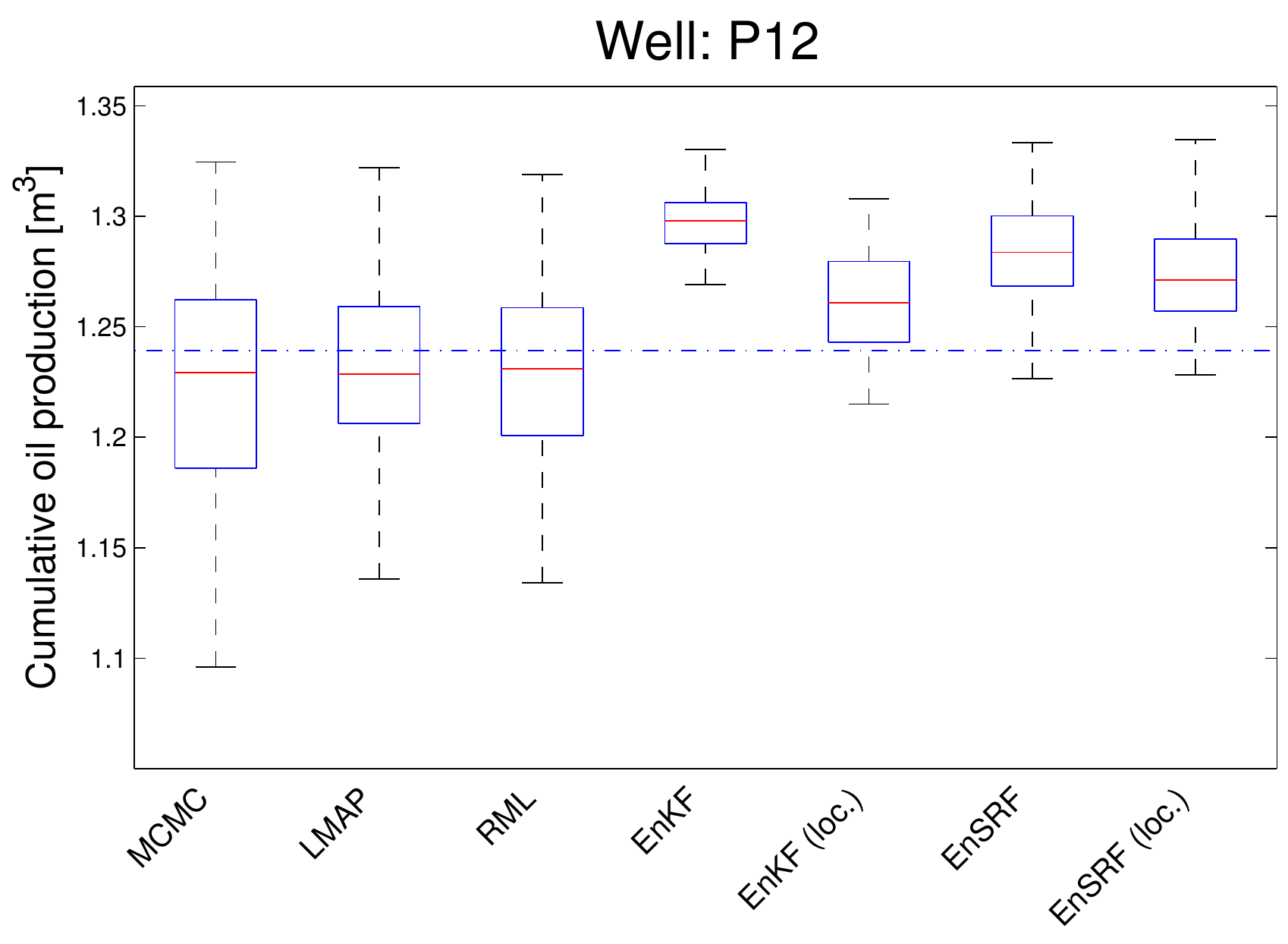}
\includegraphics[scale=0.35]{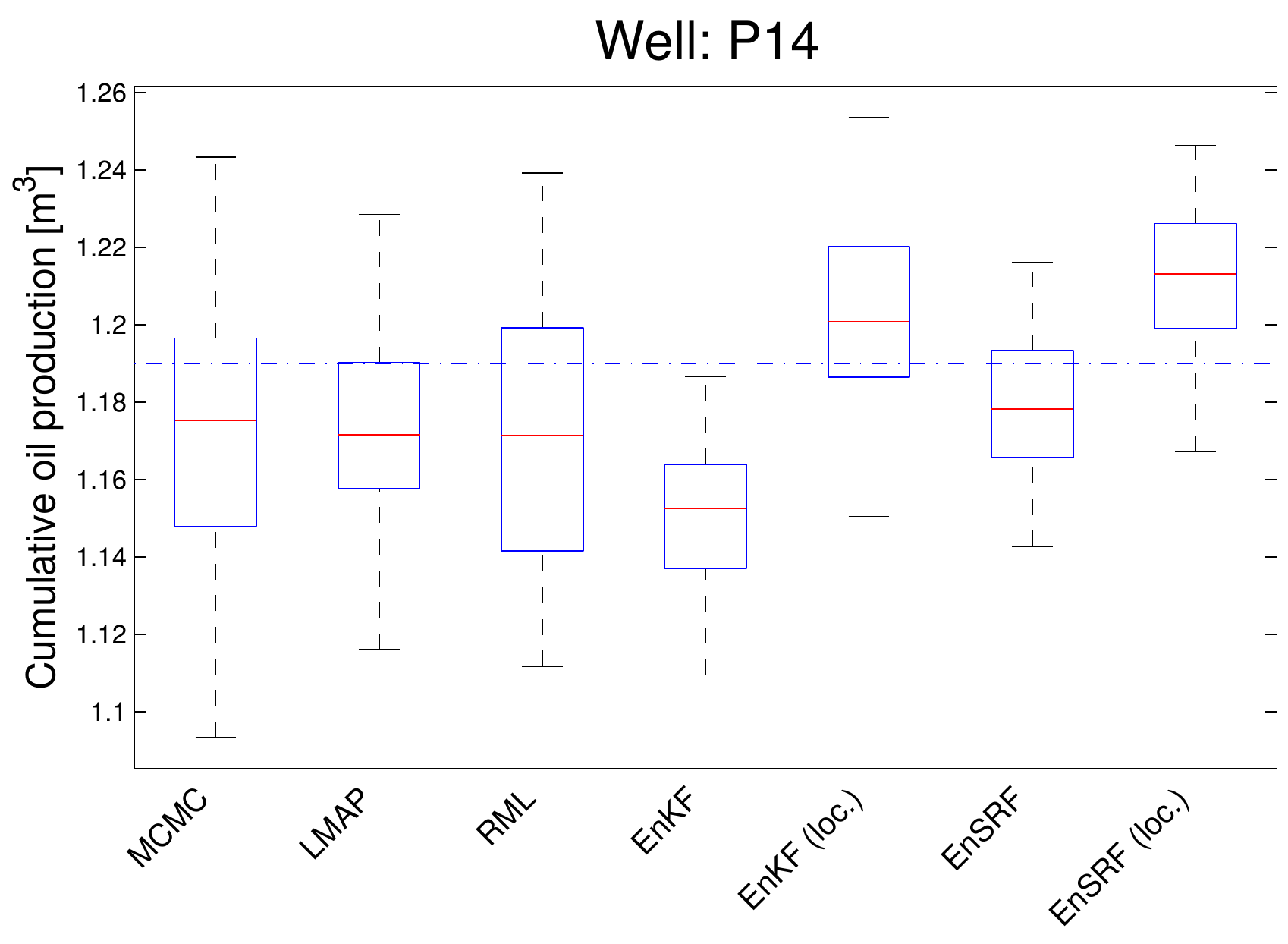}\\
\includegraphics[scale=0.35]{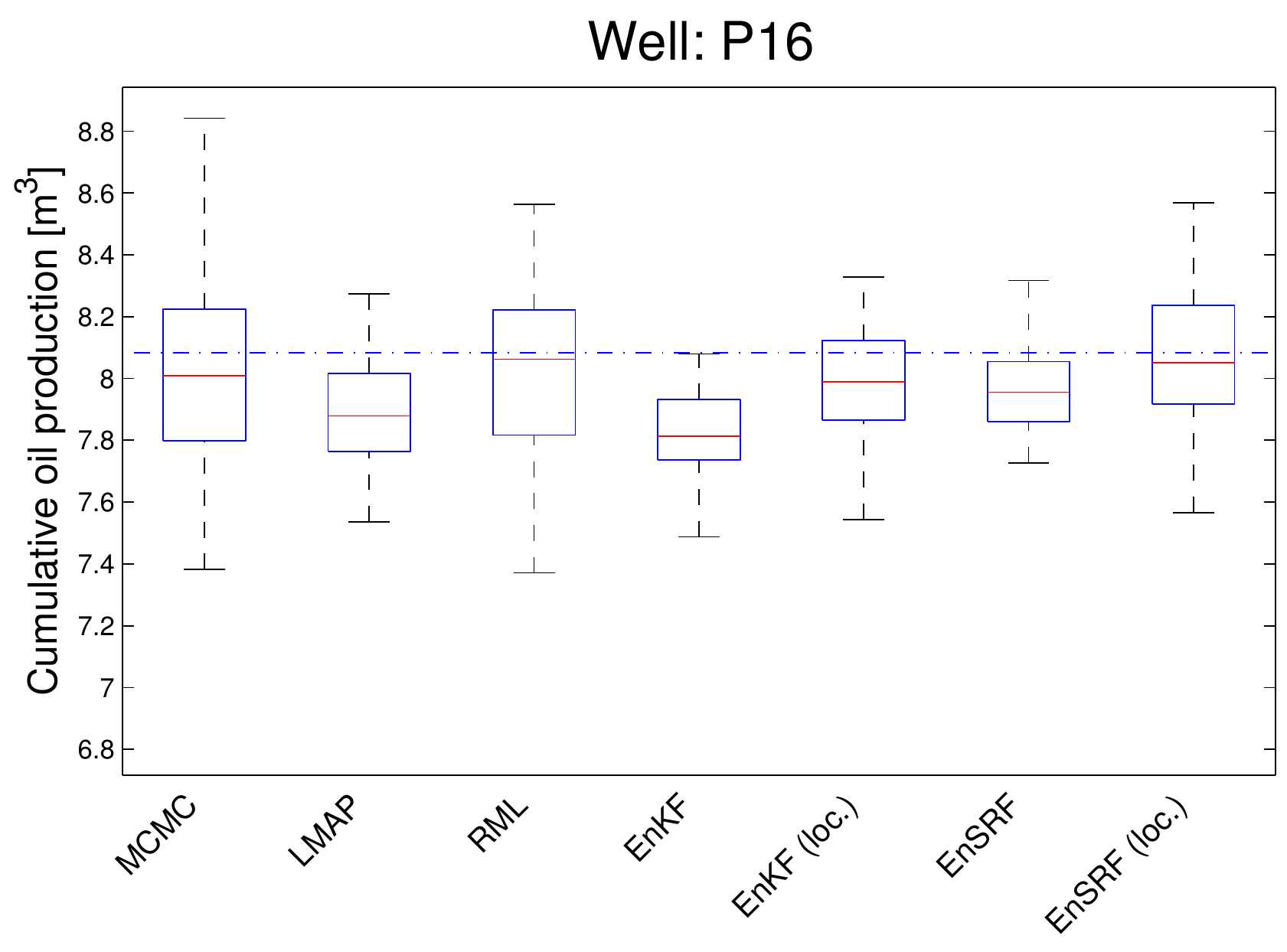}
\caption{Two-phase model (large number of wells). Distribution of cumulative oil production at wells $P_{7},P_{10},P_{12},P_{14},P_{16}$ at the final time of simulation $t=10~ \textrm{years}$}  
\label{Figure19}
\end{figure}

Among all the ensemble methods with $N_{e}=50$, RML provides the best approximation in terms of mean. Note that the approximation provided by $N(u_{MAP},C_{MAP})$ provides the best approximation in terms of combined mean and variance. Additionally, even with localization both EnKF and EnSRF provide a very poor approximation in terms of mean and variance. It is worth mentioning that RML and LMAP provided a better approximation (in terms of mean and variance) than the ensemble Kalman filter-type methods for $N_{e}=50$. In Figure \ref{Figure18} we show the total flow rates (from $P_{1}$ and $P_{5}$) and bottom-hole pressure (from $I_{1}$) simulated with permeabilities from the prior (first row), the posterior (second row), and some of the Gaussian approximations under analysis (third-sixth row). The vertical line divides the assimilation from the prediction. In this case, prediction is performed by simulating an additional $3.5$ years under the same well configuration. The red curve is computed from the posterior mean at the corresponding location. In Figure \ref{Figure19} we display the distribution of the final time cumulative oil production simulated from the posterior and the Gaussian approximation. The poor performance of EnKF and EnSRF is reflected in the poor performance at  characterizing the predicting distribution.

As we mentioned earlier, limitations of the EnKF and EnSRF arise when large number of measurements are assimilated. In the present work we are interested in the associated detrimental effect on the approximation of the posterior distribution. In order to observe that effect, we now consider application of our Gaussian approximation on a larger ensemble $N_{e}=250$. These results are presented in the second part of Table \ref{Table4}. Note that for $N_{e}=250$ the ratio of total number of wells to ensemble size is the same as 
in the previous experiment ($N_{e}=50$ and $N_{w}=5$). For $N_{en}=250$, Table \ref{Table4} indicates that RML provides again the best approximation in terms of mean. The performance of EnKF, EnSRF and their localizations are considerably improved with respect to the ones for $N_{e}=50$. As in previous examples, EnSRF with localization provides the best approximation in terms of variance. Also similarly to the previous experiments, increasing the size of the standard EnKF does not improve the approximation in terms of variance. On the other hand, in this case the variance increases with the size of ensemble. 
This can be observed from the last part of Table \ref{Table4} where EnKF was implemented for $N_{en}=1000$, $N_{en}=3000$, $N_{en}=6000$ and $N_{en}=18500$.  Note that the computational cost of EnKF for $N_{en}=18500$ coincides with the cost of our implementation of RML with $N_{en}=50$.

\begin{table}
\caption{Evaluation of Gaussian approximations for the two-phase model. Case with large number of wells.}
\label{Table4}       
\begin{tabular}{lccc}
\hline\noalign{\smallskip}
Method & Relative error &  Relative error & Computational cost\\
 &  in the mean $\epsilon_{u}$& in the variance $\epsilon_{\sigma}$ &  [Forward model runs]\\
\noalign{\smallskip}\hline\noalign{\smallskip}
MCMC			 &0.000 & 0.000 &  $5.5\times 10^{7}$\\
MAP 						&0.131 & 0.165 &  $3.5\times 10^{2}$\\
LMAP ($N_{e}=50$)			 &0.179 & 0.287 & $3.5\times 10^{2}$\\
RML ($N_{e}=50$)			& 0.169& 0.307 &  $1.85\times 10^{4}$\\
EnKF ($N_{e}=50$) 		        &0.932& 0.816 &  $5.0\times 10^{1}$\\
EnKF (localization, $N_{e}=50$)  & 0.635&0.616 &$5.0\times 10^{1}$\\
EnSRF ($N_{e}=50$) 		 &0.862 & 0.658&  $5.0\times 10^{1}$\\
EnSRF (localization, $N_{e}=50$) & 0.539 &0.471 &$5.0\times 10^{1}$\\ \hline
LMAP ($N_{e}=250$)			 &0.146 & 0.190 &  $3.5\times 10^{2}$\\
RML ($N_{e}=250$)			       & 0.121& 0.231 &  $9.25\times 10^{4}$\\
EnKF ($N_{e}=250$) 		        &0.434 & 0.166 &  $2.5\times 10^{2}$\\
EnKF (localization, $N_{e}=250$)  & 0.304&0.113&$2.5\times 10^{1}$\\
EnSRF ($N_{e}=250$) 		 &0.371 & 0.110&  $2.5\times 10^{2}$\\
EnSRF (localization, $N_{e}=250$) &0.285  &0.101 &$2.5\times 10^{2}$\\ \hline
EnKF ($N_{e}=1000$) 			& 0.243& 0.101 &$1.0\times 10^{3}$\\
EnKF ($N_{e}=3000$)			 & 0.161& 0.137 &$3.0\times 10^{3}$\\
EnKF ($N_{e}=6000$)			 &0.127 & 0.148 &$6.0\times 10^{3}$\\
EnKF (large $N_{e}=18500$)			 & 0.111& 0.154 &$1.85\times 10^{4}$\\
\noalign{\smallskip}\hline
\end{tabular}
\end{table}

\clearpage
\section{Conclusions}\label{Conclusions}

The controlled experiments from the preceding sections enable us to numerically assess the performance of widely used {\em ad hoc} Gaussian approximations with respect to their ability to
correctly reproduce the mean and variance under the true 
Bayesian posterior distribution. The true posterior is obtained 
by use of the expensive, but accurate, gold-standard preconditioned Crank-Nicolson Markov Chain Monte Carlo (pCN-MCMC) method. The forward operators associated to the reservoir models under consideration are nonlinear. Therefore, even though a Gaussian prior is considered, the posterior distribution associated to each experiment is non-Gaussian. Indeed, for all our experiments a significant discrepancy between the maximum a posteriori (MAP) estimate and the mean of the posterior distribution was obtained giving clear
numerical evidence that the posteriors for the Bayesian data assimilation problems under consideration are not Gaussian. Even in the single-phase reservoir whose associated forward operator is ``less'' nonlinear (the PDE (\ref{eq:2.1}) is linear with respect to $p$), we find that the relative error in the MAP estimator with respect to the mean of the posterior was $3\%$. For the two-phase reservoirs, this relative error was $27\%$ and $13\%$ in the case of small and large number of wells, respectively. 
Thus the problems that we study demonstrate a range of
deviations from Gaussianity in the posterior. This makes them
a suitable range of test problems for the {\em ad hoc}
algorithms, all of which can be systematically derived
in the linear Gaussian scenario, but whose accuracy in
the non-Gaussian case is, in general, unclear.

We clearly observe substantial differences in the approximation properties of the posterior distribution with respect to the
choice of method, reservoir model and well configuration. For 
all the experiments conducted here we conclude that, among all the Gaussian approximations under analysis, the linearization around the MAP (LMAP) is arguably
the best technique at reproducing the posterior distribution in terms of combined variance and mean. It is interesting to speculate why this might be so, and what it tells us about the posterior.
The LMAP algorithm is the only Gaussian approximation which 
samples from $N(u_{MAP},C_{MAP})$ where $u_{MAP}$ is the MAP estimate and $C_{MAP}$ the associated covariance matrix defined by (\ref{eq:4.1}) and (\ref{eq:4.2}), respectively. This suggests that,
out of all the Gaussian approximations considered, the
posterior distribution in all our examples can be best
approximated, in terms of mean and variance, by $N(u_{MAP},C_{MAP})$. We emphasize that this does not imply that the posterior distribution is Gaussian. Indeed, although this may be the
best approximation, errors may still be large.

We recall that all the techniques described in Section \ref{Gaussian} produce samples of the posterior distribution in the linear-Gaussian case. In other words, they sample from the exact posterior distribution $N(u_{MAP},C_{MAP})$. In our experiments, however, we observe clear differences in the approximations obtained with each of the techniques under consideration. For example, note from all our experiments that the randomized maximum likelihood (RML) provided the best approximation of the posterior in terms of the mean. In addition, in the case of single-phase, RML provides a reasonable approximation of the posterior variance (like the one obtained with LMAP). It is worth mentioning that favorable RML results for single-phase reservoirs are also reported in \cite{Oliver2}. In contrast, for the two-phase model we find examples where the error of the RML variance is the largest compared to other Gaussian approximations. These
observations are likely to be related to the higher nonlinearity in the two-phase forward operator (\ref{eq:2.14}) that results from the nonlinear PDE system (\ref{eq:2.7})-(\ref{eq:2.7B}). Due to the aforementioned higher nonlinearity, large changes in the absolute values of the log permeability field  may not necessarily correspond to large changes in the production data. In fact, production data may typically have smaller sensitivity to the permeability values far from (or in-between) the well locations. On the other hand, we recall from the RML algorithm, that each ensemble member $u_{RML}^{(j)}$ (see equation (\ref{eq:4.4})) produces a model output $G(u_{RML}^{(j)})$ that is close to the perturbed data $y^{(j)}$ while keeping $u_{RML}^{(j)}$ close to the corresponding sample from the prior $u^{(j)}$. Due to the aforementioned small sensitivity of production data to the log-permeability in some regions of the domain, it may be possible that the penalty term $\vert\vert u_{RML}^{(j)} - u^{(j)}\vert\vert$ in (\ref{eq:4.4}) may not provide sufficient constraint to avoid possible large values of $\vert  u_{RML}^{(j)} \vert$ in the aforementioned regions for which the (perturbed) production data is minimally affected by large value of permeability. Although in our controlled experiment LMAP outperformed the RML in terms of combined mean and variance, it has been reported that RML has the
advantage of approximating multimodal distribution for which $N(u_{MAP},C_{MAP})$ and therefore LMAP is suboptimal. The assessment of techniques where multimodal posterior distribution arises deserves further investigation; however it has not formed part of our studies which have been confined to problems with unimodal posteriors. Moreover, we recall that the computational cost of RML can be amortized if each ensemble member is computed in parallel. Therefore, the cost of the parallel implementation of RML equals the cost of LMAP. Note also that, for very large problems, the factorization of $C_{MAP}$ used in  (\ref{eq:4.3}) may be computationally prohibitive while the covariance of RML is computed directly from the ensemble at a negligible cost.

For each of our experiments, very poor approximations of the posterior distribution are obtained with the ensemble Kalman filter (EnKF) with a small ensemble size. However, covariance localization leads to a
significant reduction in the relative error of the mean and variance with respect to the posterior. Note for example that, in the two-phase model with small number of wells, localization reduces the relative error in the mean and the variance by a factor of two. Additionally, the  ensemble Kalman smoother (EnSRF) provides better approximations of the posterior (in terms of mean and variance) than the ones obtained with EnKF. Furthermore, in all our experiments, the ensemble generated with EnSRF with localization provides the best approximation of the posterior in terms of variance. The advantage of using covariance localization as well as using EnSRF instead of EnKF has been widely investigated in terms of reconstructing the truth and/or recovering the truth within the confidence intervals provided by the ensemble approximations. Our results offer now numerical evidence of the advantage of using covariance localization and square root filters for reconstructing the posterior distribution. The choice of covariance localization that provides optimal approximation of the posterior distribution must be further investigated.

Reducing the detrimental effect of sampling error due to the small ensemble size and the possible large amount data  is essential in practical applications where a small number of ensemble members is required to avoid high computational cost in data assimilation. Nonetheless, our results indicate that even for a large ensemble size where presumably sampling error issues are attenuated, we find that EnKF does not converge to the posterior distribution. In fact, as the ensemble sizes increased, the converged Gaussian approximation provided by EnKF resulted in errors of at least $10\%$ both in mean and variance. In addition, the approximations provided with those large size ensembles do not coincide with the approximations provided by either LMAP and RML.

In summary, our study sheds light on various aspects of the
{\em ad hoc} Gaussian approximate filters used in practice
to approximate high dimensional posterior distributions
on geological reservoir properties. The study has been made
possible by use of a fully resolved gold-standard MCMC 
computation which allows for a clear and well-founded
evaluation of the {\em ad hoc} algorithms. In our opinion
more evaluations of this kind will be beneficial in guiding
the future evolution of the {\em ad hoc} Gaussian approximate filters that are so widely used in practice.

\begin{acknowledgements}

MI, KJHL and AMS  gratefully acknowledge the support of EPSRC, ERC, ESA and  ONR for various aspects of this work.

\end{acknowledgements}

\bibliographystyle{plain}
\bibliography{MCMC_bib}   

\end{document}